      [1999/12/01 v1.4c Il Nuovo Cimento]

\documentclass[varenna]{cimento}

\usepackage{graphicx}  
\usepackage{epstopdf}
\usepackage{amsbsy}
\usepackage{amssymb}
\usepackage{amsmath}

\title{Weak gravitational lensing}
\author{H.~Hoekstra}
\institute{Leiden Observatory, Leiden University, Leiden, The Netherlands}
\begin{document}

\maketitle

\begin{abstract}
  Luminous tracers have been used extensively to map the large-scale
  matter distribution in the Universe. Similarly the dynamics of stars
  or galaxies can be used to estimate masses of galaxies and clusters
  of galaxies. However, assumptions need to be made about the
  dynamical state or how well galaxies trace the underlying dark
  matter distribution. The gravitational tidal field affects the paths
  of photons, leading to observable effects. This phenomenon,
  gravitational lensing, has become an important tool in cosmology
  because it probes the mass distribution directly. In these lecture
  notes we introduce the main relevant quantities and terminology, but
  the subsequent discussion is mostly limited to weak gravitational
  lensing, the small coherent distortion of the shapes of distant
  galaxies by intervening structures. We focus on some of the issues
  in measuring accurate shapes and review the various applications
  of weak gravitational lensing, as well as some recent results.
\end{abstract}

\section{Introduction}

The small density fluctuations in the early Universe grow into the
structures we see today under the influence of gravity.  However, most
of the matter responsible for structure formation remains
undetectable: the visible baryonic matter makes up only $\sim 16\%$ of
the total matter, the rest being dark matter. As the dark matter is
believed to be collisionless and interacts only through gravity, we
are fortunate that the process of structure formation can be simulated
fairly well using ever larger cosmological N-body simulations. Note
that this implicitly assumes that the role of baryons can be ignored,
which may not always be the case, especially for future projects. The
situation is reversed when we consider {\it observations} of the
large-scale structure in the Universe: we are forced to rely on
baryonic tracers, such as galaxies or clusters of galaxies to trace
the large-scale structure. Although these are all dark matter
dominated systems, their observable properties are heavily affected by
highly non-linear and complex baryon physics. To match the observed
abundances of galaxies and the X-ray properties of groups, AGN
feedback has been suggested as an important process. Such energetic
feedback affects both the mass function of halos and the matter power
spectrum, thus complicating the interpretation of observations using
N-body simulations as a reference.

A more direct way to relate observations to our model of structure
formation would therefore be very useful. Fortunately this is possible
thanks to the bending of light rays by massive structures. Such
inhomogeneities in the matter distribution perturb the paths of
photons that are emitted by distant galaxies. The result is equivalent
to viewing these sources through a medium with a spatially varying
index of refraction: the images appear slightly distorted and
magnified.  This phenomenon is called ``gravitational lensing''
because of the similarity with geometric optics. The amplitude of the
distortion provides a direct measure of the gravitational tidal field,
independent of the nature of the dark matter or the dynamical state of
the system of interest. The lensing signal can be used to reconstruct
the (projected) matter distribution, which is particularly useful for
the study of clusters of galaxies, as they are dynamically young and
often show signs of merging. In addition, it is possible to trace the
gravitational potential out to large radii, which is typically not
possible using dynamical methods.

If the deflection of the light rays is large enough, multiple images
of the same source can be observed: these strong lensing events
provide precise constraints on the mass on scales enclosed by these
images. The discovery of a strongly lensed QSO by \cite{Walsh79}
marked the start of a new area of research, which has grown
tremendously in recent years: in 1980 only 27 papers mentioned
``gravitational lensing'' in their abstracts, compared to 351 in 2012.
Another important milestone was the discovery of strong lensing by a
cluster of galaxies reported by \cite{Soucail87}, which is now seen
routinely, in particular in deep observations with the Hubble Space
Telescope (HST). Thanks to the large magnification provided by gravitational
lensing, such deep high-resolution observations of galaxy clusters
provide us with a unique view of the most distant galaxies.

The effects of gravitational lensing are not limited to small scales
or to high density regions. At large radii the tidal field causes a
subtle change in the shapes of galaxies, resulting in a coherent
alignment of the sources that can be measured statistically.  As the
changes in the observed images are small, it is commonly referred to
as weak gravitational lensing. The measurement of these alignments
over a large area of the sky allows for the study of the statistical
properties of the matter distribution which in turn can be used to
constrain cosmological parameters. Other applications include the
study of dark matter halos around galaxies and the study of the mass
distribution in galaxy clusters. Consequently an increasing fraction
of papers focuses on weak lensing: 2 of the 27 papers in 1980 contain
``weak lensing'' in the abstract compared to 204 out of 351 in 2012.

In these lecture notes we provide a short introduction to
gravitational lensing, with the main aim to introduce the
terminology. The focus of these notes will be on weak gravitational
lensing and its main applications. We refer readers that wish a more
thorough introduction to several excellent, comprehensive reviews.
For instance, \cite{Bartelmann01} introduces the theory of weak
gravitational lensing as well as the key cosmological parameters and
discusses the basics of shape measurements. However, the text is now
over decade old, and some of the most recent insights and results are
not discussed. \cite{Bartelmann10} derives the expressions for
gravitational lensing starting from the equation of geodesic deviation
and reviews the various applications, including microlensing, which is
not discussed in our notes. \cite{Schneider06} provides an excellent
recent introduction into the topic.

\section{Gravitational Lensing}\label{sec:lensing}

Light rays are deflected when they travel through an inhomogeneous
medium (following Fermat's principle). In geometric optics the
quantity that determines the change in the photon's path is the index
of refraction. In the case of gravitational lensing the situation is
very much the same, except that the gravitational potential replaces
the role of the index of refraction. In a cosmological setting we make
several additional assumptions. First of all, we will assume that the
gravitational field is weak (i.e. we are not considering the bending
of light rays around black holes or close to neutron stars) and that
the angles by which light rays are deflected are small. Finally, we
consider the ``thin lens'' approximation: the deflection occurs on
scales that are much smaller than the size of the Universe, i.e., the
thickness of the lens is much smaller than the distance between the
observer, lens, and source. This is a very reasonable assumption as
galaxy clusters (the most massive objects that we study) extend at
most a few Megaparsec, whereas the sources are Gigaparsecs away. We
also assume the lens is static, because the light crossing time is
short compared to the dynamical time scale.

\begin{figure}
\includegraphics[width=\hsize,bb=1 1 1100 590]{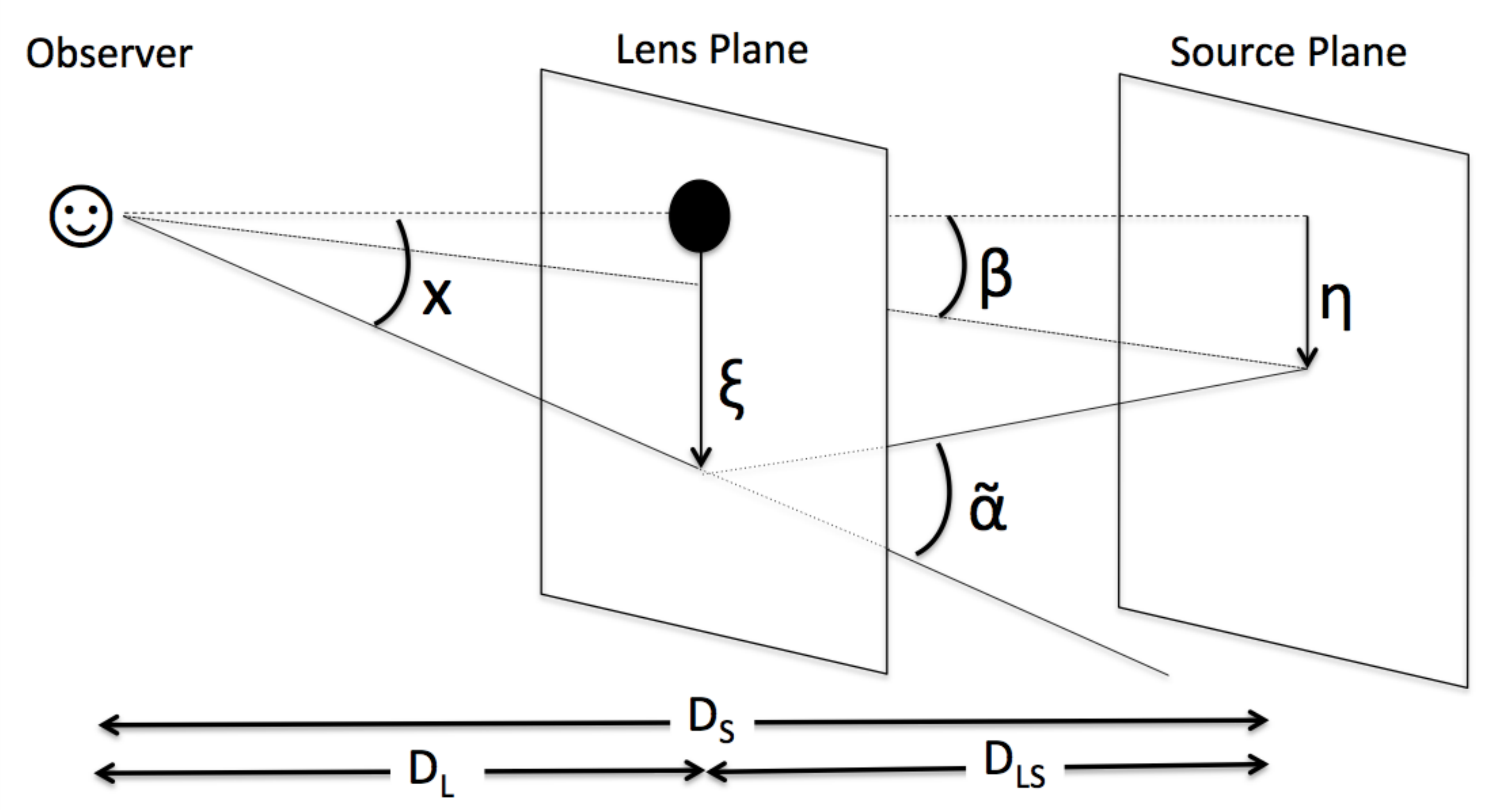}    
\caption{Geometry of the lensing problem}\label{fig:geometry}
\end{figure}

\subsection{Lens Equation}

Figure~\ref{fig:geometry} shows a schematic of the deflection of a
light ray that passes a deflector at a distance ${\mathbf \xi}$ in the
lens plane. If we project the location of the lens onto the source
plane, the source is separated by a distance ${\mathbf \eta}$. From
the Figure it is easy to see that the two positions are related
through

\begin{equation}
{\mathbf\eta}=\frac{D_S}{D_L}{\mathbf \xi} - D_{LS}{\mathbf {\tilde\alpha}}({\mathbf \xi}),
 \end{equation}
 
 \noindent where $D_S$, $D_L$ and $D_{LS}$ are the angular diameter
 distances between the observer and source, the observer and the lens,
 and the lens and the source, respectively. As we observe objects
 projected on the sky, it is more convenient to define angular
 coordinates such that ${\mathbf\eta}=D_S{\mathbf\beta}$ and
 ${\mathbf\xi}=D_L{\mathbf x}$: a source with a true position
 ${\mathbf \beta}$ will be observed at a position ${\mathbf x}$ on the
 sky. In fact, images of the source will appear at all locations
 ${\mathbf x}$ that satisfy the lens equation

\begin{equation}
{\mathbf\beta}={\mathbf x} - \frac{D_{LS}}{D_S}{\mathbf {\tilde\alpha}}(D_L{\mathbf x})
\equiv{\mathbf x} - {\mathbf \alpha}({\mathbf x}),
\end{equation}

\noindent where we defined the scaled deflection angle ${\mathbf
  \alpha}({\mathbf x})$. Although the lens equation is thus expressed
rather conveniently, the physical interpretation still requires
estimates for the lens and source redshifts. In the case of strong
gravitational lensing one can perhaps afford to obtain these through
deep spectroscopy, but in the case of weak gravitational lensing,
where the shapes of large numbers of galaxies are measured, we rely on
photometric redshift estimates instead. Although obtaining these is a
critical part for a correct interpretation of the lensing signal, in
these lectures we will discuss this issue only briefly in
\S\ref{sec:redshift}.

\subsection{Delay} As can be seen from Figure~\ref{fig:geometry} the
total path length, and thus the amount of time the deflected photon
travels, is modified. If we imagine a source that produces a burst of
light, the photons will arrive at a time

\begin{equation}
\tau=\frac{1+z_L}{H_0}\frac{D_L D_S}{D_{LS}}\left[\frac{1}{2}({\mathbf x}-{\mathbf \beta})^2-\Psi_{2D}({\mathbf x})\right],
\label{eq:tau}
\end{equation}

\noindent where the first term quantifies the increase in the path
length and the second term the ``slowing down'' as the photon travels
through the gravitational potential $\Psi_{2D}$. The latter is related
to the projected mass distribution through

\begin{equation}
\Psi_{2D}({\bf x})=\frac{1}{\pi}\int {\rm d}^2{\bf x'} \kappa({\bf x'})\ln |{\bf x-x'}|,
\end{equation}\label{eq:psi}

\noindent where the convergence $\kappa$ is the ratio of the projected
surface density $\Sigma({\bf x})$ and the critical surface density
$\Sigma_{\rm crit}$:

\begin{equation}
\kappa({\bf x})=\frac{\Sigma({\bf x})}{\Sigma_{\rm crit}},
\end{equation}
\noindent with $\Sigma_{\rm crit}$ defined as

\begin{equation}
\Sigma_{\rm crit}=\frac{c^2}{4\pi G}\frac{D_{S}}{D_{L} D_{LS}}.
\label{eq:sigmacrit}
\end{equation}

\noindent Note that, analogous to the Poisson equation, $\kappa=\frac{1}{2}\nabla^2\Psi_{2D}$.

Following Fermat's principle, images are formed at the extrema of the
light arrival surface, i.e., where $\nabla{\mathbf \tau}=0$; the
result is once more the lens equation, while demonstrating that
${\mathbf \alpha}=\nabla\Psi_{2D}$. As the actual arrival times will
typically differ between images, variations in the source brightness
will also appear at different times. Provided one has an accurate
model for the deflection potential and has measured the time delays
(as well as the redshifts for the lens and source), the remaining key
parameter in Eqn.~\ref{eq:tau} is the Hubble parameter $H_0$. Hence
time delays provide an interesting, independent approach to measure
the Hubble parameter \cite{Refsdal64,Kochanek02}. The main limitation
of this method is the uncertainty in the mass distribution. Recent
studies (e.g., \cite{Suyu13}) have taken care to model the
contributions from additional deflectors, but \cite{Schneider13}
pointed out that degeneracies in the mass model may ultimately limit
the accuracy.

\subsection{Deflection} In most applications of gravitational lensing
the actual deflection is not observed, because the true position of
the source is unknown. As a consequence only the effects of
differential deflection (=distortion) can be measured. Furthermore the
surface brightness is preserved. However, CMB photons are also
deflected and as a result the observed temperature $\tilde T({\bf x})$
is in fact (see e.g., \cite{Lewis06}, for a review):

\begin{equation}
\tilde T({\bf x})=T({\bf x}+\nabla\Psi)\approx T({\bf x})+\nabla\Psi \cdot \nabla T({\bf x}).
\end{equation}

Hence gravitational lensing changes the statistical properties of the
temperature fluctuations, resulting in a smoothing of the acoustic
peaks in the power spectrum (e.g.,
\cite{Seljak96,Lewis06}). Furthermore, the measurement of the 4-point
function can be used to measure the lensing signal directly and map
the mass distribution, albeit on very large scales
(e.g.,\cite{Bernardeau97}). This signal has been recently detected at
in ground-based experiments (\cite{SPTlens,
  ACTlens}) and by {\it Planck} from space (\cite{Plancklens}).

\subsection{Differential Deflection}
\label{sec:shear}

Analogous to the case of the CMB sky the observed surface brightness
of a distant galaxy is remapped. If the deflection angle and its
spatial variation are small (compared to the extent of the source),
this mapping can be linearized:

\begin{equation}
f^{\rm obs}({\bf x})=f^{\rm s} [{\mathbf \beta}({\bf x})]\approx f^{\rm s}({\cal A}{\bf x}),
\end{equation}

\noindent with $f^{\rm s}({\bf x})$ the unlensed surface brightness
and $\cal A$ the Jacobian of the transformation.  This regime is
commonly referred to as ``weak gravitational lensing''. The distortion
matrix can be written in terms of derivatives of the deflection angle,
and thus in terms of second derivatives of the deflection potential:

\begin{equation}
{\cal A}=\delta_{ij}-\frac{\partial^2 \Psi}{{
\partial x_i\partial x_j}}=
\left(
    \begin{array}{cc}
        1-\kappa-\gamma_1 & -\gamma_2 \\
        -\gamma_2        & 1-\kappa+\gamma_1 \\
    \end{array}
\right)=(1-\kappa)
\left(
    \begin{array}{cc}
        1-g_1 & -g_2 \\
        -g_2        & 1+g_1 \\
    \end{array}
\right),
\label{eq:distort}
\end{equation}

\noindent where we introduced the complex shear ${\bf \gamma}\equiv\gamma_1+i\gamma_2$, and defined the reduced shear $g_i=\gamma_i/(1-\kappa)$. The shear is related to the deflection potential through

\begin{equation}
\gamma_{1}=\frac{1}{2}\left(\frac{\partial^2 \Psi}{\partial x_1^2}
-\frac{\partial^2 \Psi}{\partial x_2^2}\right)\hspace{1em}{\rm and}\hspace{1em}
\gamma_2=\frac{\partial^2\Psi}{\partial x_1\partial x_2},
\label{eq:shear}
\end{equation}

\noindent The effect of the remapping by ${\cal A}$ is to transform a
circular source into an ellipse with axis ratio $\sim (1-|g|)/(1+|g|)$
and position angle $\alpha=0.5\arctan(g_2/g_1)$; the reduced shear
describes the anisotropic distortion of a source. If $\kappa\ll 1$
(i.e., the weak lensing regime), the observable $g_i\sim \gamma_i$. In
addition, the source is magnified by a factor

\begin{equation}
\mu=\frac{1}{\det \cal{A}}=\frac{1}{(1-\kappa)^2-|\gamma|^2}=
\frac{1}{(1-\kappa)^2(1-|g|^2)},
\end{equation}

\noindent boosting the observed flux by the same amount. To first
order, the magnification depends on the convergence only; i.e.
$\kappa$ describes the isotropic distortion of a source (contraction
or dilation). Both the shearing and magnification of sources are
observable effects, although both are quite different in terms of
techniques and systematics. The measurements are, however, statistical
in nature, because the signal cannot be inferred from individual
sources.
 
The measurement of the shear involves measuring the shapes of galaxies
and we will discuss some of the practical issues below. The effect of
weak gravitational lensing is to change the unlensed ellipticity of a
galaxy, $\epsilon_{\rm orig}=(a-b)/(a+b)$, where $a$ is the semi-major
axis, and $b$ the semi-minor axis, to an observed value:

\begin{equation}
\epsilon^{\rm obs}=\frac{\epsilon^{\rm orig}+g}{1+g^*\epsilon^{\rm orig}}\approx\epsilon^{\rm orig}+\gamma,
\end{equation}

\noindent where the asterisk denotes the complex conjugate. Only if
the shear is larger than $\epsilon^{\rm orig}$ does the observed
ellipticity of a single galaxy provide a useful estimate of the
shear. A population of intrinsically round sources would therefore be
ideal, but unfortunately real galaxies have an average intrinsic
ellipticity of $\sim 0.25$ per component (e.g., \cite{Leauthaud07}).
Instead the lensing signal is inferred by averaging over an ensemble
of sources, under the assumption that the unlensed orientations are
random. Although this assumption has been adequate up to now, the
increased precision of future weak lensing studies requires that the
contribution from intrinsic alignments is taken into account.

Intrinsic alignments are expected to arise because of tidal torques,
or alignments of the angular momentum, when galaxies form during the
collapse of a filament. For instance it has been known for a while
that the central galaxies in clusters point towards each other
(\cite{Binggeli82}). In general one expects the orientation of central
galaxies to be correlated with the large scale mass distribution they
formed in. Early numerical simulations also show strong alignments
between halos (e.g., \cite{Croft00, Heavens00}), albeit with
amplitudes that are already ruled out by observations. This highlights
the difficulty in predicting the alignment signal, because the outer
regions of a halo are more easily aligned than the inner,
baryon-dominated regions, which are what we actually observe. For
instance initial alignments between the angular momentum of baryons
and the halo can be erased through the violent process of galaxy
formation. Hence, the study of intrinsic alignments, although a source
of bias for cosmic shear studies, provides a unique probe of baryon
physics. Although further progress is expected from hydrodynamic
numerical simulations, direct observational constraints are
needed. This requires precise redshift information for the galaxies in
question. Fortunately high-quality photometric redshifts can also be
used to this end.

The intrinsic alignment signal has been detected for early type
galaxies (e.g., \cite{Mandelbaum06, Hirata07,
  Joachimi11,Heymans13}). The alignment of late type galaxies is
believed to arise from alignments in the angular momentum, and has
proven more difficult to constrain observationally. Finally the local
torques may align satellite galaxies within a halo such that they
preferentially point towards the host (\cite{Pereira08}). Some
tentative detections of such radial alignment have been reported (see
\cite{Schneider13b} for a discussion of results), but significant
progress is expected in the coming years thanks to the combination of
large imaging surveys with adequate spectroscopic observations.

The intrinsic alignments between galaxies that are physically
associated can be readily reduced by avoiding correlating the shapes
of galaxies at similar redshifts and using cross-correlations between
galaxies with different redshifts instead. However, an additional
consequence of intrinsic alignments was found by \cite{Hirata04}. It
arises because galaxy shapes may be correlated with the surrounding
density field (e.g. the radial alignments of satellite galaxies
mentioned above). As this density field is also the source of the
lensing signal the combination of a radial alignment at the lens
redshift and a tangential alignment of the sources, leads to a
suppression of the lensing signal. As this leads to anti-correlations
between shapes of galaxies at different redshifts it cannot be easily
corrected for. The signal can be suppressed, but only with a great
loss in precision. Therefore the current approach is to model the
intrinsic alignment as part of the cosmological analysis (e.g.,
\cite{Bernstein09, Joachimi10}).
 
As is the case for the shear, we cannot measure the actual
magnification of a single object because the intrinsic flux and size
are typically unknown. Instead, the signal can be inferred from the
change in the source number counts, which is determined by the balance
between two competing effects. On the one hand the actual volume that
is surveyed is reduced, because the solid angle behind the cluster is
enlarged. On the other hand the fluxes of the sources in this smaller
volume are boosted, thus increasing the limiting magnitude. As a
consequence, the net change in source surface density depends not only
on the mass of the lens, but also on the steepness of the intrinsic
luminosity function of the sources. If it is steep, the increase in
limiting magnitude wins over the reduction in solid angle, and an
excess of sources is observed. If the number counts instead are
shallow, a reduction in the source number density is observed.

One interesting recent application of magnification is the study of
the lensing signal around high-redshift lenses using Lyman break
galaxies (LBGs) as sources at $z>3$. Although these galaxies are too
faint and too small to have their shapes determined accurately, their
magnitudes can still be measured. The feasibility of this approach was
demonstrated by \cite{Hildebrandt09} who studied the signal around
lenses at intermediate redshifts. Magnification studies are
particularly interesting to study high redshift lenses, because the
number density of sources for which shapes can be measured decreases
with redshift (especially for ground-based data). For instance,
\cite{Hildebrandt11} measured the mass of $z\sim 1$ clusters of
galaxies and recently \cite{Hildebrandt13} were able to constrain the
masses of a sample of distant submillimetre galaxies.
 
\subsection{Mass reconstructions}

The shear and convergence can both be derived from the second
derivatives of the deflection potential. If we consider the definition
of $\Psi_{2D}$ given by Eqn.~\ref{eq:psi} and compute the shear
following Eqn.~\ref{eq:shear} we find that the shear can be written as
a convolution of the convergence with a kernel $\chi({\bf x})$:

\begin{equation}
\gamma({\bf x})=\frac{1}{\pi}\int {\mathrm d}^2{\bf x'}\chi({\bf x-x'}) \kappa({\bf x'}),
\label{eq13}
\end{equation}

\noindent where the convolution kernel $\chi({\bf x})$ is given by

\begin{equation}
\chi({\bf x})=\frac{x_2^2-x_1^2-2ix_1 x_2}{|{\bf x}|^4}.
\label{eq14}
\end{equation}

It is also possible to express the surface density in terms of the
observable shear by inverting this expression (as was first shown by
\cite{KS93}). The resulting "mass reconstruction" $\hat\kappa$ is
given by

\begin{equation}
\hat\kappa({\bf x})-\kappa_0=\frac{1}{\pi}\int {\mathrm d}^2{\bf x'} \chi^*({\bf x-x'})\gamma({\bf x'}),\label{eq:ks}
\end{equation}

\noindent where the constant $\kappa_0$ shows that the surface density
can only be recovered up to a constant.  This reflects the fact that a
constant $\kappa$ does not cause a shear. In fact one can show that a
transformation $\kappa'=\lambda\kappa+(1-\lambda)$ leaves the reduced
shear unchanged. This complication is known as the mass-sheet
degeneracy (\cite{Gorenstein88}).  Note that $\kappa({\bf x})$ is a
real function, which may not be the case for the actual mass
reconstruction. A non-negligible imaginary component of $\hat\kappa$
is a sign of systematics in the data and thus can be used to identify
problems with the data analysis.

The possibility to reconstruct the surface mass density is clearly an
important application of weak gravitational lensing: it allows us to
directly map the total mass distribution, which is particularly
interesting in the case of merging clusters of galaxies. Note,
however, that Eqn.~\ref{eq:ks} cannot be used in practice as
evaluating it requires data out to infinity. Additional complications
are the fact that shear is only sampled at the locations of the
sources and the fact that we observe the reduced shear rather than the
shear itself. Several solutions have been proposed, such as finite
field inversions (e.g., \cite{Seitz96,Squires96}) or maximum
likelihood methods (e.g., \cite{Bartelmann96, Seitz98}). A question
that has renewed interest in this area is whether the statistics of
the mass reconstruction can be used for cosmological parameter
estimation. For instance \cite{Waerbeke13} detected the second-,
third- and fourth-order moment of the distribution of the convergencem
map.  Although comparison with simulations suggested fair agreement,
more work is needed to understand the biases that might arise.

\begin{figure}
\centering
\hbox{%
\includegraphics[width=0.53\hsize]{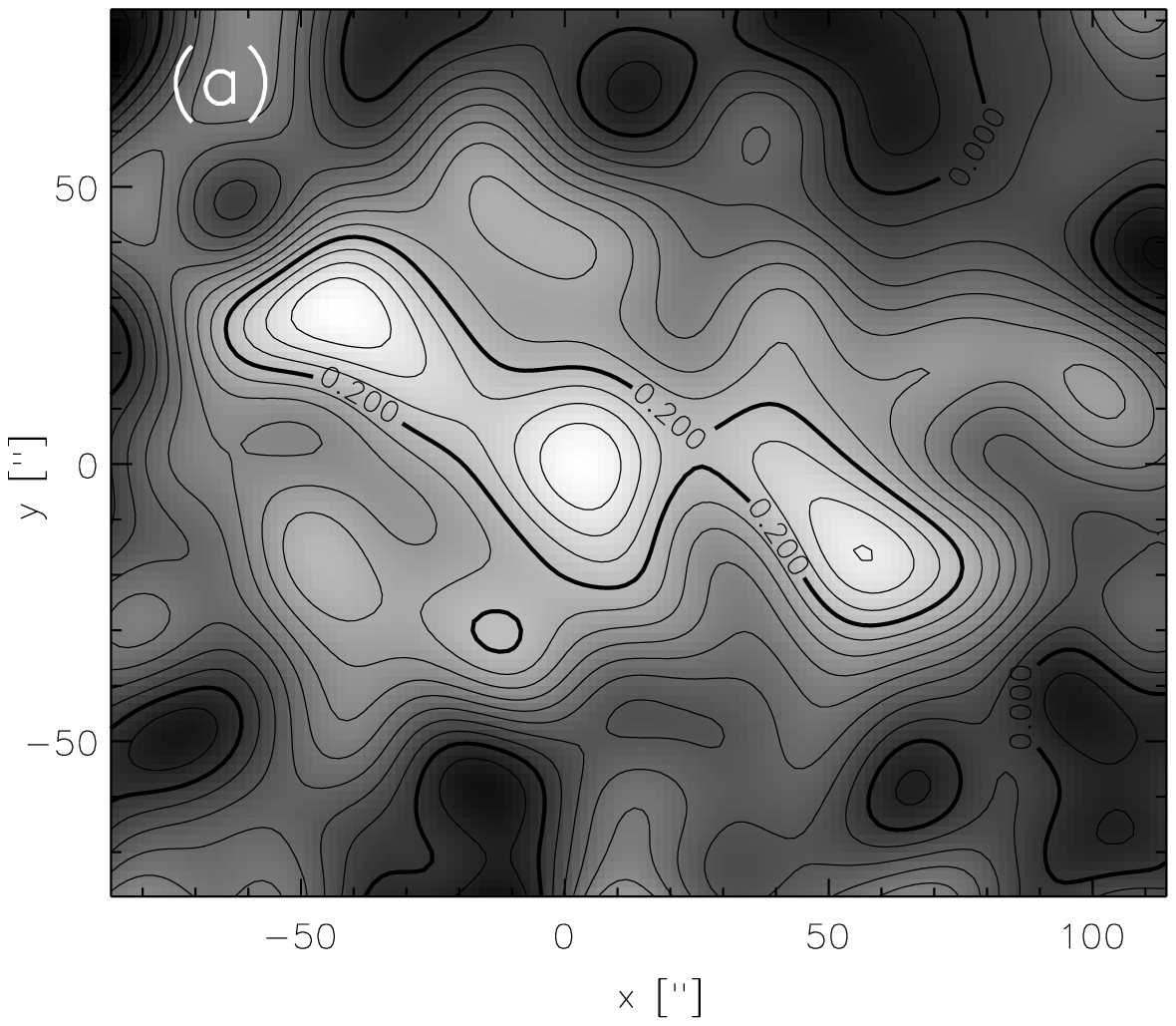}    
\includegraphics[width=0.47\hsize]{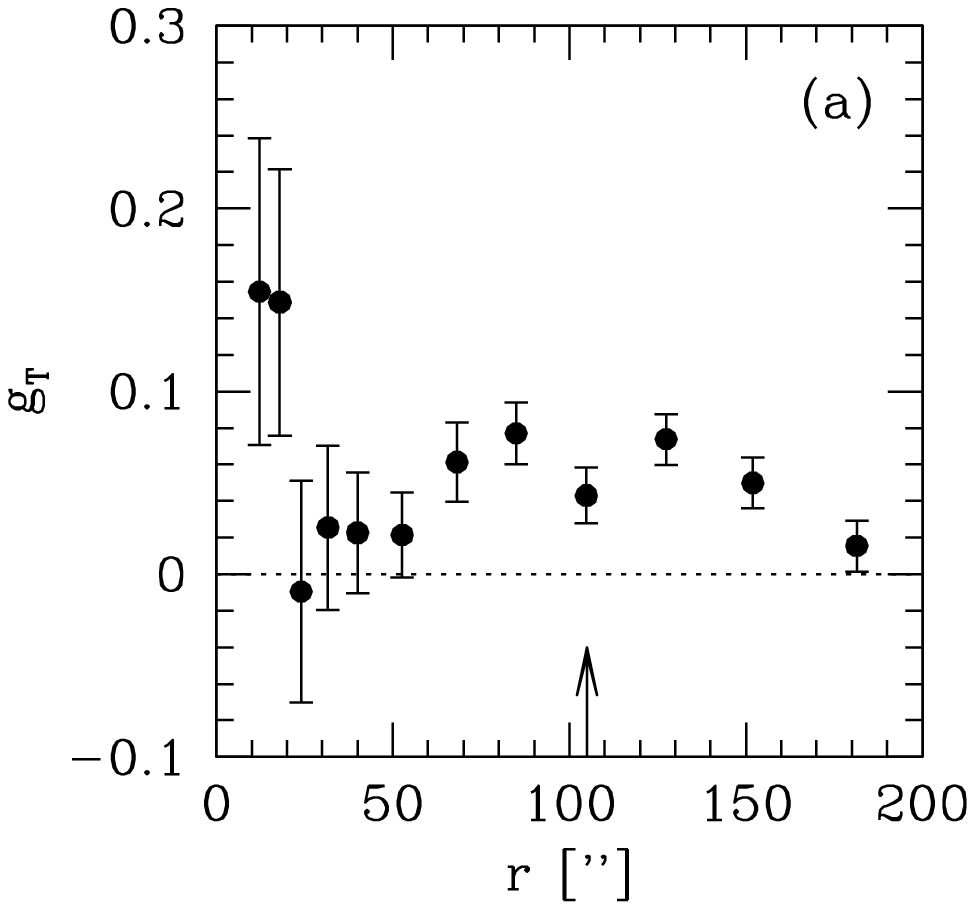}}
\caption{{\it Left panel:} mass reconstruction of the central region of the $z=0.83$ cluster of galaxies
MS1054-03 from \cite{Hoekstra00} demonstrating that the complex mass distribution is not well described by a simple parametric model. {\it Right panel:} the azimuthally averaged tangential distortion as a function of distance to the cluster
center. The drop in signal at $r~20-50''$ before rising again is the result of the complex mass distribution seen in the mass reconstruction.}\label{fig:massrec}
\end{figure}

Figure~\ref{fig:massrec} shows an example of the usefulness of a mass
reconstruction: the central regions of the $z=0.83$ cluster MS1054-03
show significant substructure based on a weak lensing analysis of deep
HST observations by \cite{Hoekstra00}. It is clear that the mass
distribution cannot be described by a simple model, nor can one assume
that the cluster is dynamically relaxed.

One of the best-known weak lensing results is the mass reconstruction
of the ``Bullet cluster'' 1ES 0657-558 by \cite{Clowe06}. This is an
advanced stage of a merger of two clusters of galaxies, in which the
lower mass cluster has already passed through the core of the main
body. The mass reconstruction, based on deep HST data, shows a
significant separation of the baryons (in the form of the hot X-ray
emitting gas) and the dark matter, as inferred from weak lensing.  Due
to its collisional nature the gas has been slowed down in the
collision, whereas the dark matter halos have passed through each
other without significant interaction. What makes this a particularly
important observation is that the significant displacement between the
baryonic mass and the peaks in the lensing map cannot be (easily)
explained without dark matter.

\subsection{Mass estimates}

As the calculation of the mass reconstruction involves smoothing the
data, the results are not ideal if one wants to measure
masses. Instead one can fit a parameterized model to the observed
shear field, but as we saw, this may not be accurate in the case of
merging systems. It is, however, possible to obtain an estimate of the
projected mass within an aperture, with relatively few assumptions
about the actual mass distribution. It makes use of the fact that for
{\it any} mass distribution the azimuthally averaged tangential shear
$\langle\gamma_T\rangle$ as a function of radius from the cluster
center measures a contrast in surface density

\begin{equation}
\langle\gamma_T\rangle(r)=\frac{\bar\Sigma(<r)-\bar\Sigma(r)}{\Sigma_{\rm crit}}=\bar\kappa(<r)-\bar\kappa(r),
\end{equation}

\noindent where the tangential shear is defined as

\begin{equation}
\gamma_T=-(\gamma_1\cos 2\phi + \gamma_2 \sin 2\phi),
\end{equation}

\noindent where $\phi$ is the azimuthal angle with respect to the
lens. For this reason the tangential shear is a very convenient way to
represent the lensing signal around a single massive lens (e.g. a
galaxy cluster) or a sample of lenses (galaxies or groups). In
addition to the tangential shear, one can also define a cross shear
$\gamma_\times$, which is equivalent to rotating sources by 45 degrees
and measuring the corresponding tangential shear. The azimuthally
averaged $\gamma_\times$ should vanish in the absence of systematics
(similar to the imaginary part of the mass reconstruction discussed
earlier).

Due to the substructure in the central regions of MS1054-03, the
resulting tangential shear, shown in the right panel of
Figure~\ref{fig:massrec}, does not decline monotonically.  Hence the
tangential shear profile cannot always be interpreted by fitting
simple parameterized models, such as an isothermal sphere or NFW
profile, to the data.  Similarly at large radii the lensing signal
around a sample of galaxies is sensitive to their clustering
properties. A correct interpretation thus becomes more involved,
although the halo-model has been quite successful in this regard (see
\S\ref{sec:halo model}). In the case of galaxy clusters one can use
aperture mass statistics, such as the $\zeta_c$ estimator proposed by
\cite{Clowe98}

\begin{equation}
\zeta_c(r_1)=2\int_{r_1}^{r_2}d\ln r\langle\gamma_t\rangle+
\frac{2r_{\rm max}^2}{r_{\rm max}^2-r_2^2} \int_{r_2}^{r_{\rm max}}
d\ln r \langle\gamma_t\rangle,
\end{equation}
 
\noindent which can be expressed in terms of the mean dimensionless
surface density interior to $r_1$ relative to the mean surface density
in an annulus from $r_2$ to $r_{\rm max}$:

\begin{equation}
\zeta_c(r_1)=\bar\kappa(r'<r_1)-\bar\kappa(r_2<r'<r_{\rm max}).
\end{equation}

Although the surface density in the outer annulus should be small for
results based on wide field imaging data, it cannot be completely
ignored. Its value can be estimated using parametric models resulting
in a rather weak model dependence of the projected mass estimate. The
resulting estimate for the projected mass in the aperture is thus
rather robust, but comparison to other (baryonic) proxies for the mass
is typically not as straightforward, as this does require assumptions
about the geometry of the cluster.

\section{Measuring galaxy shapes}

The typical change in ellipticity due to gravitational lensing is much
smaller than the intrinsic shape of the source, even in the case of
clusters of galaxies. Although this can be dealt with by averaging the
shapes of many galaxies, the shear signal can be overwhelmed by
instrumental effects, which may be difficult to assess on an
object-by-object basis. Hence the study of algorithms that can
accurately determine the shapes of faint galaxies has been a major
part of the development of weak gravitational lensing as a key tool
for cosmology.

\begin{figure}
\centering
\hbox{%
\includegraphics[width=\hsize]{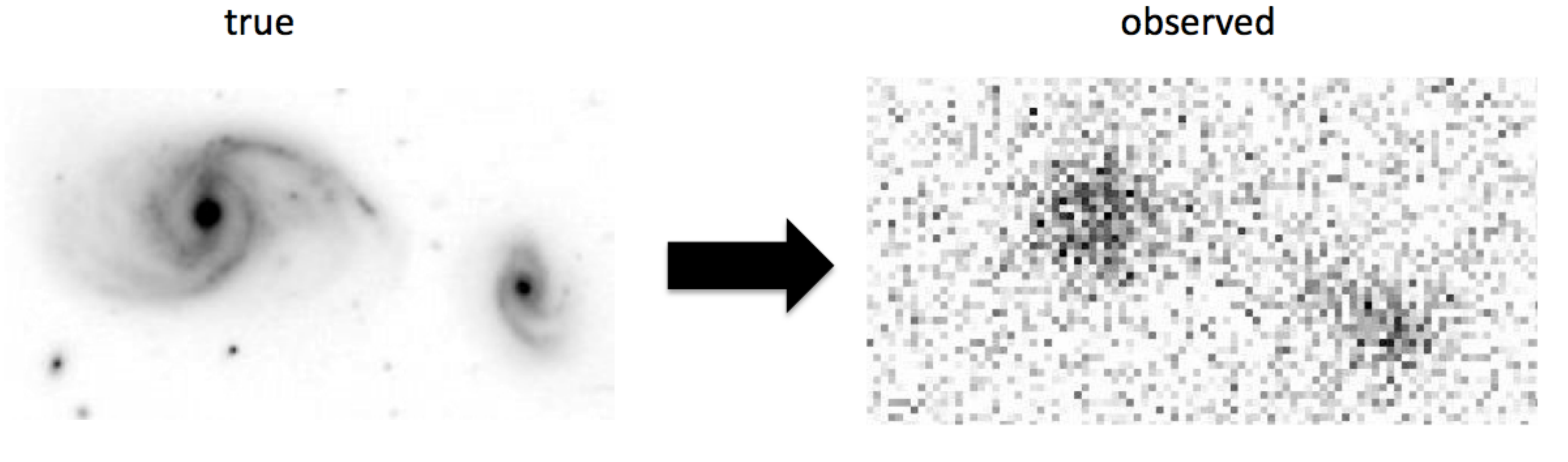}}
\caption{We need to infer accurate information about the shape of the
  true surface brightness distribution (left) from images that have
  been corrupted by various sources of bias, such as pixellation,
  seeing and noise. Given a good description of the instrumental
  effects it is possible to simulate their effects and thus examine
  the performance of shape measurement
  algorithms.}\label{fig:observations}
\end{figure}

The problem is highlighted by comparing the "true" image of an object
to the "observed" version shown in Figure~\ref{fig:observations}. The
main source of bias that needs to be corrected for is the blurring of
the images by the point spread function (PSF). Unless the pixels are
large with respect to the PSF, the pixellation is not a major source
of concern. As it is easier to measure properties when the noise is
low, the S/N ratio is another key parameter determining how well
shapes can be measured. In the case of space-based observations, due
to the combination of low sky background and radiation damage, charge
transfer inefficiency may also be an important effect (e.g.,
\cite{Massey13}).

One approach to recover the true galaxy shapes is to adopt a suitable
model of the surface brightness distribution. An estimate of the
lensing signal is obtained by shearing the model and convolving it
with the PSF, comparing the result to the observed image until a best
fit is found. A model for the PSF is typically obtained by analyzing
the shapes of a sample of stars in the actual data.  An important
advantage of this approach is that instrumental effects can be
incorporated in a Bayesian framework. As the modeling requires many
calculations and thus is computationally expensive, the use of
model-fitting algorithms has only recently become more prominent. Some
examples of this approach are {\tt lensfit} (\cite{Miller13}) which
was used to analyze the CFHTLenS data, and {\tt im3shape}
(\cite{Zuntz13}). There are challenges as well: the model needs to
accurately describe the surface brightness of the galaxies, while
having a limited number of parameters in order to avoid fitting the
noise. A model that is too rigid will lead to model bias (e.g.,
\cite{Bernstein10}), as does a model that is too flexible.

Galaxy shapes can also be quantified by computing the moments of the
galaxy images. Such methods have been applied to data extensively,
especially the method proposed by \cite{KSB}. The shapes can be
quantified by the polarization

\begin{equation}
e_1=\frac{I_{11}-I_{22}}{I_{11}+I_{22}},~{\rm and}~e_2=\frac{2 I_{12}}{I_{11}+I_{22}},\label{eq:polarization}
\end{equation}

\noindent where the quadrupole moments $I_{ij}$ are given by

\begin{equation}
I_{ij}=\frac{1}{I_0}\int {\rm d}^2{\mathbf x}~x_i x_j W({\mathbf x}) f({\mathbf x}),
\end{equation}

\noindent where $f({\mathbf x})$ is the observed galaxy image,
$W({\mathbf x})$ a suitable weight function to suppress the noise and
$I_0$ the weighted monopole moment. It is also convenient to define
$R^2=I_{11}+I_{22}$ as a measure of the size of the galaxy. Both
model-fitting and moment-based methods are used to measure the weak
lensing signal and further development is ongoing. In these notes we
continue with a closer look at the use of moments, because it is
somewhat easier to see how the results are impacted by instrumental
effects.

\subsection{Observational distortions}

The observed moments are changed by the blurring of the PSF: the PSF
has a width which leads to rounder images and typically is
anisotropic, which leads to a preferred orientation. If that were not
enough, noise in the images leads to additional biases. The various
sources of bias can be grouped into two kinds: a multiplicative bias
$m$ that scales the shear, and an additive bias $c$ that reflects
preferred orientations that are introduced. The observed shear and
true shear are thus related by (e.g., \cite{STEP1}):

\begin{equation}
\gamma_i^{\rm obs}=(1+m)\gamma_i^{\rm true}+c,
\end{equation}

\noindent where we implicitly assumed that the biases are the same for
both shear components. The additive bias is a major source of error
for cosmic shear studies because the PSF patterns can overwhelm the
lensing signal. Studies of clusters and galaxies use the tangential
shear averaged using many lens-source pairs, and much of the additive
biases tend to average away. As we discuss below it is possible to
test how well the correction for additive bias has performed, but the
estimate of the multiplicative bias requires image simulations.

If it were possible to ignore the effects of noise in the images, we
could use unweighted moments. In this case the correction for the PSF
is straightforward as the corrected moments are given by

\begin{equation}
I^{\rm true}_{ij}=I^{\rm obs}_{ij}-I^{\rm PSF}_{ij},
\end{equation}

\noindent i.e. one only needs to subtract the moments of the PSF from
the observed moments. The result provides an unbiased estimate of the
polarization, but to convert the result into a shear still requires
knowledge of the unlensed ellipticity (distribution), although this
could be established iteratively from the data.

From a pure statistical perspective it is more efficient to image
large areas of the sky rather than take deep images of smaller regions
(e.g., \cite{Amara07}). Hence the images of the sources are typically
noisy and unweighted moments cannot be used. The optimal estimate is
obtained by matching the weight function to the size (and shape) of
the galaxy image. However, the correction for the change in moments
due to both the weight function and the PSF is no longer simple, but
involves higher order moments of the surface brightness distribution,
which themselves are affected by noise (e.g., \cite{Massey13,
  Semboloni13a}). Note that limiting the expansion in moments is
analogous to the model bias in model-fitting approaches.

\cite{Massey13} present a detailed breakdown of the various sources of
bias that affect weak lensing analyses and interested readers are
encouraged to examine the findings of that paper. For instance, in the
simplified case of unweighted moments, the change in the observed
ellipticity $\hat\epsilon$ due to small errors in the PSF size
$(\delta R^2_{\rm PSF})$ or PSF ellipticity $(\delta\epsilon_{\rm
  PSF})$ can be expressed as

\begin{equation}
\hat{\epsilon}_{\rm gal} \approx \epsilon_{\rm gal} +\frac{\partial\epsilon_{\rm gal}}{\partial(R^2_{\rm PSF})}
\delta(R^2_{\rm PSF})+
\frac{\partial\epsilon_{\rm gal}}{\partial\epsilon_{\rm PSF}}\delta\epsilon_{\rm PSF},
\end{equation}

\noindent which can be written as

\begin{equation}
\hat{\epsilon}_{\rm gal} \approx \left[1+\frac{\delta(R^2_{\rm PSF})}{R^2_{\rm gal}}\right]\epsilon_{\rm gal} -
\left[{\frac{R^2_{\rm PSF}}{R^2_{\rm gal}}\delta\epsilon_{\rm PSF}+
\frac{\delta(R^2_{\rm PSF})}{R^2_{\rm gal}}\epsilon_{\rm PSF}}\right].
\end{equation}

The first term shows the multiplicative bias caused by errors in the
PSF size, relative to the galaxy size. The second term corresponds to
the additive bias and is determined by errors in the PSF model
$\delta\epsilon_{\rm PSF}$ and residuals in the correction for the PSF
anisotropy (last term). However, the PSF is not the only source of
bias, especially when considering weighted moments and hence the
expression for the multiplicative bias (idem for the additive one)
becomes more involved when more effects are included (see
\cite{Massey13} for details). In particular new contributions arise
that are related to the correction method (method bias). As is already
clear from the expression reproduced above, a small PSF is important
in order to minimize the biases. Although small PSF anisotropy is
preferable, a good model of the PSF size and shape is critical.

\begin{figure}
\centering
\hbox{%
\includegraphics[width=0.45\hsize]{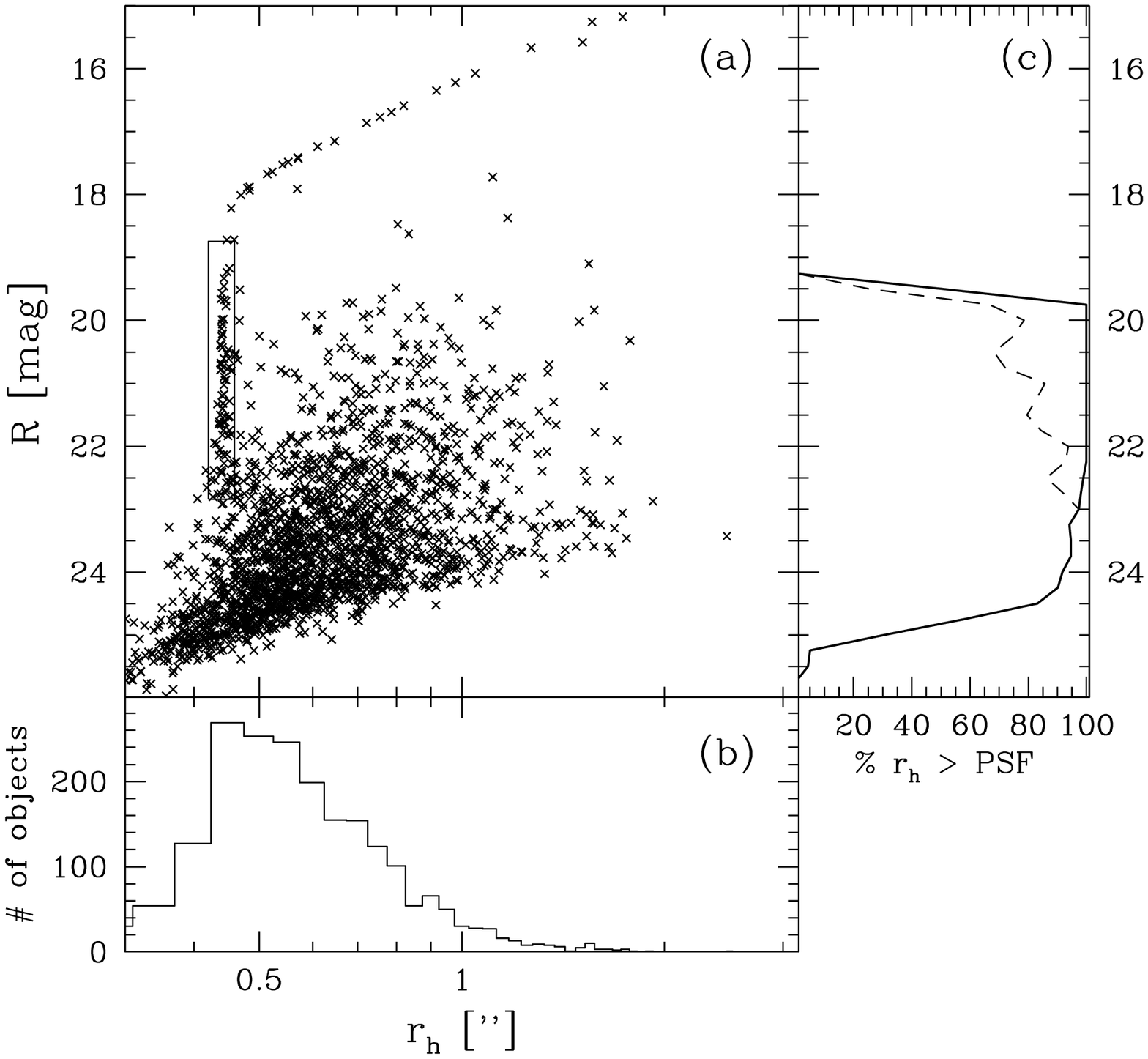}    
\includegraphics[width=0.55\hsize]{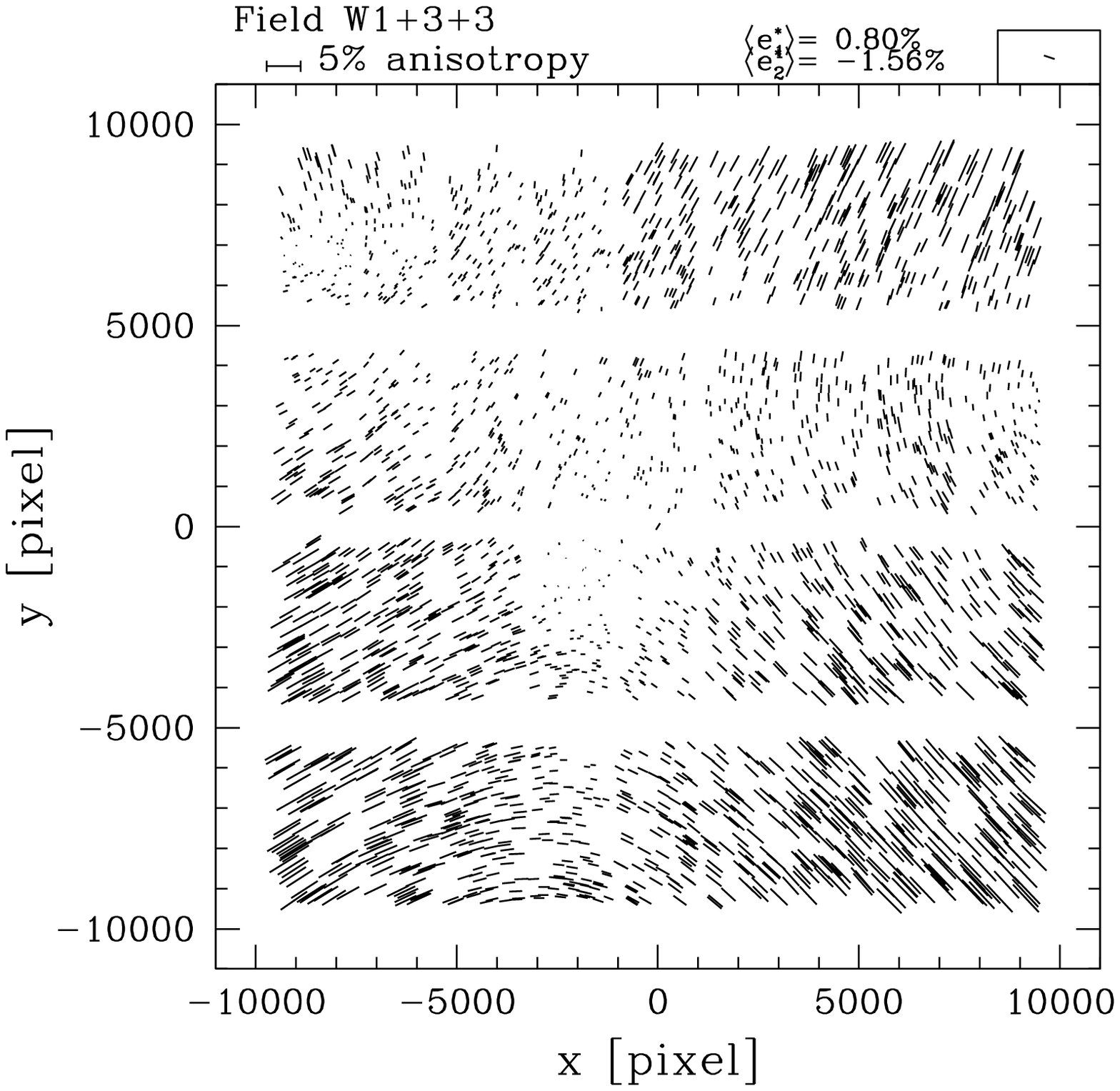}}
\caption{{\it Left panel:} Plot of the apparent magnitude versus
  half-light radius for RCS data from \cite{Hoekstra02}. The rectangle
  indicates the sample of stars that can be used to model the PSF
  variation. Brighter stars saturate and their observed sizes increase
  as can be seen as well.  {\it Right panel:} An example of the
  pattern of PSF anisotropy for MegaCam on CFHT from
  \cite{Hoekstra06}. The sticks indicate the direction of the major
  axis of the PSF and the length is proportional to the PSF
  ellipticity. A coherent pattern across the field-of-view is clearly
  visible.}\label{fig:psf}
\end{figure}

\subsection{PSF model}

Although much effort has been spent on improving the correction for
the PSF, without an accurate model for the spatial variation of the
PSF, the resulting signal will nonetheless be biased (e.g.,
\cite{Hoekstra04a}). In ground-based data the PSF changes from
exposure to exposure due to changing atmospheric conditions and
gravitational loads on the telescope. The PSF of HST observations
changes due to the change in thermal conditions as it orbits around
the Earth. If a sufficient number of stars can be identified in the
images, these can be used to model the PSF. As can be seen from
Figure~\ref{fig:psf} stars can be identified by plotting magnitude
versus size. As the stars are the smallest objects, they occupy a
clear vertical locus. If the star saturates, charge leaks into
neighboring pixels and the size increase which explains the trail
towards larger sizes, whereas the size and shape estimates become
noisy for faint stars. After selecting a sample of suitable stars (not
too bright such that they are saturated but also not too faint such
that they cannot be separated from faint, small sources) the resulting
PSF pattern can be modeled. The right panel in Figure~\ref{fig:psf}
shows an example for MegaCam on the Canada-France-Hawaii Telescope
(CFHT), which shows a coherent pattern across the field-of-view.

Most studies to date fit an empirical model to the measurements of a
sample of stars to capture the spatial variation. As the PSF pattern
is determined by (inevitable) misalignments in the optics, the overall
pattern varies relatively smoothly. However, to efficiently image
large areas of sky, observations are carried out using mosaic
cameras. For instance, Megacam on CFHT consists of 36
chips. Misalignments and flexing of the chips will lead to small
additional PSF patterns on the scale of the chips. A single low-order
model fit to the full focal plane cannot capture these small scale
variations.  Instead a low-order (typically second-order) polynomial
is used for each chip, but this may lead to over-fitting on small
scales due to the limited number of stars per chip. \cite{Hamana13}
examined whether a model based on the typical optical distortions can
be used to model the PSF pattern of the Subaru telescope (also see
\cite{Schechter11}). The global pattern, which varies from exposure to
exposure, can indeed be described fairly well. By combining many
observations, one can also try to account for the misalignments of the
individual chips.
 
To obtain deeper images and to fill in the gaps between the chips,
exposures are dithered and combined into a stack. As the observing
conditions typically vary between exposures, the combined PSF pattern
becomes very complicated (especially at the location of the chip
gaps). It is important to account for this, for instance by modeling
the PSF of each exposure and keeping track which PSFs contribute to
the stack at a particular location \cite{Jee13}. Alternatively one
can model each exposure, which is the approach used in
\cite{Miller13}.

The true shapes of galaxies should not correlate with the PSF pattern,
although chance alignments of the shear and the PSF may occur. This
enables an important test of the fidelity of the correction for PSF
anisotropy: we can measure the correlation between the corrected
galaxy shapes and the PSF ellipticity, the star-galaxy
correlation. The detection of a significant correlation points to an
inadequate correction, which may be due to the method itself or the
PSF model.  A detailed discussion and application of this test is
presented in \cite{Heymans12}. Importantly, this test does not depend
on cosmology, while being very sensitive to one of the most dominant
sources of bias in cosmic shear studies. In the analysis of the
CFHTLenS data, presented in \cite{Heymans12} this was used to identify
and omit fields that showed significant systematics.

\subsection{Image Simulations}

\begin{figure}
\centering
\hbox{%
\includegraphics[width=\hsize,bb=35 250 550 730,clip]{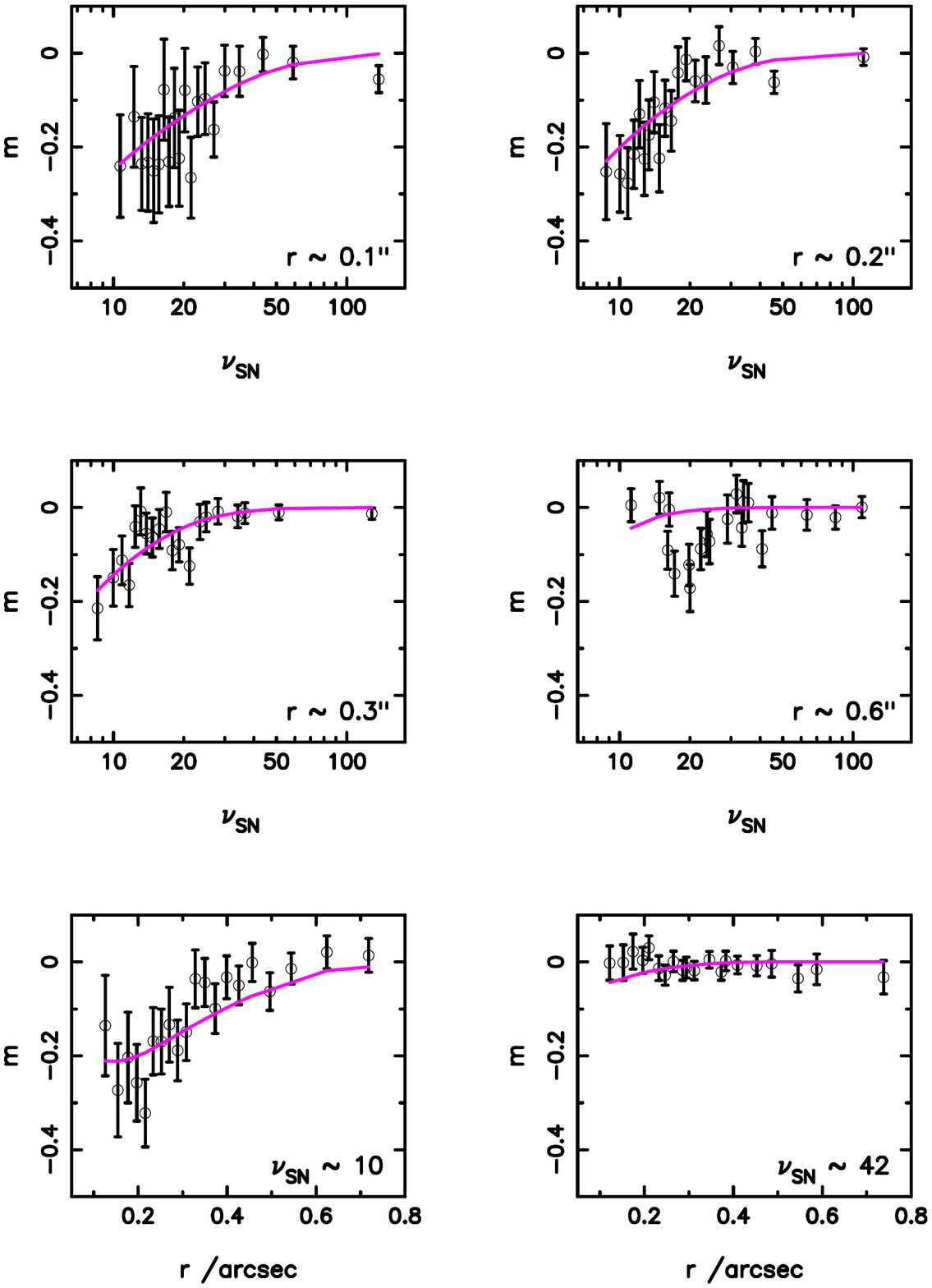}}
\caption{Multiplicative bias $m$ as a function of signal-to-noise
  ratio $\nu_{\rm SN}$ for galaxies with different sizes measured by
  {\tt lensfit} from \cite{Miller13}. The solid curves are the result
  of a fitting function matched to a wide range in size and
  S/N.}\label{fig:miller}
\end{figure}

The star-galaxy correlation function is a sensitive test of the
correction for additive bias, but no such test exists for the
multiplicative bias. Instead the performance of an algorithm needs to
be determined by applying it to simulated data where the input shear
is known. As an example Figure~\ref{fig:miller} shows the recovered
multiplicative bias $m$ for {\tt lensfit} from \cite{Miller13}.  The
bias increases as the signal-to-noise ratio $\nu_{\rm SN}$
decreases. The bias also depends on the size of the of the sources,
with a larger bias for smaller objects. These results can be used to
find an empirical correction as a function of the object properties,
thus reducing the bias to an acceptable level.

Several community-wide blind studies have been carried out to provide
a bench-mark for the performance of weak lensing studies. The first of
these was the Shear TEsting Programme (STEP;\cite{STEP1}), which
simulated observations using simple models for the galaxies. This was
followed by STEP2 (\cite{STEP2}) using synthetic galaxies based on HST
data. These results showed a range in multiplicative biases, with some
methods finding biases as small as $\sim 1\%$. These initiatives were
followed by GREAT08 (\cite{GREAT08}) and GREAT10 (\cite{GREAT10}) which
were more systematic investigations of the sources of bias.

As the bias depends on the galaxy properties, in particular the S/N
and size, it is important that image simulations match the
observations. If this is not the case, the resulting bias may not be
representative for the actual data. For instance the lack of faint
galaxies in STEP1 leads to an underestimate of the bias compared to
that of actual data. For instance new studies find that one needs to
include galaxies at least 1.5 magnitudes fainter than the limiting
magnitude of the source sample. As mentioned earlier, the bias in
shape measurement algorithms also depends on the intrinsic ellipticity
distribution (also see \cite{Melchior12}). Hence, in order to examine
the performance of the algorithms, the fidelity of the simulations
themselves is critical. With the more stringent requirements of future
experiments, this is an important area of development.

\section{Lensing by Clusters}

The first attempt to detect a weak lensing signal was made by
\cite{Tyson84} who stacked a sample of galaxies, with shapes measured
from photographic plates. However, it was the advent of CCD cameras
with their improved performance and the possibility to correct for
systematics that allowed for the first detections of the lensing
signal. As the amplitude of the lensing signal is proportional to the
mass of the lens, galaxy clusters provided a natural target for these
pioneering studies. The first detection of the lensing signal, around
the massive cluster Abell~1689, was presented by \cite{Tyson90}. The
cluster studies carried out in the '90s demonstrated the feasibility
of weak lensing and thus paved the way for the other applications
of weak lensing, in particular the cosmic shear.

Cluster samples are increasing rapidly thanks to large surveys at
various wavelengths that aim to use the evolution in the number
density of clusters to constrain cosmological parameters. To do so,
however, requires the calibration of the scaling relations between the
observed properties and the cluster masses. As numerical simulations
cannot (yet) capture the full effects of the complex baryon physics,
and dynamical methods are biased, weak lensing masses provide
important information. Another application, which we already
encountered when we discussed the usefulness of mass reconstructions,
is the study of merging clusters. A recent review of lensing studies
of galaxy clusters can be found in \cite{Hoekstra13}.

\begin{figure}
\centering
\hbox{%
\includegraphics[width=0.47\hsize,bb=1 1 570 570,clip]{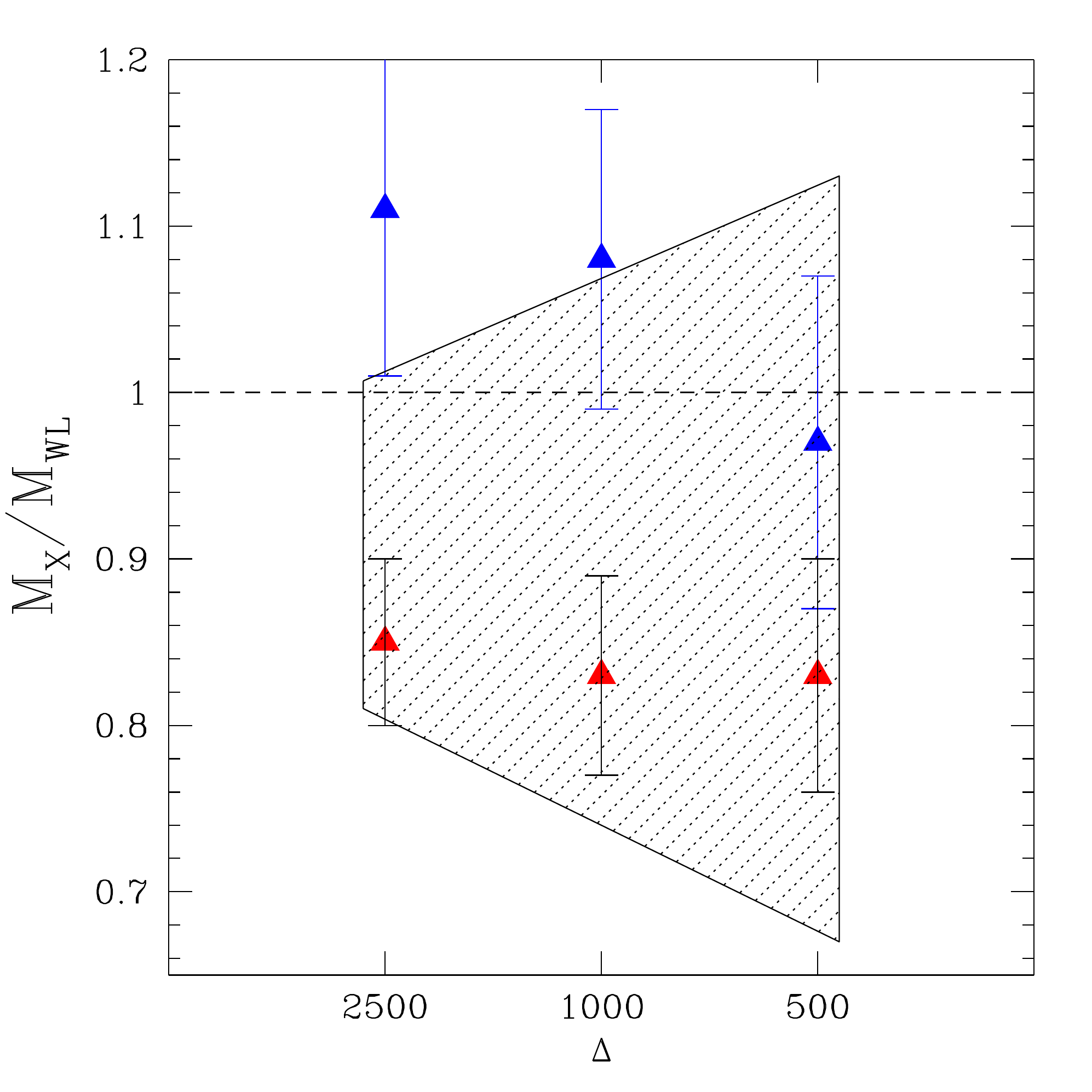}    
\includegraphics[width=0.53\hsize]{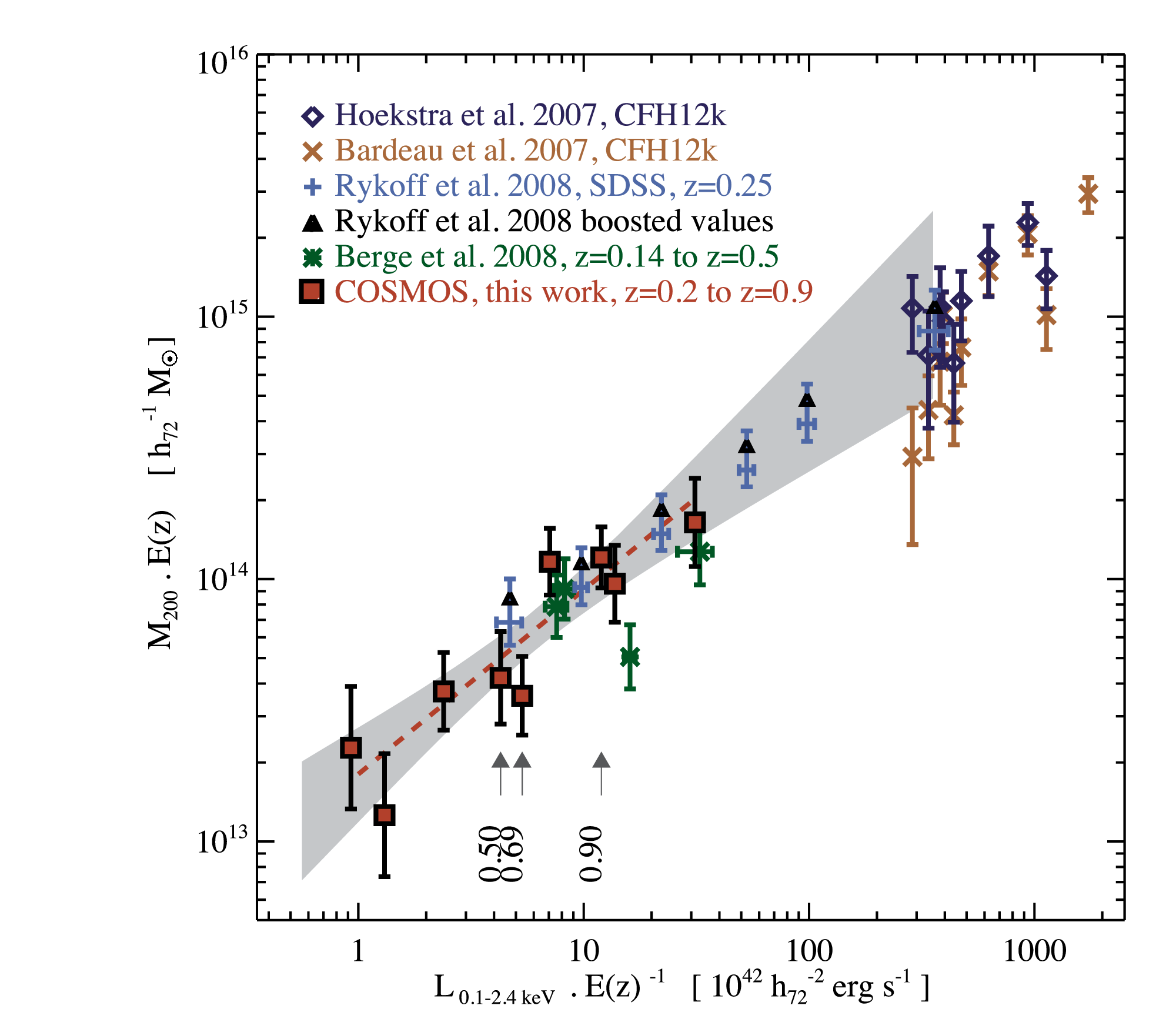}}
\caption{{\it Left panel:} the ratio of the hydrostatic mass, derived
  from X-ray observations, to the weak lensing mass as a function of
  density contrast for cool-core clusters (blue triangles) and
  non-cool-core systems (red triangles) reproduced from
  \cite{Mahdavi13}. The shaded region shows the predictions from
  numerical simulations. {\it Right panel:} from \cite{Leauthaud10}
  showing the scaling relation between lensing mass and X-ray
  luminosity for stacked galaxy groups in the COSMOS field and
  literature results for massive clusters.}\label{fig:cluster}
\end{figure}

Several studies have recently released results for samples of $\sim
50$ well-known massive clusters (\cite{Hoekstra12,Okabe10,
  Okabe13,Applegate12}).  Such large samples allow us to compare
scaling relations and examine their intrinsic scatter. These results
may help to identify shortcomings of numerical simulations. For instance
\cite{Kravtsov06} found that $Y_X=M_{\rm gas}T_X$, i.e. the product of
gas mass and X-ray temperature, showed the smallest scatter in their
hydrodynamic simulations. \cite{Mahdavi13}, however, found that the
gas mass itself has the smallest intrinsic scatter, although they note
that the scatter in $Y_X$ is the same for different subsamples of
clusters, which may be a useful property for cluster
surveys. Numerical simulations have long suggested that hydrostatic
cluster masses are biased low because of turbulent motions in the
gas. Weak lensing studies have now confirmed this, and an example is
presented in the left panel of Figure~\ref{fig:cluster}, showing the
results from \cite{Mahdavi13}.

Individual masses can only be determined for the most massive clusters
of galaxies, thus significantly limiting the mass range that can be
studied. However, the lensing signal is directly proportional to the
mass, which allows us to stack the signals of lower mass systems if we
sort these by baryonic properties. This is useful, because the scaling
relations for low mass systems are affected more by feedback
processes. The right panel of Figure~\ref{fig:cluster} shows the
results of an analysis of the stacked weak lensing signal around
groups discovered in the COSMOS field. \cite{Leauthaud10} presented a
comparison of the lensing mass and the X-ray luminosity. Complementing
the results with literature results from massive clusters, the results
suggest that a single power-law relation can describe the
observations. Stacking also allows one to trade depth against survey
area, with the SDSS results presented by \cite{Johnston07} being a
nice example; although the SDSS data are too shallow to detect a
significant signal for individual clusters, the large number of
targets allowed for precise measurements over a wide range in cluster
richness.

\section{Lensing by Galaxies}
\label{sec:ggl}

Although the large dynamical mass for the Coma cluster provided some
of the earliest indications of the existence of dark matter,
measurements of the rotation curves of spiral galaxies in the '70s
showed that galaxies must be surrounded by massive dark matter
halos. To understand how galaxies form and evolve, it is thus
important to be able to relate the baryonic properties to the total
mass (see e.g. \cite{Courteau13} for a review). However, most
dynamical methods to constrain galaxy masses are limited to relatively
small scales. To complicate matters further, those scales are baryon
dominated, thus complicating the connection to predictions from
numerical simulations. For that purpose it would be
useful to constrain the relation between the observable baryonic
properties and the virial mass.

This is possible using weak gravitational lensing, but only for
ensembles of galaxies. The study of the lensing signal induced by
galaxies is commonly referred to as ``galaxy-galaxy'' lensing. The first
attempt to measure this signal was made by \cite{Tyson84}, allowing
them to place upper limits on the galaxy masses. The first detection
was made over a decade later by \cite{Brainerd96} using deep CCD
imaging.  As the lensing signal is much lower than that of galaxy
clusters, the study of the galaxy-galaxy lensing signal benefits
tremendously from having large lens samples. Therefore the Sloan
Digital Sky Survey has made a big impact in this area of
research. Although the SDSS data are relatively shallow, the large
survey area provides the large number of lens-source pairs needed to
measure the lensing signal with high precision. Another benefit is the
availability of redshifts for a large sample of lenses. This was used
by various groups to study the relation between halo mass and stellar
mass or luminosity (e.g., \cite{Mckay01,Mandelbaum05}).

\begin{figure}
\centering
\hbox{%
\includegraphics[width=0.9\hsize]{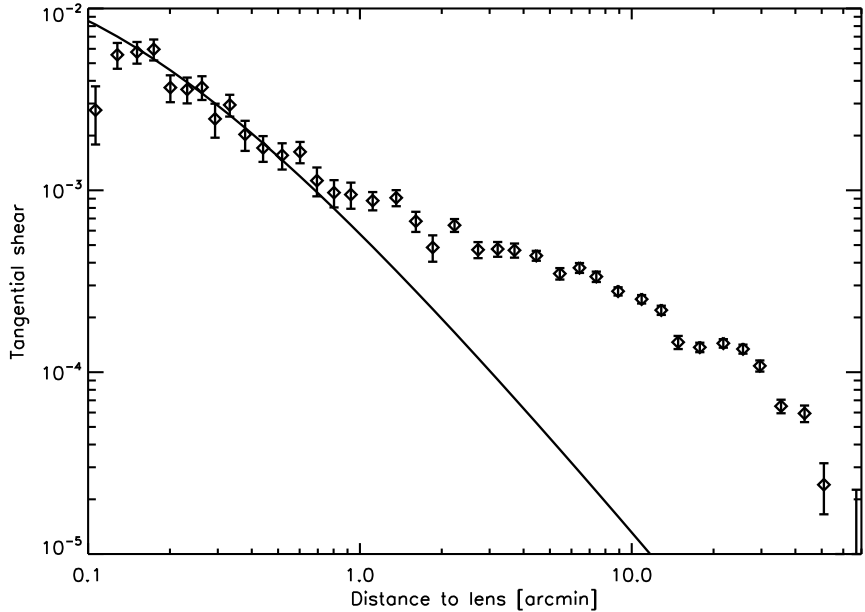}} 
\caption{The ensemble averaged tangential shear as a function of
  distance around lenses with $19.5<m_r<21.5$ using shape measurements
  of sources with $22<m_r<24$ from the RCS2 \cite{vanUitert11}.  For
  reference the best fit NFW profile is shown, which provides a good
  fit to the data on scales $<1'$ ($\sim 300$ kpc). At larger radii
  the clustering of the lenses leads to a significant increase in the
  observed lensing signal.}\label{fig:gglensing}
\end{figure}

On sufficiently small scales the lensing signal is dominated by the
dark matter halos surrounding the lenses, but on larger scales we
measure the combined signal of many halos. If lenses were distributed
randomly, these contributions would average away, but in the real
universe galaxies cluster and the interpretation of the resulting
``galaxy-mass cross-correlation function'' is more complicated. This can
be seen in Figure~\ref{fig:gglensing} (from \cite{vanUitert11}), which
shows the lensing signal around a sample of photometrically selected
lenses as a function of distance from the lens. For reference the best
fit NFW profile is shown; it describes the data well on small scales,
but on scales larger than $\sim 300$ kpc the lensing signal is much
larger than the signal from a single halo. Hence to interpret the
signal over the full range in scale it is important to be able to
include the effects of clustering.

One way to do so is to use the observed locations of the lenses and to
assume that all the mass is associated with the lenses. This allows
one to predict the shear field, which can be compared to the data
using a maximum likelihood approach, first proposed by
\cite{Schneider97}. The parameters of the adopted density profile,
such as an NFW model or a truncated isothermal sphere, can thus be
constrained. For instance, \cite{Hoekstra04b} used this approach to
find constraints on the extent of dark matter halos around
galaxies. The clustering of the lenses is naturally taken into
account, but it is more difficult to account for the expected
diversity in density profiles. For instance, in high density regions,
such as clusters or groups of galaxies, the dark matter halos of the
galaxies are stripped due to the tidal field of the host halo. We
therefore need to distinguish between central galaxies and satellite
galaxies (which are sub-halos in a larger halo). This is an important
limitation of the maximum likelihood method and galaxy-galaxy studies
now largely interpret the data using the halo model, which we discuss
in more detail below. An alternative approach, employed in
\cite{Hoekstra05} is to limit the analysis to central galaxies.

Rather than interpreting the galaxy-mass cross-correlation to
constrain the properties of dark matter halos, one can relate the
signal to that of the clustering of galaxies and the underlying dark
matter distribution. Galaxies trace the dark matter, but they are
typically biased tracers, i.e. $\delta_g(x)=b(x) \delta_m(x)$. For
instance, early type galaxies trace the highest density regions and
thus cluster more strongly than spiral galaxies. The cross-correlation
coefficient $r$ describes how well the galaxy and matter density field
are correlated. On large scales the galaxy and the dark matter
distributions are well correlated and the biasing is (close to)
linear, and the galaxy power spectrum is simply $b^2$ times the matter
power spectrum $P(k)$. Including constraints from galaxy-galaxy
lensing, which measures the product $b\times r$, it is possible to
measure the bias parameters (\cite{Hoekstra02b,Sheldon04}). Such
constraints are useful to better interpret clustering measurements.

Galaxy lensing also provides new opportunities to test gravity on
cosmological scales. \cite{Reyes10} combined SDSS galaxy lensing
measurements with clustering and redshift space distortion results
from the redshift survey to test deviations from General Relativity on
cosmological scales. The redshift space distortions are sensitive to a
combination of the growth rate and the bias, whereas the combination
of clustering and galaxy lensing can determine the bias, under the
reasonable assumption that the cross-correlation coefficient $r=1$ on large
scales. The precision of the measurement was merely limited by the
precision with which the galaxy lensing and redshift space distortions
could be determined, and further progress is expected as larger data sets
are used.

\subsection{Halo Model}

The study of the bias parameters discussed above makes use of the fact
that the galaxy-mass cross-correlation function can be expressed in
terms of the matter power spectrum (e.g, \cite{Guzik01}). This power
spectrum, and thus the lensing signal can be computed rather well
using the so-called ``halo model'' (\cite{Seljak00,Cooray02}). In this
approach we assume that the dark matter distribution can be described
by the clustering of halos that have mass-dependent density profiles.
This is supported by the results of numerical simulations that find
that dark matter halos are well-described by NFW profiles (although
more general profiles are used in the analysis of data). The
clustering properties of dark matter halos as a function of their mass
is fairly well understood. To compute the galaxy lensing signal we
also need a prescription of the way galaxies populate the dark matter
halos.

To do so, we need to distinguish between galaxies located at the
centers of the halos ({\it centrals}) and galaxies that are located in
the sub-halos throughout the main halo ({\it satellites}). For a given
sample of lenses, a fraction $\alpha$ of the galaxies is satellite,
the rest being central. Although one could try to identify these
specifically, for instance by assigning membership to groups or
clusters, most lensing studies treat the distinction statistically,
where the satellite fraction $\alpha$ is simply a parameter to be
determined. In addition to specifying the density profiles of the main
halos, we also need to make assumptions about the distribution of
satellites within the halo and their density profiles. A typical
assumption is that the galaxies trace the density distribution of the
main halo. The halos of the satellites are tidally stripped, but
observational constraints are still limited (but see e.g.,
\cite{Limousin07}). Inspired by numerical simulations, most recent
studies assume that the halos are truncated and have lost half their
mass (\cite{Mandelbaum05}).

The number of galaxies in a halo is described by the halo occupation
distribution (HOD). Typical choices are $N_{\rm gal}(M_{\rm
  halo})\propto M_{\rm halo}$ above some cute-off mass. Alternatively
one can use the conditional luminosity function $\Phi_x(L|M)$, where
$x$ indicates central or satellite galaxies (e.g., \cite{Cacciato13})

\begin{equation}
\langle N_x|M\rangle=\int_{L_{\rm min}}^{L_{\rm max}}\Phi_x(L|M){\rm d}L,
\end{equation}

\noindent for a sample of lenses with luminosities $L_{\rm
  min}<L<L_{\rm max}$. Note that selecting lenses based on observable
properties, such as luminosity and stellar mass, corresponds to a
wider range in halo masses. This is because for a given luminosity (or
stellar mass) the corresponding halo mass is described by a log-normal
distribution. As a result, the width of this distribution needs to be
specified or measured, in order to interpret the results
(e.g. \cite{Tasitsiomi04, Cacciato13}). The combined lensing signal,
which is to be compared to the data, is the sum of the contributions
from the central and satellite galaxies
$\gamma_T=(1-\alpha)\gamma_T^{\rm cen}+\alpha\gamma_T^{\rm
  sat}$. These contributions themselves consist of several terms which
we discuss now, starting with the centrals.

\begin{figure}
\centering
\hbox{%
\includegraphics[width=0.57\hsize,trim=0cm -2cm 0cm 1cm,clip=true]{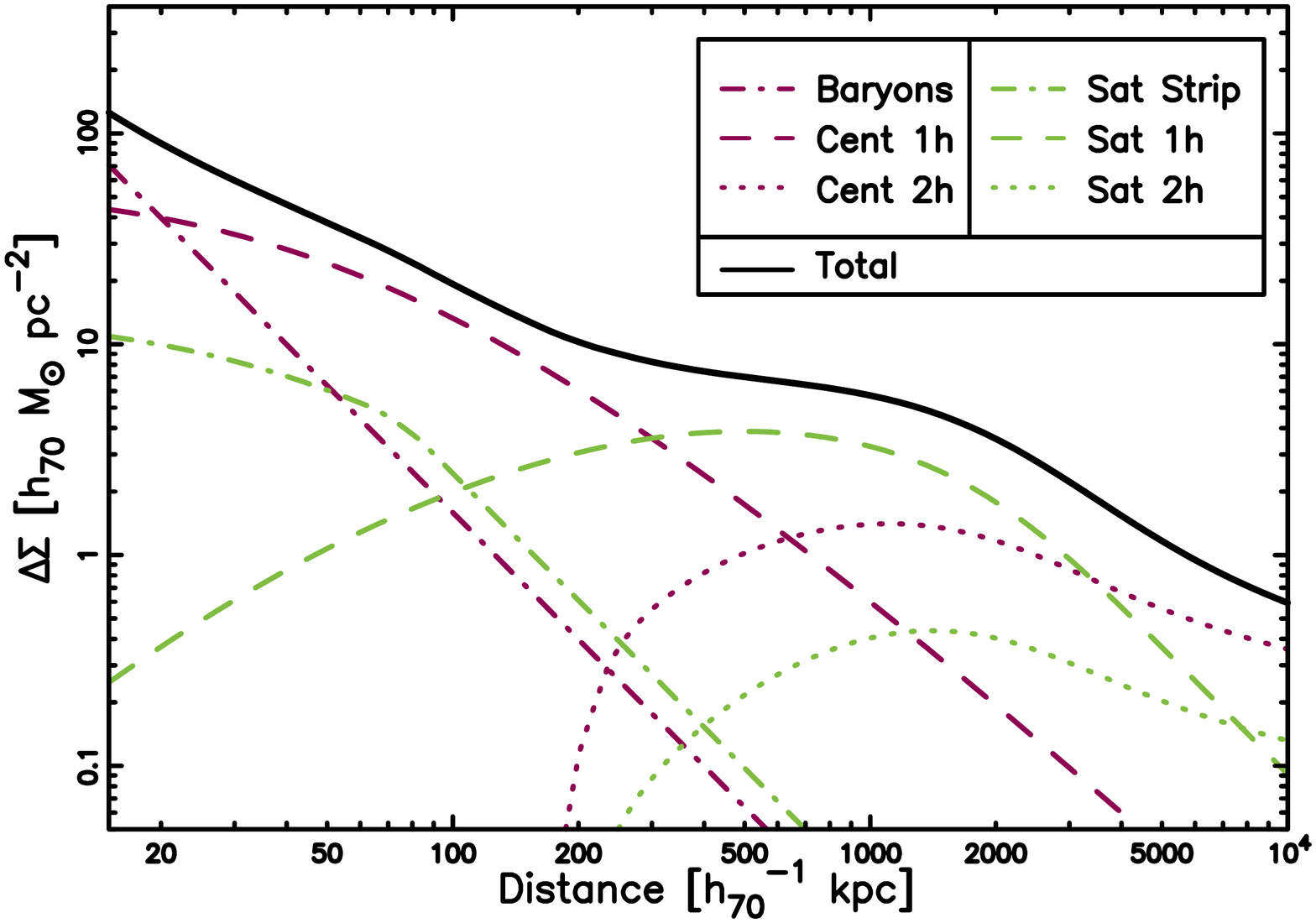}
\includegraphics[width=0.43\hsize,trim=1cm 2cm 12cm 2cm, clip=true]{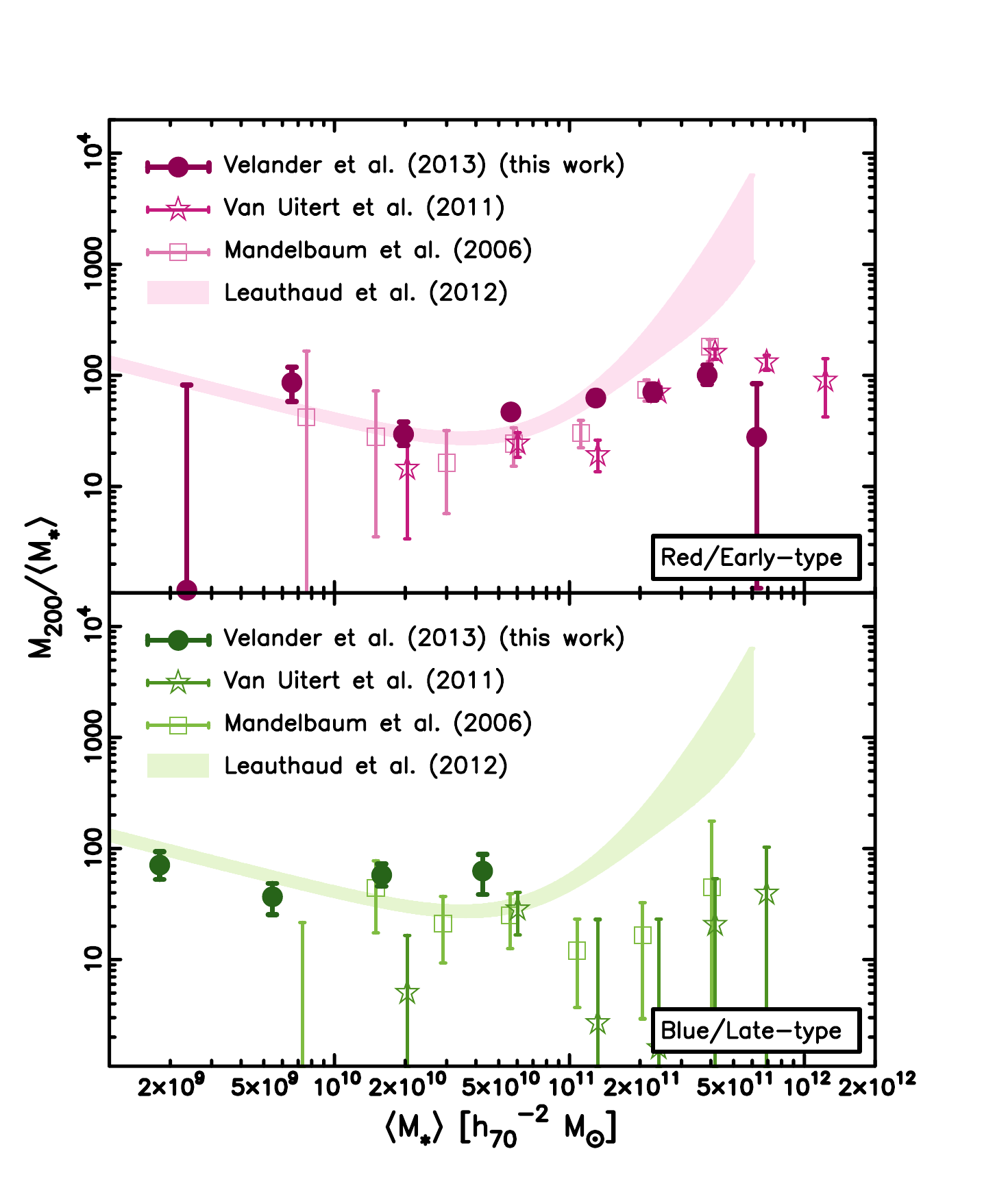}}
\caption{{\it Left panel:} Example from \cite{Velander13} showing how the various components of the halo 
contribute to the model lensing signal in the case of a halo at $z=0.5$ with mass $M_{200}=10^{12}M_\odot$,
a stellar mass of $5\times 10^{10}M_\odot$ and a satellite fraction $\alpha=0.2$. The purple lines indicate
quantities that correspond to centrals and the green lines are for satellites. {\it Right panel:} Comparison
of lensing mass as a function of stellar mass for four independent data sets (as indicated) from \cite{Velander13}. 
The top shows results for red lenses and the bottom shows results for blue lenses. The results agree fairly
well, where we note that the exact splits between samples differ between studies, which might explain some of the difference. }\label{fig:velander}
\end{figure}

The lensing signal around central galaxies consists of two
components. First of all, the lensing signal from the halo the galaxy
resides in. This signal, the 1-halo term $\gamma_{T,1h}^{\rm cen}$,
dominates on small scales as is highlighted in
Figure~\ref{fig:gglensing}. On larger scales the contribution from
neighboring halos becomes dominant, which is described by the 2-halo
term $\gamma_{T,2h}^{\rm cen}$. The terms simply add to give the total
signal from central galaxies. The calculation of the 1-halo term
requires the adoption of a halo density profile, but the 2-halo term
is more involved. It requires the power spectrum to account for the
correlation between the central galaxy and the dark matter
distribution in nearby halos (expressions can be found in
e.g.~\cite{Mandelbaum05, vanUitert11, Cacciato13}). The detailed
outcome depends on how these terms are computed, but the overall
principles are pretty much the same. The main difference is how one
deals with the transition from the single halo to multiple halo
signal. This is because the dark matter halos of nearby halos cannot
overlap, and different approaches to account for this have been
proposed. More work is needed to improve the predictions in this
critical regime.

The lensing signal around satellite galaxies consists of three
terms. The simplest is the signal from the sub-halo in which the
galaxy resides, although the result, $\gamma_{T,trunc}^{\rm sat}$
depends on the adopted tidally stripped density profile. There is also
a contribution arising from the central halo in which the galaxy is
located. As the satellites are mis-centered, calculating
$\gamma_{T,1h}^{\rm sat}$ involves the convolution of the halo density
profile with the spatial distribution of satellites. The last term,
which is relevant on large scales, is the contribution from
neighboring haloes, $\gamma_{T,2h}^{\rm sat}$. These three terms are
added to obtain the satellite signal.

The left panel of Figure~\ref{fig:velander} shows an example of the
signal predicted by the halo model and the various components from
\cite{Velander13}. In addition to the components introduced above,
\cite{Velander13} include a point mass contribution from the
stellar mass contained in the galaxies. The result shows that the
resulting signal shows features, which can be used to establish the
satellite fractions. The right panel of Figure~\ref{fig:velander}
shows a comparison of four different measurements (from
\cite{Mandelbaum06, Leauthaud12, vanUitert11, Velander13}) where the
lenses have been split into blue and red samples (although we note
that the detailed splits differ between studies, causing some
differences). The results agree fairly well, although there are
differences at the high mass end. These measurements will improve
dramatically thanks to larger surveys. Of particular interest is the
possibility to probe the evolution of galaxy properties, which only
now is becoming possible. These early results show that galaxy-galaxy
lensing is an important complement to galaxy clustering and luminosity
function studies to place observational constraints on models of
galaxy formation.

\subsection{Halo Shapes}

In addition to a specific average density profile, numerical
simulations of cold dark matter show that dark matter halos are
triaxial, with a typical ellipticity, in projection, of $\sim 0.3$
(e.g., \cite{Dubinski91, Jing02, Hayashi07}). The inner regions are
believed to be significantly affected by baryonic processes, and
predictions are rather uncertain. On larger scales, which are best
constrained by the simulations, although we note that
\cite{Kazantzidis04} and {Kazantzidis10} have argued that baryonic
effects lead to more oblate halos at all radii, weak lensing provides
a unique way to constrain the halo shapes.

Although the precision of galaxy-galaxy lensing measurements has
improved dramatically over the past years, constraints on the halo
shapes are more difficult to obtain. This is because we need to
measure the azimuthal variation in the galaxy lensing signal.  For a
singular isothermal ellipsoid, \cite{vanUitert12} showed that the
signal-to-noise ratio with which the azimuthal variation can be
determined is proportional to the mean halo ellipticity, $\epsilon_h$,
and the S/N of the (isotropic) galaxy-galaxy lensing signal:

\begin{equation}
(S/N)_{\rm ani}=\frac{0.15}{\sqrt{2}}{\left(\frac{\epsilon_h}{0.3}\right)}(S/N)_{\rm iso}.
\end{equation}

\noindent Hence in this best-case scenario, the precision is an order
of magnitude lower than that of the galaxy lensing signal itself.  The
signal-to-noise ratio is reduced further because the signal declines
relatively quickly with distance from the lens for more realistic
density profiles, such as the NFW model. Furthermore, the light
distribution of the lens is used as the reference frame to measure the
azimuthal variation and any misalignment between the halo and the
light will reduce the expected signal.

The azimuthally averaged galaxy lensing signal is robust against
biases in the correction for PSF anisotropy, but this is not the case
for the measurement of halo shapes: an imperfect correction may lead
to correlations in the shapes of the lens and the sources, which
changes the signal, although the impact is fairly modest
(\cite{Hoekstra04}). Lensing by structures at lower redshift can also
cause correlations in the shapes. These contributions to the signal
can be suppressed by subtracting the cross-shear from the measurements
(\cite{Mandelbaum06a,vanUitert12}), but at the expense of another
factor $\sqrt{2}$ reduction in S/N.

Several studies have attempted to measure this signal, but the results
remain inconclusive. \cite{Hoekstra04} claim to have detected an
alignment between the light distribution and the lensing signal
through a maximum likelihood analysis of their data.  These results
are in agreement with the results from \cite{Parker07} who compared
the tangential shear in quadrants. However, \cite{Mandelbaum06a} did
not detect a signal for their full sample of SDSS lenses, but a
tentative signal was observed for bright early type lenses. Similarly,
\cite{vanUitert12} did not detect a significant anisotropy signal,
even though they analyzed an area that was $\sim 20$ times larger than
\cite{Hoekstra04}. An open question is whether this difference is due
to the use of the maximum likelihood approach by \cite{Hoekstra04},
which should be more sensitive to the signal, although the
interpretation may be more difficult. Significant progress is only
possible with the next generation of imaging surveys, with their
reduced errors and photometric redshift information for the lenses and
the sources.

\section{Lensing by Large-Scale Structure}

Thus far we examined how weak gravitational lensing can be used to
study the dark matter halos of galaxies and clusters. Such studies
provide new observational constraints on models of galaxy evolution
and provide an important approach to calibrate cluster masses.
However, perhaps the most prominent application is weak lensing by
large-scale structure, commonly referred to as ``cosmic
shear''. Compared to the other applications it is more sensitive to
observational biases. The desire to measure the cosmic shear signal
with high accuracy is therefore driving much of the development in the
field of weak lensing, which in turn benefits the other applications.

The cosmic shear signal is a direct measure of the projected matter
power spectrum, which simplifies the interpretation of the signal in
principle, although we recall the aforementioned intrinsic
alignments. Furthermore, as we discuss in \S\ref{sec:powerspectrum} the
predictions for the matter power spectrum remain
uncertain. Nonetheless cosmic shear is considered one of the most
powerful probes to study the distribution and growth of large-scale
structure in the Universe. Since the first detections, reported in
early 2000 (\cite{Bacon00, Kaiser00, Waerbeke00, Wittman00}), the
statistical power of the surveys has increased exponentially. We
present some of the recent results from the Canada-France-Hawaii
Telescope Legacy Survey (CFHTLS) in \S\ref{sec:results}, but note that
new surveys, which are an order of magnitude larger, are already
underway. Importantly, the promise of weak lensing as a key
cosmological probe was recognized by the European Space Agency with
the selection of {\it Euclid} (\cite{Euclid}), scheduled for launch in
2020.

In the case of lensing by large-scale structure the deflection of
light rays is not caused by a single deflector, but rather by all the
inhomogeneities between the source and the observer. Here we only
discuss the essentials and refer the interested reader to more
detailed discussions in \cite{Bartelmann01} or
\cite{Schneider06}. Recent reviews of the cosmological applications of
weak lensing can be found in \cite{Hoekstra08} and \cite{Munshi08}.

We assume that each deflection is small. Consequently, the distortion
can be approximated by the integral along the unperturbed light ray
rather than the actual path. This so-called Born approximation has
been compared to full ray-tracing calculations and is found to be
accurate to a few percent and allows us to define a deflection
potential for a source at radial distance $\chi$ analogous to the
thin-lens case:

\begin{equation}
  \Psi({\bf x},\chi)=\frac{2}{c^2}\int_0^\chi {\rm d}\chi' \frac{f_K(\chi-\chi')}{f_K(\chi)f_K(\chi')}\Phi(f_K(\chi'){\bf x},\chi'),
\end{equation}

\noindent where the integral is over the lens coordinates $\chi'$ and
$f_K(\chi)$ indicates the angular diameter distances. The convergence
and shear can be computed from this potential, just as for the single
lens plane case. The use of the 3D Poisson equation in co-moving
coordinates allows us to relate an overdensity $\delta$ to
$\nabla^2\Phi$ as

\begin{equation}
\nabla^2\Phi=\frac{3H_0^2\Omega_m}{2a}\delta,
\end{equation}

\noindent where $a$ is the scale factor. This allows us to compute the
convergence $\kappa({\bf x},\chi)$:

\begin{equation}
\kappa({\bf x},\chi)=\frac{3H_0^2 \Omega_m}{2c^2}
\int_0^\chi {\rm d}\chi' \frac{f_K(\chi-\chi')f_K(\chi')}{f_K(\chi)}
\frac{\delta(f_K(\chi'){\bf x},\chi')}{a(\chi')}.
\end{equation}

\noindent If we consider an ensemble of sources with a redshift
distribution $n(z){\rm d}z=p(\chi){\rm d}\chi$ the effective
convergence is given by

\begin{equation}
  \kappa({\bf x})=\int_0^{\chi_{\rm max}}{\rm d}\chi p(\chi)\kappa({\bf x},\chi)=\frac{3H_0^2 \Omega_m}{2c^2}
  \int_0^{\chi_{\rm max}} {\rm d}\chi' g(\chi')f_K(\chi')\frac{\delta(f_K(\chi'){\bf x},\chi')}{a(\chi')},\label{eq:kappa_eff}
\end{equation}

\noindent i.e., the contribution from each lens plane is weighed by the
lens efficiency factor

\begin{equation}
g(\chi')=\int_{\chi'}^{\chi_{\rm max}}{\rm d}\chi p(\chi)\frac{f_K(\chi-\chi')}{f_K(\chi)}.\label{eq:redshift}
\end{equation}

\noindent We can now compute the convergence power spectrum that
arises from the power spectrum of density fluctuations, which in turn
is determined by the initial cosmological conditions that we wish to
determine. The convergence power spectrum $P_\kappa(\ell)$ is given by
(see e.g., \cite{Bartelmann01,Schneider06} for details):

\begin{equation}
P_\kappa(\ell)=\frac{9 H_0^4 \Omega_m^2}{4c^4} \int_0^{\chi_{\rm max}} 
{\rm d}\chi \frac{g^2(\chi)}{a^2(\chi)} 
P_\delta\left({\frac{\ell}{f_K(\chi)},\chi}\right).\label{eq:pkappa}
\end{equation}

\noindent One complication is the fact that cosmic shear probes
relatively small scales $(100<\ell<5000)$ where density fluctuations
are non-linear. As a consequence, the power spectrum on those scales cannot
be computed directly, but instead needs to be inferred from numerical
simulations. We revisit this issue in \S\ref{sec:powerspectrum}.

So far we have only discussed the two-point statistics of the
convergence field. However, as we show below the shear power spectrum
is the same as that of the convergence (in the flat sky
approximation). There is nonetheless a small complication because we
cannot measure the shear, but only the reduced shear $g$
(see~\S\ref{sec:shear}). Hence the relation between the observables and
the underlying convergence field is more involved, but ray-tracing
through cosmological simulations allow us to compute the necessary
corrections (e.g., \cite{Hilbert09}).

To demonstrate that the shear and convergence power spectra are
identical, we express the Fourier transforms of the convergence and
shear in terms of the deflection potential (recall that a derivative
$\partial/\partial x_j$ in real space corresponds to a multiplication
by $-i \ell_j$ in Fourier space). Hence the Fourier transforms of the
convergence, $\hat\kappa(\ell)$ and the shear $\hat\gamma(\ell)$ are
given by

\begin{equation}
\hat\kappa(\ell)=-\frac{\ell^2}{2}\hat\Psi(\ell),~{\rm and}~\hat\gamma(\ell)=-\frac{\ell_1^2-\ell_2^2+2i \ell_1\ell_2}{2}\hat\Psi(\ell),
\end{equation}

\noindent which once more shows that the shear and convergence are
related (cf. Eqns~\ref{eq13} and \ref{eq14}):

\begin{equation}
\hat\gamma(\ell)=\frac{\ell_1^2-\ell_2^2+2i \ell_1\ell_2}{\ell^2}\hat\kappa(\ell)=e^{2i\phi}\hat\kappa(\ell),
\end{equation}

\noindent which implies that 

\begin{equation}
\langle\gamma(\ell)\gamma^*(\ell')\rangle=\langle\kappa(\ell)\kappa^*(\ell')\rangle=(2\pi)^2\delta(\ell-\ell')P_\kappa(\ell).
\end{equation}

In practice the situation is more complicated, because shapes cannot
be measured reliably at every position. In particular bright stars
saturate the detector, causing blooming and bleeding
trails. Furthermore reflections in the optics give rise to halos and
scattered light, which may prevent accurate measurements in some
areas. As a result the actual survey geometry can be very
complicated. This is very relevant as the holes created by bright
stars have sizes of $\sim$ arcminutes, the scale where cosmic shear is
most sensitive. Such a mask leads to the mixing of modes and therefore
the observed power spectrum needs to be corrected, which increases the
uncertainties further.

Alternatively one can consider other two-point statistics to quantify
the cosmic shear signal. Of particular interest is the ellipticity
correlation function, because it can be computed directly from a
catalog of objects with shape measurements, and is insensitive to the
mask. The mask, however, does change the number of pairs at certain
distances, and thus the measurement errors. This does still complicate
the calculation of the covariance matrix which is needed to interpret
the cosmological nature of the signal.

\begin{figure}
\centering
\hbox{%
\includegraphics[width=0.55\hsize,trim=0cm 0cm 0cm 0cm,clip=true]{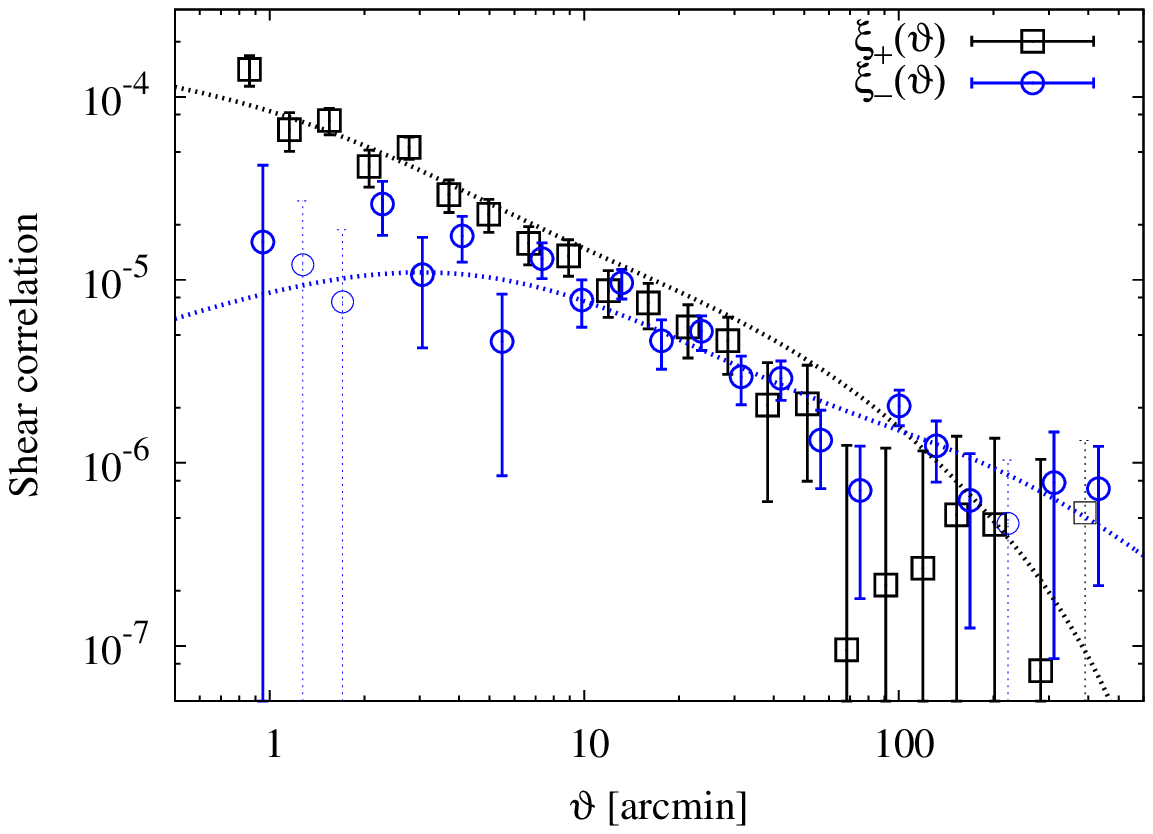}
\includegraphics[width=0.45\hsize,trim=0cm 0cm 0cm 0cm, clip=true]{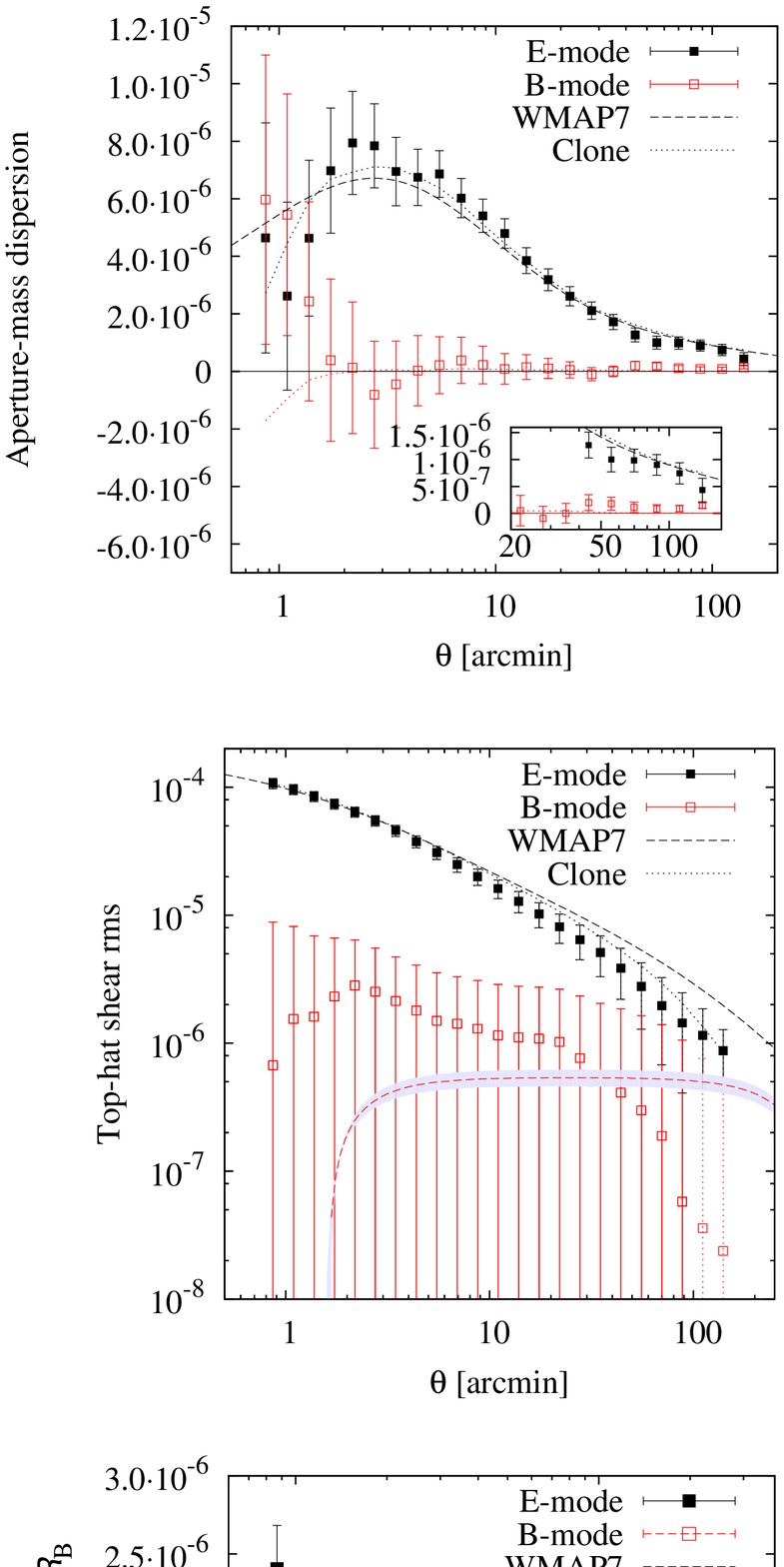}}
\caption{{\it Left panel:} Ellipticity correlation functions $\xi_+$
  and $\xi_-$ measured by CFHTLenS from \cite{Kilbinger13}.  The
  dashed lines indicate the predictions for a WMAP7 cosmology. {\it
    Right panel:} Aperture mass variance from \cite{Kilbinger13} which
  shows the E-mode in black and the B-mode in red. The latter is
  consistent with 0, suggesting that (additive) systematics have been
  removed successfully.}\label{fig:kilbinger}
\end{figure}

Although one could simply compute the correlation functions of the two
shear component $\gamma_1$ and $\gamma_2$, it is more convenient to
compute the correlation between the tangential $(\gamma_T)$ and cross
$(\gamma_\times)$ components with respect to the line connecting a
pair of galaxies and combine them into $\xi_+(\theta)$ and
$\xi_-(\theta)$, defined as:

\begin{equation}
\xi_\pm(\theta)=\langle\gamma_T(\theta_1)\gamma_T(\theta_2)\rangle\pm\langle\gamma_\times(\theta_1)\gamma_\times(\theta_2)\rangle,
\end{equation}

\noindent where $\theta=|\theta_1-\theta_2|$, hence the ensemble
average depends only on the angular separation between the
sources. The left panel in Figure~\ref{fig:kilbinger} shows
measurements for the ellipticity correlation functions by
\cite{Kilbinger13} from the weak lensing analysis of CFHTLS data by
CFHTLenS, which is discussed in more detail in
\S\ref{sec:results}. For reference, predictions for a WMAP7 cosmology
are also indicated. These can be computed from the convergence power
spectrum because

\begin{equation}
\xi_{+/-}(\theta)=\frac{1}{2\pi}\int_0^\infty {\rm d}\ell J_{0/4}(\ell \theta)P_\kappa(\ell),
\end{equation}

\noindent where $J_i$ is the $i-$th order Bessel function of the first
kind. Note that the Bessel function does not vanish quickly, and
consequently a large range in $\ell$ contributes to the correlation
function at a given scale $\theta$. 

The shear field is derived from the scalar deflection potential, which
is a single (real) function of position. The fact that we measure two
correlation functions implies there is redundancy in the data, as a
single function should have sufficed. We use a similar argument in the
case of mass reconstructions: in this case the imaginary part of the
mass map does not arise from lensing, but from systematics (the same
is true for the azimuthally averaged tangential shear). It is
therefore useful to examine how such a separation can be done in the
case of the cosmic shear signal. As the signal for
over-(under)densities gives rise to a tangential (radial) shear
pattern, analogous to the polarizations of the electric and magnetic
field, the lensing induced signal is referred to as the "E-mode",
whereas the cross component is given by the "B-mode"
(\cite{Crittenden02, Schneider02}).

The derivation in \cite{Schneider02} is particularly nice, as it
clearly makes the link to the mass reconstruction case, and we refer
the interested reader to this paper. It is possible to express E and
B-mode power spectra in terms of the observed ellipticity correlation
functions:

\begin{equation}
P_{E/B}(\ell)=\pi\int_0^\infty {\rm d}\theta \theta [\xi_+(\theta)J_0(\ell \theta)\pm \xi_-(\theta)J_4(\ell \theta)],
\end{equation}

\noindent where the E-mode corresponds to the sum of the terms and the
B-mode to the difference. It is also possible to separate the
correlation functions themselves into $\xi_{E\pm}(\theta)$ and
$\xi_{B\pm}(\theta)$, but as \cite{Schneider02} point out, to compute
these requires knowledge of $\xi_-(\theta)$ to arbitrarily large, and
of $\xi_+(\theta)$ to arbitrarily small scales. This is not possible
in the case of real data as neither can be observed. One solution is
to account for this limited range in scales when predicting the
signal, or instead using the model predictions to account for the lack
of data.

This problem can be avoided by using aperture mass statistics,
$\langle M_{\rm ap}^2\rangle(\theta)$, instead \cite{Schneider98}.
In this case the separation into E- and B-modes can be performed using
correlation functions that extend to twice the scale of interest.
Another advantage is that the smoothing of the convergence power
spectrum is less severe. For this reason, cosmic shear studies often
present results for the aperture masses. The right panel of
Figure~\ref{fig:kilbinger} shows the results for the CFHTLenS analysis
from \cite{Kilbinger13}. A clear E-mode signal is detected (black
points), whereas the B-mode signal (red points) is consistent with 0,
suggesting that (additive) systematics have been removed successfully.

In addition to the ellipticity correlation function and the aperture
mass statistics, several other options have been proposed, either
because of their ease of use, or because they allow for a cleaner
E/B-mode separation. These are discussed in \cite{Kilbinger13} who
also provide expressions for their relation to the matter power
spectrum.

\subsection{Photometric redshifts and tomography}
\label{sec:redshift}

\begin{figure}
\centering
\hbox{%
\includegraphics[width=0.49\hsize,trim=0cm 0cm 0cm 0cm,clip=true]{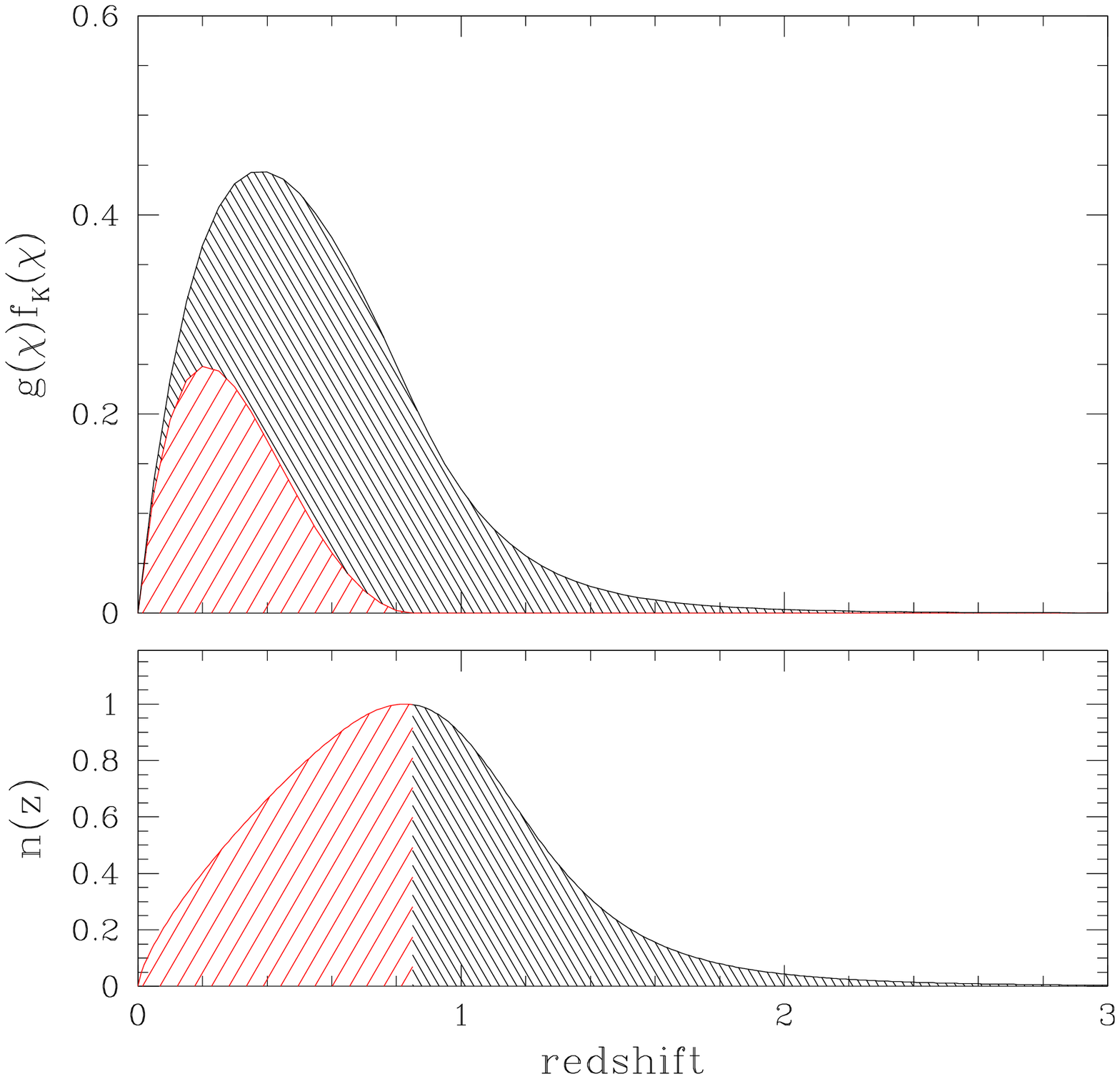}
\includegraphics[width=0.51\hsize,trim=0cm 0cm 0cm 0cm, clip=true]{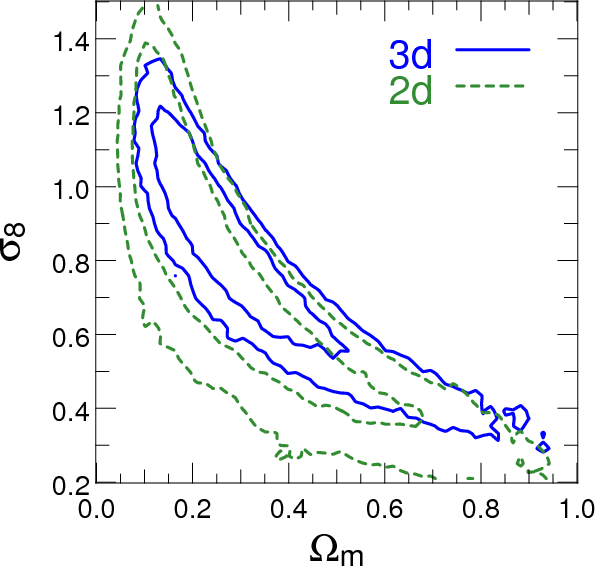}}
\caption{{\it Left panel:} The lensing kernel $g(\chi)f_K(\chi)$ as a
  function of lens redshift for the source redshift distributions
  shown in the bottom panel (see text for details.) Even though the
  sources have been split into a low ($z<0.85$) and high redshift bin
  $(z>0.85)$ as indicated, the lensing kernels overlap.  {\it Right
    panel:} Constraints on $\Omega_m$ and $\sigma_8$ from
  \cite{Schrabback10}. The green contours show the results when the
  individual source redshifts are not used, whereas the blue contours
  show the results from a tomographic analysis. Although an
  improvement can be observed, the gain is limited because of cosmic
  variance due to the limited area covered by the 2 deg$^2$ COSMOS
  survey}\label{fig:tomography}
\end{figure}

It is evident from Eqn.~\ref{eq:pkappa} that the cosmological
interpretation of the cosmic shear signal requires accurate knowledge
of the source redshifts.  The lensing kernel in
Eqn.~\ref{eq:kappa_eff}, $g(\chi)f_K(\chi)$ (or $D_L D_{LS}/D_S$)
peaks roughly halfway between the observer and the source, as can be
seen from the left panel in Figure~\ref{fig:tomography}. The source
redshift distribution, shown in the lower panel is described by

\begin{equation}
n(z)=n_0\frac{z^a}{z^b+c},
\end{equation}

\noindent where $n_0$ is a normalization. For the results in
Figure~\ref{fig:tomography} we used $a=0.723$, $b=6.772$ and
$c=2.282$, which yields a mean redshift of 0.85; this choice of
parameters provides a good approximation for the redshift distribution
of the CFHTLS data (\cite{Benjamin07}). The kernel is broad, and the
lensing signal is sensitive to a large range in redshift. It does have
the benefit that source redshifts need not be known with high
precision: photometric redshifts are sufficient, although it is
important that the mean is unbiased and the outlier fractions are
known well. This avoids the need for spectroscopic redshifts for the
individual sources, but note that the sources are very faint, which
pushes the determination of photometric redshifts to the
limits. Reliable measurements require good photometry in multiple
bands and it is therefore important to correct for differences in the
data quality \cite{Hildebrandt13}. 

Nonetheless large samples of galaxies with spectroscopic redshifts are
needed to train and calibrate photometric redshift algorithms. This is
because biases in the mean and higher order moments of the redshift
distributions directly lead to biases in the inferred cosmological
parameters. This is particularly critical for future
projects. \cite{Newman13} provide an excellent overview of the needs
for dark energy experiments. This aspect of cosmic shear is arguably
as important as the fidelity of shape measurement techniques, but has
not been examined as thoroughly, perhaps largely due to the paucity of
large spectroscopic surveys. This is, however, changing rapidly and several
approaches to improve photometric redshift estimates are being explored.

The first cosmic shear results were based on observations in a single
filter and the interpretation relied on redshift distributions
obtained from deeper data (\cite{Hoekstra02,Waerbeke05}), but current
surveys obtain multi-color data. Both the Deep Lens Survey (DLS;
\cite{Jee13}) and CFHTLenS (\cite{Heymans12}) use observations in
five optical bands, which allowed them to determine photometric
redshifts for individual sources. This not only helps the
interpretation of the signal, but it also allows for the study of the
lensing signal as a function of source redshift. Rather than
considering the full redshift distribution, one can define redshift
bins and compute the corresponding lensing signal, as well as cross
correlations between bins. As first shown in \cite{Hu99} this cosmic
shear {\it tomography} greatly improves the statistical power of a
survey. The right panel of Figure~\ref{fig:tomography} shows the
results from the analysis of the 2 deg$^2$ COSMOS survey by
\cite{Schrabback10}. An improvement in the constraints can be seen,
although the results are limited by cosmic variance due to the
relatively small survey area. A consequence of the broad lensing
kernel is that the lensing signals for the bins are highly correlated,
and the statistical gain saturates quickly (see e.g., Figure~3 in
\cite{Ma06}), and $\sim 5$ bins are sufficient. One does need to
account for intrinsic alignments, which are more pronounced in the
narrower tomographic redshift bins (\cite{Heymans13}).  The modeling
of the intrinsic alignment signal as a function of redshift is therefore a
reason to actually increase the number of bins for future projects
(e.g. \cite{Joachimi10}).

\subsection{Predictions for the matter power spectrum}
\label{sec:powerspectrum}

In addition to accurate shape measurements and photometric redshift
estimates, the cosmological interpretation of the lensing signal
requires accurate predictions for the matter power spectrum.  On the
largest scales the linear power spectrum can be computed for a given
set of initial conditions. Perturbation theory can be used to compute
corrections due to structure growth (see, \cite{Bernardeau02} for a
review; also see \cite{Crocce06}) with an accuracy of a per cent for
$k< 0.2~h~{\rm Mpc}^{-1}$ (e.g., \cite{Crocce12, Taruya12}), depending
on redshift. This approach, however, breaks down as structures become
too overdense, collapse and virialize.

Cosmic shear surveys, however, are most sensitive to structures that
are much smaller, such as groups of galaxies, i.e. very non-linear
structures. The signal-to-noise ratio of the cosmic shear signal peaks
on angular scales of $5-10$ arcminutes, or physical scales of $\sim
1$~Mpc (see e.g., \cite{Huterer05, Semboloni11b}). Restricting the
analysis to large scales does not necessarily solve the problem,
because the observed two-point ellipticity correlation functions are
still sensitive to small scale structures projected along the
line-of-sight. This can be avoided using a full 3D cosmic shear
analysis (see \cite{Castro05,Kitching11b} for details), but using only
large scales increases the statistical uncertainties dramatically due
to cosmic variance, and thus is not a viable solution. Therefore the
non-linear power spectrum needs to be computed to optimally extract
the cosmological information from the measurements.

Currently only N-body simulations allow us to capture non-linear
structure formation. In principle this can be done with good accuracy,
provided the simulations are started with adequate initial conditions,
with a large simulation volume, good time stepping and high mass
resolution. For instance, \cite{Heitmann10} obtained an accuracy of
$\sim 1\%$ out to $k\sim 1~h~{\rm Mpc}^{-1}$ for a gravity-only
simulation. N-body simulations have also been used to calibrate
recipes that allow one to compute the matter power spectrum from the
linear power spectrum, such as \cite{PD96}, or the popular{\it
  halofit} from \cite{Smith03}. Although these methods were sufficient
for the interpretation of early cosmic shear studies, given their
accuracy of $5-10\%$, the next generation of projects require percent
level accuracy. One way forward is to run a suite of cosmological
simulations spanning a range of cosmological parameters. These can be
used to create an emulator of the power spectrum. For instance
\cite{Lawrence10} describe the ``Coyote Universe'', which comprises
nearly 1000 N-body simulations, which can be used to predict the
non-linear power spectrum out to $k\sim 1~h~{\rm Mpc}^{-1}$.

Another complication is that to infer cosmological parameters using a
maximum likelihood method requires a covariance matrix. The non-linear
growth of structure leads to correlations between scales, and the
resulting mode coupling moves information from lower order moments of
the density field to higher order moments. As a result the covariance matrix
is not diagonal and the information content of two-point statistics is
reduced (e.g. \cite{Semboloni07,Kiessling11}). Hence large numbers of
numerical simulations are required for precise estimates of the
covariance, in order to avoid biases in its inversion
(\cite{Hartlap07}).

Thus far we only discussed N-body simulations, in which only gravity
acts on the collisionless particles. This greatly simplifies the
problem and allows for efficient calculations with high mass
resolution over large volumes (e.g. \cite{Springel05, Crocce10}).
Although most of the matter in the Universe is indeed believed to be
in the form of collisionless cold dark matter, baryons still represent
about 17 percent of the total matter content. The distribution of
baryons largely follows the underlying dark matter density field, and
thus the results from N-body simulations should resemble the matter
power spectrum fairly well. Nonetheless, differences in the spatial
distribution of baryons with respect to the dark matter may result in
significant changes (i.e. more than a few percent) in the actual
matter power spectrum.

\begin{figure}
\centering
\hbox{%
\includegraphics[width=0.49\hsize,trim=0cm 0cm 0cm 0cm,clip=true]{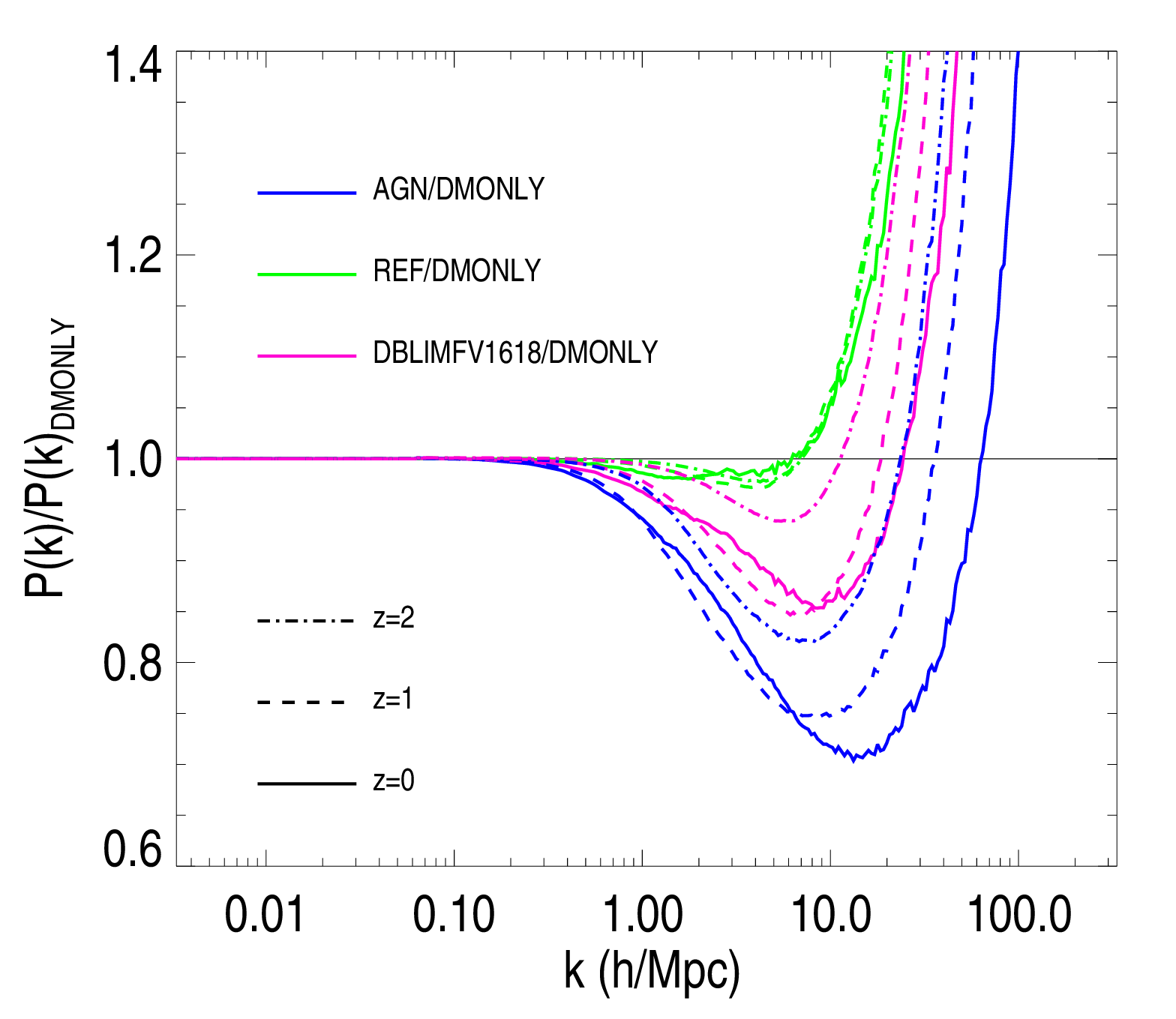}
\includegraphics[width=0.51\hsize,trim=0cm 0cm 0cm 0cm, clip=true]{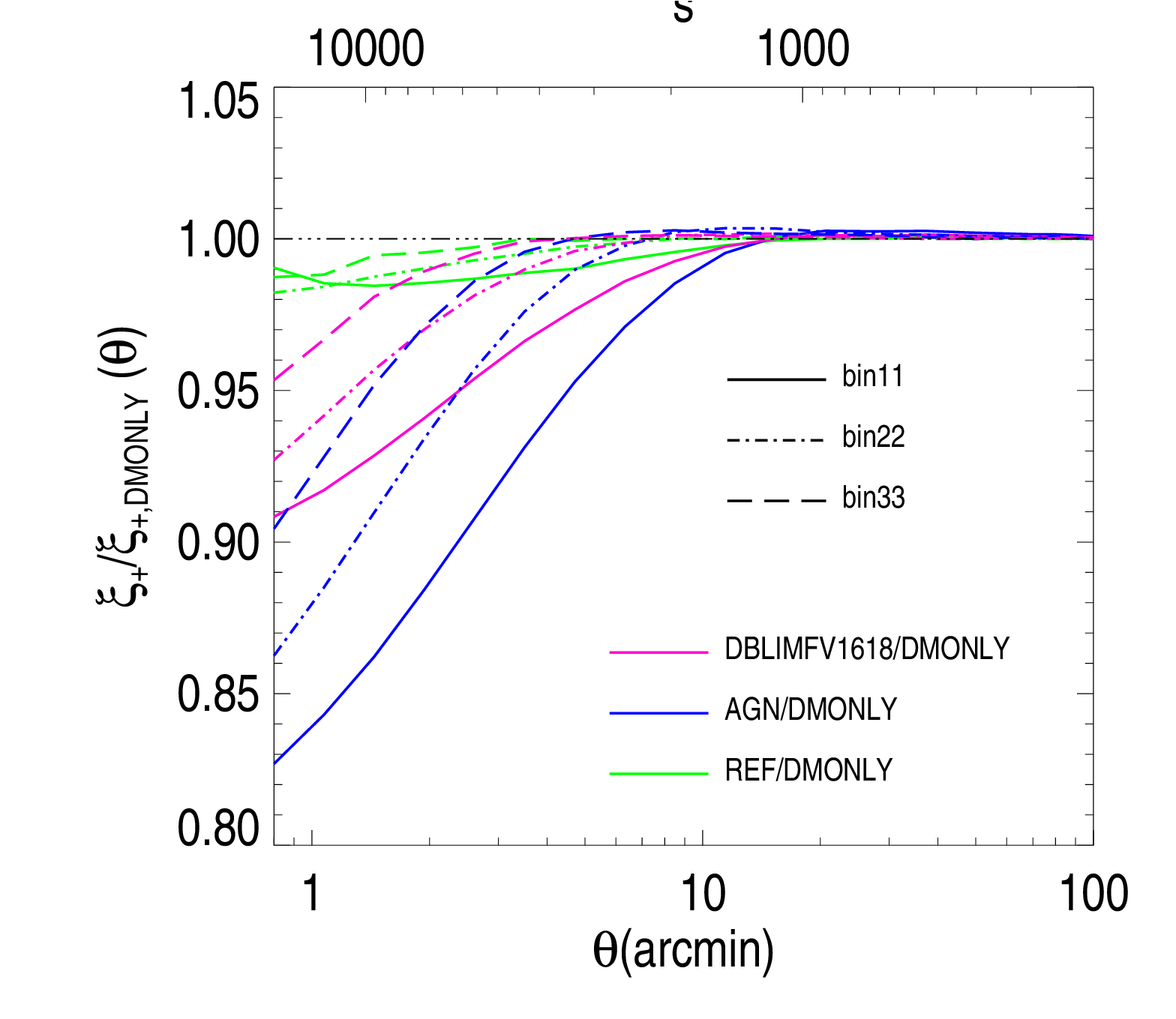}}
\caption{{\it Left panel:} Ratio of the power spectrum for different
  feedback models and the dark matter only case for three different
  redshifts from \cite{Semboloni11b} based on the results from
  \cite{vanDaalen11}. Of these models the AGN (blue curves) matches
  observations best. {\it Right panel:} The resulting change in the
  ellipticity correlation function $\xi_+$ for three redshift bins
  (0-0.6; 0.6-1.2;1.2-3.4), which is significant on scales where the
  cosmic shear S/N ratio is maximal ($s\sim$ few arcminutes). Ignoring
  the effects of baryon physics will therefore lead to biased
  cosmological parameter estimates.}\label{fig:feedback}
\end{figure}

Hence an accurate prediction for the matter power spectrum should
include baryon physics. Various physical processes affect the
distribution of baryons, such as radiative cooling, star formation and
feedback from supernovae and active galactic nuclei
(AGN). Prescriptions for these processes can be implemented in
cosmological hydrodynamic simulations. As such simulations are much
more expensive than N-body simulations,  it has only recently become
possible to simulate interesting cosmological volumes with reasonable
resolution. It is, however, not clear which processes need to be
included to reproduce observations. For this reason the accuracy of
the results from hydrodynamic simulations is still under discussion,
although we note that some recent simulations appear to match several
key observables quite well (\cite{McCarthy10}).

\begin{figure}
\centering
\hbox{%
\includegraphics[width=0.5\hsize,trim=0cm 0cm 0cm 0cm,clip=true]{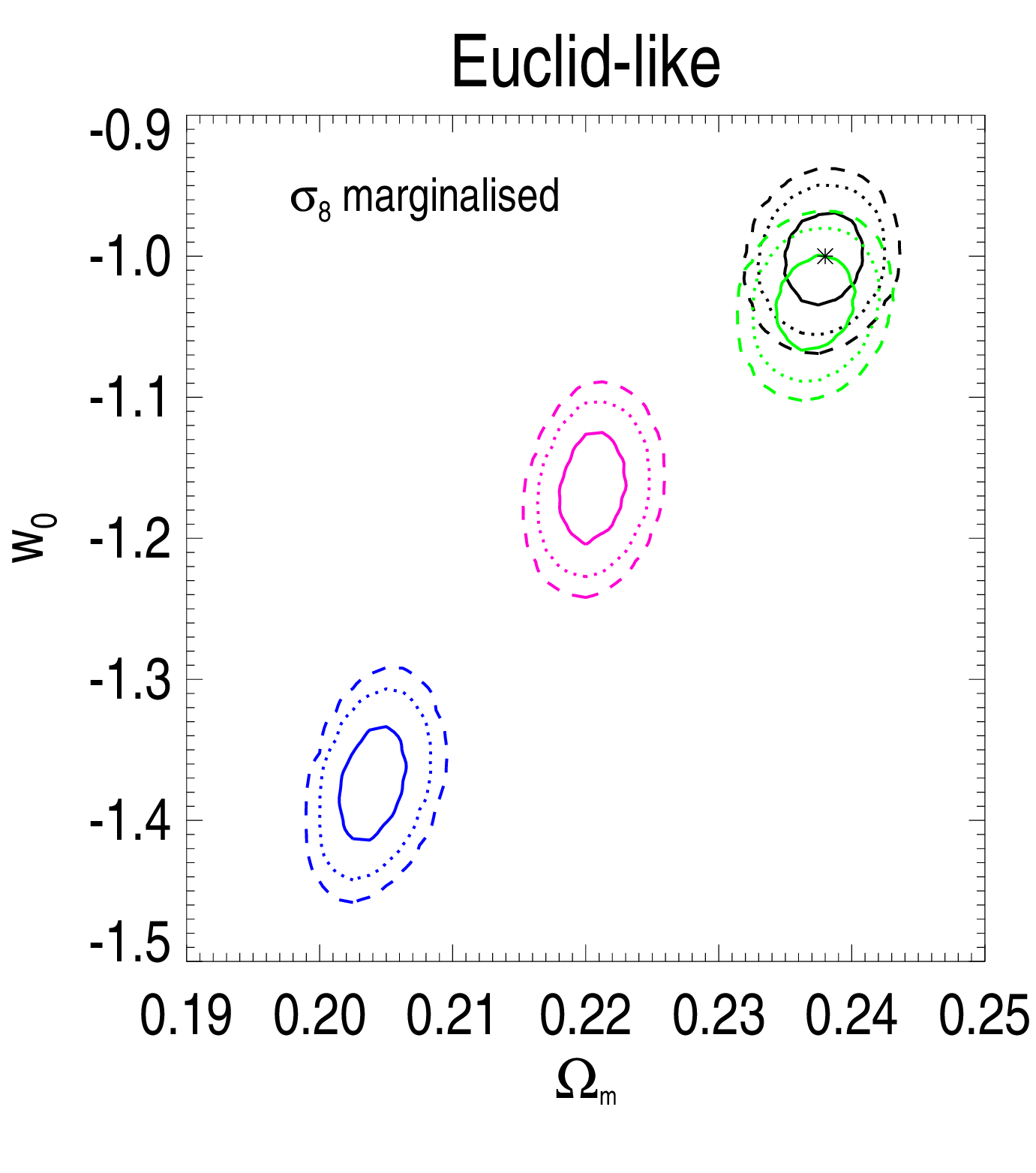}
\includegraphics[width=0.5\hsize,trim=0cm 0cm 0cm 0cm, clip=true]{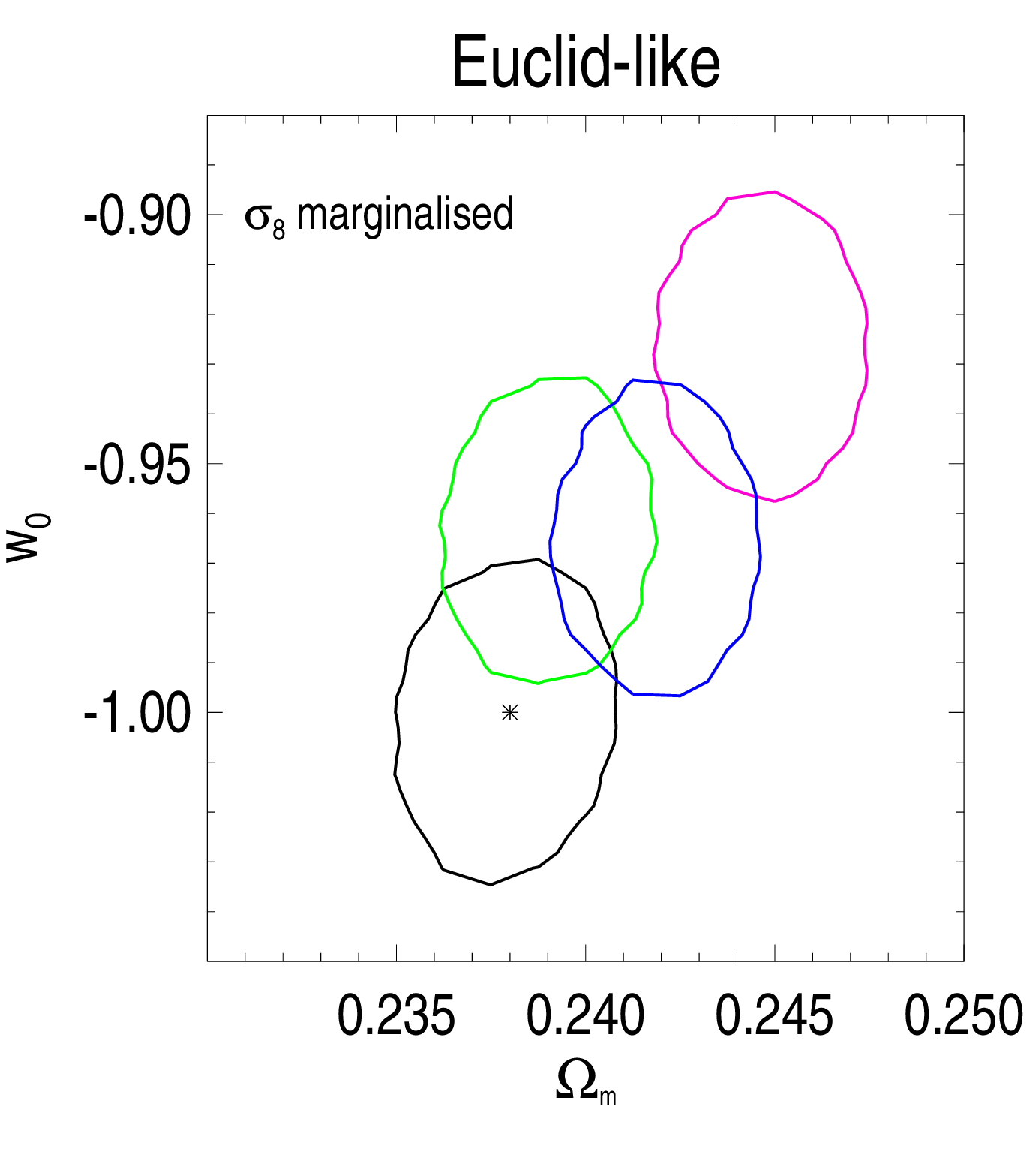}}
\caption{{\it Left panel:} Biases in recovered cosmological parameters
  (for a {\it Euclid}-like survey) for three different feedback models
  if the effects of baryon physics are ignored in the analysis (from
  \cite{Semboloni11b}). The dot indicates the true parameter
  values. {\it Right panel:} The biases can be largely corrected for
  using a simple halo model that takes into account the different
  spatial distribution of baryons and dark
  matter.}\label{fig:feedback_correct}
\end{figure}

The OverWhelmingly Large Simulations project (OWLS; \cite{Schaye10})
produced a large number of simulations where various parameters that
govern the feedback processes were varied. \cite{vanDaalen11} used
these results to examine the impact on the matter power spectrum. They
found that different feedback processes can lead to rather large
changes in the results. These results were used by \cite{Semboloni11b}
to examine the impact on cosmic shear studies and the left panel of
Figure~\ref{fig:feedback} shows the ratio of the matter power spectrum
and the dark matter only results. The AGN model has the strongest
feedback and leads to a suppression of $\sim 10\%$ at
$k\sim 1~h~$Mpc$^{-1}$, but is also believed to match observations best
(\cite{McCarthy10}). The right panel shows the corresponding change in
the ellipticity correlation function $\xi_+(\theta)$, which is
substantial on scales of a few arcminutes. The left panel of
Figure~\ref{fig:feedback_correct} shows that the resulting
cosmological parameter estimates are biased if baryon physics is
ignored.

Our knowledge of the various feedback processes is still incomplete
and we therefore cannot use these simulations to interpret the cosmic
shear signal. Furthermore, hydrodynamic simulations are too expensive
to compute covariance matrices, which would also be needed.  Therefore
several approaches have been suggested to parametrize our ignorance.
\cite{Bernstein09} proposed to include an additional contribution,
described by Legendre polynomials, to the power spectrum and
marginalize over the nuisance parameters. A similar approach was
suggested by \cite{Kitching11a}. These approaches reduce the
statistical precision of a cosmic shear survey. Although one would
like to obtain unbiased constraints on cosmological parameters, one
should generally avoid reducing biases at the expense of
precision. For instance \cite{Zentner08} estimate that the degradation
in dark energy constraints may be as much as $\sim 30\%$. Experiments
are costly, and marginalizing over nuisance parameters should
therefore be used as a last resort.

Instead it is useful to examine whether it is possible to capture the
effects of baryon physics on the matter power spectrum more
effectively. \cite{Semboloni11b} attempted to do so using a halo model
approach, in which the baryons and stars were treated separately from
the dark matter distribution: the distribution of stars were described
by point masses, whereas the baryons followed a profile observed in
X-ray observations. Importantly, feedback processes not only affect
the power spectrum, but also affect scaling relations between halo
mass and the stellar and gas mass. These can be observed, and thus
used to constrain the halo model parameters. \cite{Semboloni11b}
fitted the simulated scaling relations and used the results to compute
corrections to the dark matter power spectrum. The right panel in
Figure~\ref{fig:feedback_correct} shows that, in principle, this
relatively simple approach can reduce the biases in cosmological
parameters significantly.

So far we have discussed only two-point statistics, mainly because
those have been measured with high precision, and because the
interpretation is relatively straightforward. However, as mentioned
above, structure formation increases the power of higher order
moments, and thus additional information can be extracted from e.g.,
the bispectrum (three-point statistics). For instance, the measurement
of the three-point cosmic shear signal can break degeneracies between
cosmological parameters
(e.g. \cite{Bernardeau97,Waerbeke99,Vafaei10}), and a first reliable
detection of the signal was reported in \cite{Semboloni11a}.

Compared to second-order statistics, higher-order statistics probe
even smaller scales. Hence, baryon physics is expected to be more
relevant. This was investigated by \cite{Semboloni13b}, who showed
that this is indeed the case. Interestingly, \cite{Semboloni13b} found
that feedback affects the two- and three-point signal differently,
opening up the possibility of testing the fidelity of the adopted
feedback model by requiring consistency. The combination of second-
and third-order statistics can also reduce the sensitivity to
multiplicative biases, as was explored in \cite{Huterer06}. Despite
this promise, relatively little work is has been done to improve predictions
for the bispectrum.

\subsection{Results from CFHTLenS}
\label{sec:results}

Since the first reported detections of the cosmic shear signal, the
surveyed areas have increased dramatically in size. The
signal-to-noise ratio with which the signal can be measured has been
increasing exponentially, and is projected to do so for another
decade. Importantly, much effort has been spent on reducing and
correcting for systematic errors. Another important development has
been the availability of photometric redshift information for the
sources, thanks to multi-color data.

We discuss future developments next, but in this section we highlight
some results from the CFHTLenS team\cite{Heymans12}, who analysed the
CFHTLS data. This survey was a large project to image 154 deg$^2$ in
five optical filters with the CFHT. The survey comprises of four
fields that range in area from $23-64$ deg$^2$. Thanks to the queue
scheduling of CFHT the image quality in the $i'-$band used for the
lensing analysis is excellent. Nonetheless dealing with the various
systematics has proven challenging. The data are presented in
\cite{Erben13} and the lensing analysis is described in detail in
\cite{Heymans12}, including the various tests that were performed.
The resulting catalogs are publicly available from {\tt
  http://cfhtlens.org}. Importantly, the analysis was done in a way
that is blind to the cosmological parameters in order to avoid
confirmation bias. Image simulations were used to calibrate the shape
measurement algorithm {\tt lensfit}, described in \cite{Miller13}.

The resulting ellipticity correlation function from \cite{Kilbinger13}
was already shown in Figure~\ref{fig:kilbinger}, with a B-mode signal
consistent with zero. The large area covered by the survey allows for
a unique reconstruction of the large-scale mass distribution. These
results are presented in \cite{Waerbeke13} and showed a nice
correlation between the peaks (and throughs) of the matter
distribution and the light distribution.  Some of the galaxy-galaxy
lensing results were already discussed in \S\ref{sec:ggl}.

\begin{figure}
\centering
\hbox{%
\includegraphics[width=0.5\hsize,trim=0cm 0cm 0cm 0cm,clip=true]{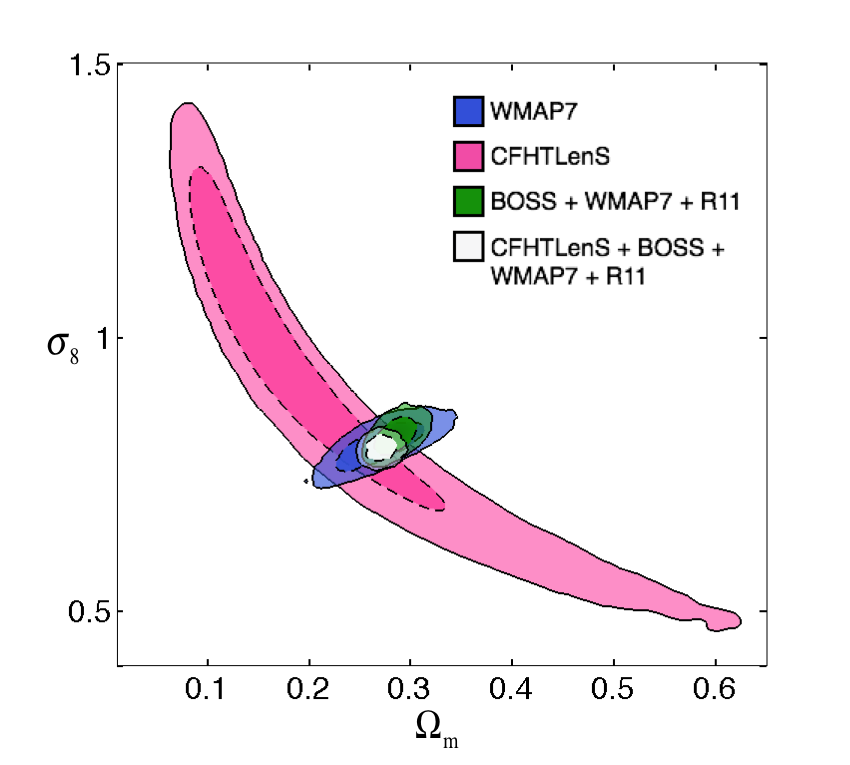}
\includegraphics[width=0.5\hsize,trim=0cm 0cm 0cm 0cm, clip=true]{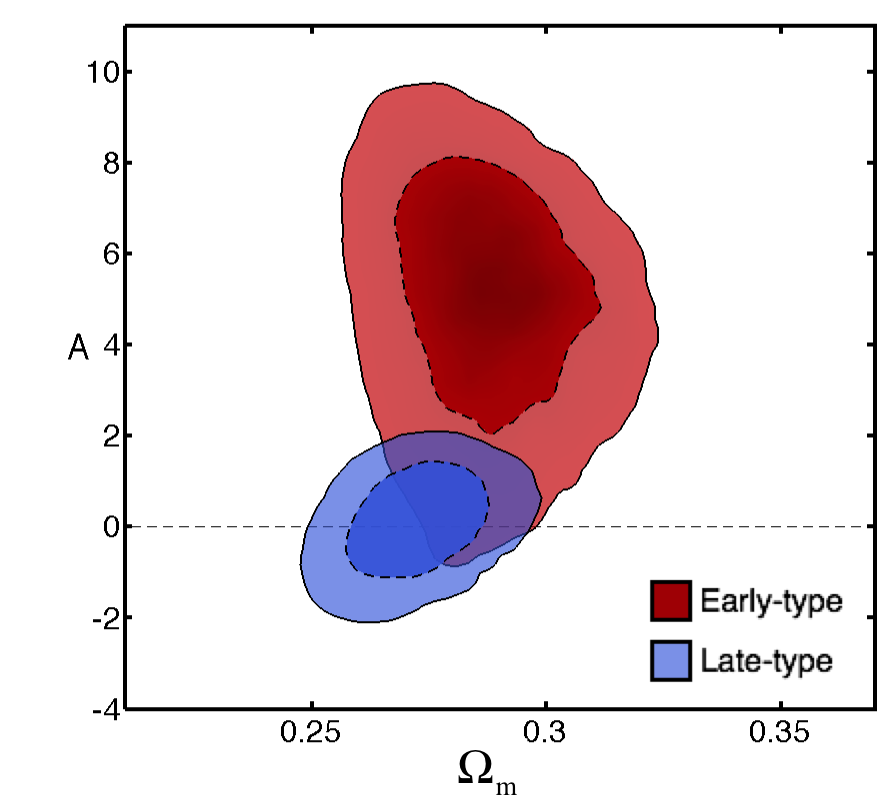}}
\caption{{\it Left panel:} Joint constraints on $\Omega_m$ and
  $\sigma_8$ from a tomographic weak lensing analysis by CFHTLenS
  (from \cite{Heymans13}).  The analysis accounted for the
  contamination by intrinsic alignments by adopting a simple alignment
  model, for which the amplitude was a free parameter. {\it Right
    panel:} The constraints on the amplitude of the intrinsic
  alignment signal and $\Omega_m$ for early- and late-type galaxies.
  A significant signal is measured for the former (from
  \cite{Heymans13}).}\label{fig:heymans}
\end{figure}

The determination of the photometric redshifts is described in
\cite{Hildebrandt12}, who showed that the results are reliable in the
range $0.2<z_{\rm phot}<1.3$, with a scatter after outlier rejection
(the outlier rate is less than 10\%) of $0.03<\sigma_z/(1+z)<0.06$.
The fidelity of the redshift probability distribution functions was
examined in \cite{Benjamin13} using measurements of the angular
correlation function between bins in photometric redshifts. If the
true redshifts of the bins do not overlap, then no significant
correlation function should be observed (magnification could introduce
a small signal though). Comparison of the clustering signal with that
expected from the summed photometric redshift probability
distributions showed good agreement.

\cite{Benjamin13} determined constraints on $\Omega_m$ and $\sigma_8$
by splitting the source samples into two broad redshift bins. This
allowed them to ignore the spurious signal caused by intrinsic
alignments. A more finely binned tomographic analysis, with the
sources sorted into six redshift bins, was presented in
\cite{Heymans13}. For such narrow bins, intrinsic alignments start to
contribute to the observed ellipticity correlation functions. To
account for this \cite{Heymans13} adopted a simple alignment model and
determined the amplitude of the contaminating signal. The resulting
constraints on $\Omega_m$ and $\sigma_8$ are reproduced in the left
panel of Figure~\ref{fig:heymans}. The right panel shows the
constraints on the amplitude of the alignment signal and $\Omega_m$.
\cite{Heymans13} found that the intrinsic alignment signal depends on
galaxy type: for late-type galaxies the amplitude was found to be
consistent with zero, but a significant signal was obtained for early
type galaxies. Another nice application was presented in
\cite{Simpson13}, who compared the cosmic shear results to
measurements from the WiggleZ redshift survey (\cite{Blake12}) to
observationally test modified gravity models.  No deviation from
general relativity was observed, but we note that statistical
uncertainties are large. Nonetheless this study demonstrated the
complementarity of cosmic shear surveys with redshift surveys, and the
potential for future projects, which we will discuss next.

\section{Outlook \& Conclusions}

The results from the CFHTLenS have demonstrated the potential of weak
gravitational lensing to study the properties of dark matter halos
around galaxies and to constrain cosmological parameters (also see
\cite{Choi12, Jee13} for results from the Deep Lens Survey). The small
survey area limited the precision of the constraints, but did provide
an invaluable testbed to prepare for the next generation of surveys,
which have already started taking data.

Of these, the Kilo Degree Survey (KiDS) started observations in 2012
with an anticipated completion of the 1500 deg$^2$ survey in 2016.
Importantly the combination of excellent image quality and 9-band
photometry, including sufficiently deep NIR data allows for the
determination of good photometric redshifts. The Dark Energy Survey
(DES) started in 2013 and will image 5000 deg$^2$, some of which will
overlap with KiDS. Finally the new HyperSuprimeCam on Subaru will be
used to image 1500 deg$^2$, but significantly deeper than the other
two. Although the image quality of Subaru is excellent, the PSF is
still too large to make optimal use of the additional galaxies, that
will be mostly unresolved.

These projects will survey areas that are an order of magnitude
larger than the CFHTLS. However, to learn more about the nature of dark
energy and to study deviations from General Relativity, even larger
areas of sky need to be imaged, to greater depth. From the ground the
Large Synoptic Survey Telescope (LSST) is being developed. It will
image the observable sky every week, and by combining the sharpest
images into a deep stack, the weak lensing signal can be measured over
more than 10,000 deg$^2$ (see \cite{LSST} for an overview of the many
science applications). It will, however, be challenging to model and
correct the ground-based PSF sufficiently well to reach the
statistical limit without significant biases.

The much smaller and stable PSF that can be achieved in space allows
{\it Euclid} to measure the cosmic shear signal from two billion
galaxies, surveying 15,000 deg$^2$, while observational biases are
subdominant, as was shown in \cite{Euclid}. A detailed examination of
the weak lensing design is presented in \cite{Cropper13}.  Thanks to
its excellent image quality {\it Euclid} will also discover many
strong lensing systems. 

With a range of new surveys underway, or commencing very soon, the
prospects for further progress in weak lensing are excellent. In these
notes we touched upon the various applications, but also highlighted
some of the key areas where further progress is needed. In particular
we need to improve the modeling of the matter power spectrum as well
as the bispectrum, develop a better understanding of intrinsic
alignments and continue work on PSF modeling and shape measurement
algorithms. Although they may be challenging, these all present
excellent research opportunities.

\acknowledgments I would like to thank the organizers and staff of the
Fermi School for a memorable meeting. I thank Fabian K{\"o}hlinger and
Jeroen Franse for a careful reading of the manuscript. I also
acknowledge support from NWO Vidi grant 639.042.814.

\bibliographystyle{varenna}
\bibliography{hoekstra}

\begin{thebibliography}{100}
\expandafter\ifx\csname url\endcsname\relax\def\url#1{\texttt{#1}}\fi
\expandafter\ifx\csname urlprefix\endcsname\relax\def\urlprefix{URL }\fi

\bibitem{Walsh79}
\NAME{{Walsh} D., {Carswell} R.~F. \atque {Weymann} R.~J.},
  \IN{\nat}{279}{1979}{381}.

\bibitem{Soucail87}
\NAME{{Soucail} G., {Fort} B., {Mellier} Y. \atque {Picat} J.~P.},
  \IN{\aap}{172}{1987}{L14}.

\bibitem{Bartelmann01}
\NAME{{Bartelmann} M. \atque {Schneider} P.}, \IN{\physrep}{340}{2001}{291}.

\bibitem{Bartelmann10}
\NAME{{Bartelmann} M.}, \IN{Classical and Quantum Gravity}{27}{2010}{233001}.

\bibitem{Schneider06}
\NAME{{Schneider} P.}, \TITLE{{Part 3: Weak gravitational lensing}}, in proc.
  of \TITLE{Saas-Fee Advanced Course 33: Gravitational Lensing: Strong, Weak
  and Micro}, edited by \NAME{{G.~Meylan, P.~Jetzer, P.~North, P.~Schneider,
  C.~S.~Kochanek, \& J.~Wambsganss}} 2006, pp. 269--451.

\bibitem{Refsdal64}
\NAME{{Refsdal} S.}, \IN{\mnras}{128}{1964}{307}.

\bibitem{Kochanek02}
\NAME{{Kochanek} C.~S.}, \IN{\apj}{578}{2002}{25}.

\bibitem{Suyu13}
\NAME{{Suyu} S.~H., {Auger} M.~W., {Hilbert} S., {Marshall} P.~J., {Tewes} M.,
  {Treu} T., {Fassnacht} C.~D., {Koopmans} L.~V.~E., {Sluse} D., {Blandford}
  R.~D., {Courbin} F. \atque {Meylan} G.}, \IN{\apj}{766}{2013}{70}.

\bibitem{Schneider13}
\NAME{{Schneider} P. \atque {Sluse} D.}, \IN{ArXiv e-prints}{}{2013}{}.

\bibitem{Lewis06}
\NAME{{Lewis} A. \atque {Challinor} A.}, \IN{\physrep}{429}{2006}{1}.

\bibitem{Seljak96}
\NAME{{Seljak} U.}, \IN{\apj}{463}{1996}{1}.

\bibitem{Bernardeau97}
\NAME{{Bernardeau} F.}, \IN{\aap}{324}{1997}{15}.

\bibitem{SPTlens}
\NAME{{van Engelen} A., {Keisler} R., {Zahn} O., {Aird} K.~A., {Benson} B.~A.,
  {Bleem} L.~E., {Carlstrom} J.~E., {Chang} C.~L., {Cho} H.~M., {Crawford}
  T.~M., {Crites} A.~T., {de Haan} T., {Dobbs} M.~A., {Dudley} J., {George}
  E.~M., {Halverson} N.~W., {Holder} G.~P., {Holzapfel} W.~L., {Hoover} S.,
  {Hou} Z., {Hrubes} J.~D., {Joy} M., {Knox} L., {Lee} A.~T., {Leitch} E.~M.,
  {Lueker} M., {Luong-Van} D., {McMahon} J.~J., {Mehl} J., {Meyer} S.~S.,
  {Millea} M., {Mohr} J.~J., {Montroy} T.~E., {Natoli} T., {Padin} S., {Plagge}
  T., {Pryke} C., {Reichardt} C.~L., {Ruhl} J.~E., {Sayre} J.~T., {Schaffer}
  K.~K., {Shaw} L., {Shirokoff} E., {Spieler} H.~G., {Staniszewski} Z., {Stark}
  A.~A., {Story} K., {Vanderlinde} K., {Vieira} J.~D. \atque {Williamson} R.},
  \IN{\apj}{756}{2012}{142}.

\bibitem{ACTlens}
\NAME{{Das} S., {Sherwin} B.~D., {Aguirre} P., {Appel} J.~W., {Bond} J.~R.,
  {Carvalho} C.~S., {Devlin} M.~J., {Dunkley} J., {D{\"u}nner} R.,
  {Essinger-Hileman} T., {Fowler} J.~W., {Hajian} A., {Halpern} M.,
  {Hasselfield} M., {Hincks} A.~D., {Hlozek} R., {Huffenberger} K.~M., {Hughes}
  J.~P., {Irwin} K.~D., {Klein} J., {Kosowsky} A., {Lupton} R.~H., {Marriage}
  T.~A., {Marsden} D., {Menanteau} F., {Moodley} K., {Niemack} M.~D., {Nolta}
  M.~R., {Page} L.~A., {Parker} L., {Reese} E.~D., {Schmitt} B.~L., {Sehgal}
  N., {Sievers} J., {Spergel} D.~N., {Staggs} S.~T., {Swetz} D.~S., {Switzer}
  E.~R., {Thornton} R., {Visnjic} K. \atque {Wollack} E.}, \IN{Physical Review
  Letters}{107}{2011}{021301}.

\bibitem{Plancklens}
\NAME{{Planck Collaboration}, {Ade} P.~A.~R., {Aghanim} N., {Armitage-Caplan}
  C., {Arnaud} M., {Ashdown} M., {Atrio-Barandela} F., {Aumont} J.,
  {Baccigalupi} C., {Banday} A.~J. \atque et~al.}, \IN{ArXiv
  e-prints}{}{2013}{}.

\bibitem{Leauthaud07}
\NAME{{Leauthaud} A., {Massey} R., {Kneib} J.-P., {Rhodes} J., {Johnston}
  D.~E., {Capak} P., {Heymans} C., {Ellis} R.~S., {Koekemoer} A.~M., {Le
  F{\`e}vre} O., {Mellier} Y., {R{\'e}fr{\'e}gier} A., {Robin} A.~C.,
  {Scoville} N., {Tasca} L., {Taylor} J.~E. \atque {Van Waerbeke} L.},
  \IN{\apjs}{172}{2007}{219}.

\bibitem{Binggeli82}
\NAME{{Binggeli} B.}, \IN{\aap}{107}{1982}{338}.

\bibitem{Croft00}
\NAME{{Croft} R.~A.~C. \atque {Metzler} C.~A.}, \IN{\apj}{545}{2000}{561}.

\bibitem{Heavens00}
\NAME{{Heavens} A., {Refregier} A. \atque {Heymans} C.},
  \IN{\mnras}{319}{2000}{649}.

\bibitem{Mandelbaum06}
\NAME{{Mandelbaum} R., {Hirata} C.~M., {Ishak} M., {Seljak} U. \atque
  {Brinkmann} J.}, \IN{\mnras}{367}{2006}{611}.

\bibitem{Hirata07}
\NAME{{Hirata} C.~M., {Mandelbaum} R., {Ishak} M., {Seljak} U., {Nichol} R.,
  {Pimbblet} K.~A., {Ross} N.~P. \atque {Wake} D.},
  \IN{\mnras}{381}{2007}{1197}.

\bibitem{Joachimi11}
\NAME{{Joachimi} B., {Mandelbaum} R., {Abdalla} F.~B. \atque {Bridle} S.~L.},
  \IN{\aap}{527}{2011}{A26}.

\bibitem{Heymans13}
\NAME{{Heymans} C., {Grocutt} E., {Heavens} A., {Kilbinger} M., {Kitching}
  T.~D., {Simpson} F., {Benjamin} J., {Erben} T., {Hildebrandt} H., {Hoekstra}
  H., {Mellier} Y., {Miller} L., {Van Waerbeke} L., {Brown} M.~L., {Coupon} J.,
  {Fu} L., {Harnois-D{\'e}raps} J., {Hudson} M.~J., {Kuijken} K., {Rowe} B.,
  {Schrabback} T., {Semboloni} E., {Vafaei} S. \atque {Velander} M.},
  \IN{\mnras}{432}{2013}{2433}.

\bibitem{Pereira08}
\NAME{{Pereira} M.~J., {Bryan} G.~L. \atque {Gill} S.~P.~D.},
  \IN{\apj}{672}{2008}{825}.

\bibitem{Schneider13b}
\NAME{{Schneider} M.~D., {Cole} S., {Frenk} C.~S., {Kelvin} L., {Mandelbaum}
  R., {Norberg} P., {Bland-Hawthorn} J., {Brough} S., {Driver} S., {Hopkins}
  A., {Liske} J., {Loveday} J. \atque {Robotham} A.},
  \IN{\mnras}{433}{2013}{2727}.

\bibitem{Hirata04}
\NAME{{Hirata} C.~M. \atque {Seljak} U.}, \IN{\prd}{70}{2004}{063526}.

\bibitem{Bernstein09}
\NAME{{Bernstein} G.~M.}, \IN{\apj}{695}{2009}{652}.

\bibitem{Joachimi10}
\NAME{{Joachimi} B. \atque {Bridle} S.~L.}, \IN{\aap}{523}{2010}{A1}.

\bibitem{Hildebrandt09}
\NAME{{Hildebrandt} H., {van Waerbeke} L. \atque {Erben} T.},
  \IN{\aap}{507}{2009}{683}.

\bibitem{Hildebrandt11}
\NAME{{Hildebrandt} H., {Muzzin} A., {Erben} T., {Hoekstra} H., {Kuijken} K.,
  {Surace} J., {van Waerbeke} L., {Wilson} G. \atque {Yee} H.~K.~C.},
  \IN{\apjl}{733}{2011}{L30}.

\bibitem{Hildebrandt13}
\NAME{{Hildebrandt} H., {van Waerbeke} L., {Scott} D., {B{\'e}thermin} M.,
  {Bock} J., {Clements} D., {Conley} A., {Cooray} A., {Dunlop} J.~S., {Eales}
  S., {Erben} T., {Farrah} D., {Franceschini} A., {Glenn} J., {Halpern} M.,
  {Heinis} S., {Ivison} R.~J., {Marsden} G., {Oliver} S.~J., {Page} M.~J.,
  {P{\'e}rez-Fournon} I., {Smith} A.~J., {Rowan-Robinson} M., {Valtchanov} I.,
  {van der Burg} R.~F.~J., {Vieira} J.~D., {Viero} M. \atque {Wang} L.},
  \IN{\mnras}{429}{2013}{3230}.

\bibitem{KS93}
\NAME{{Kaiser} N. \atque {Squires} G.}, \IN{\apj}{404}{1993}{441}.

\bibitem{Gorenstein88}
\NAME{{Gorenstein} M.~V., {Shapiro} I.~I. \atque {Falco} E.~E.},
  \IN{\apj}{327}{1988}{693}.

\bibitem{Seitz96}
\NAME{{Seitz} S. \atque {Schneider} P.}, \IN{\aap}{305}{1996}{383}.

\bibitem{Squires96}
\NAME{{Squires} G. \atque {Kaiser} N.}, \IN{\apj}{473}{1996}{65}.

\bibitem{Bartelmann96}
\NAME{{Bartelmann} M.}, \IN{\aap}{313}{1996}{697}.

\bibitem{Seitz98}
\NAME{{Seitz} S., {Schneider} P. \atque {Bartelmann} M.},
  \IN{\aap}{337}{1998}{325}.

\bibitem{Waerbeke13}
\NAME{{Van Waerbeke} L., {Benjamin} J., {Erben} T., {Heymans} C., {Hildebrandt}
  H., {Hoekstra} H., {Kitching} T.~D., {Mellier} Y., {Miller} L., {Coupon} J.,
  {Harnois-D{\'e}raps} J., {Fu} L., {Hudson} M., {Kilbinger} M., {Kuijken} K.,
  {Rowe} B., {Schrabback} T., {Semboloni} E., {Vafaei} S., {van Uitert} E.
  \atque {Velander} M.}, \IN{\mnras}{433}{2013}{3373}.

\bibitem{Hoekstra00}
\NAME{{Hoekstra} H., {Franx} M. \atque {Kuijken} K.}, \IN{\apj}{532}{2000}{88}.

\bibitem{Clowe06}
\NAME{{Clowe} D., {Brada{\v c}} M., {Gonzalez} A.~H., {Markevitch} M.,
  {Randall} S.~W., {Jones} C. \atque {Zaritsky} D.},
  \IN{\apjl}{648}{2006}{L109}.

\bibitem{Clowe98}
\NAME{{Clowe} D., {Luppino} G.~A., {Kaiser} N., {Henry} J.~P. \atque {Gioia}
  I.~M.}, \IN{\apjl}{497}{1998}{L61}.

\bibitem{Massey13}
\NAME{{Massey} R., {Hoekstra} H., {Kitching} T., {Rhodes} J., {Cropper} M.,
  {Amiaux} J., {Harvey} D., {Mellier} Y., {Meneghetti} M., {Miller} L.,
  {Paulin-Henriksson} S., {Pires} S., {Scaramella} R. \atque {Schrabback} T.},
  \IN{\mnras}{429}{2013}{661}.

\bibitem{Miller13}
\NAME{{Miller} L., {Heymans} C., {Kitching} T.~D., {van Waerbeke} L., {Erben}
  T., {Hildebrandt} H., {Hoekstra} H., {Mellier} Y., {Rowe} B.~T.~P., {Coupon}
  J., {Dietrich} J.~P., {Fu} L., {Harnois-D{\'e}raps} J., {Hudson} M.~J.,
  {Kilbinger} M., {Kuijken} K., {Schrabback} T., {Semboloni} E., {Vafaei} S.
  \atque {Velander} M.}, \IN{\mnras}{429}{2013}{2858}.

\bibitem{Zuntz13}
\NAME{{Zuntz} J., {Kacprzak} T., {Voigt} L., {Hirsch} M., {Rowe} B. \atque
  {Bridle} S.}, \IN{\mnras}{434}{2013}{1604}.

\bibitem{Bernstein10}
\NAME{{Bernstein} G.~M.}, \IN{\mnras}{406}{2010}{2793}.

\bibitem{KSB}
\NAME{{Kaiser} N., {Squires} G. \atque {Broadhurst} T.},
  \IN{\apj}{449}{1995}{460}.

\bibitem{STEP1}
\NAME{{Heymans} C., {Van Waerbeke} L., {Bacon} D., {Berge} J., {Bernstein} G.,
  {Bertin} E., {Bridle} S., {Brown} M.~L., {Clowe} D., {Dahle} H., {Erben} T.,
  {Gray} M., {Hetterscheidt} M., {Hoekstra} H., {Hudelot} P., {Jarvis} M.,
  {Kuijken} K., {Margoniner} V., {Massey} R., {Mellier} Y., {Nakajima} R.,
  {Refregier} A., {Rhodes} J., {Schrabback} T. \atque {Wittman} D.},
  \IN{\mnras}{368}{2006}{1323}.

\bibitem{Amara07}
\NAME{{Amara} A. \atque {R{\'e}fr{\'e}gier} A.}, \IN{\mnras}{381}{2007}{1018}.

\bibitem{Semboloni13a}
\NAME{{Semboloni} E., {Hoekstra} H., {Huang} Z., {Cardone} V.~F., {Cropper} M.,
  {Joachimi} B., {Kitching} T., {Kuijken} K., {Lombardi} M., {Maoli} R.,
  {Mellier} Y., {Miller} L., {Rhodes} J., {Scaramella} R., {Schrabback} T.
  \atque {Velander} M.}, \IN{\mnras}{432}{2013}{2385}.

\bibitem{Hoekstra02}
\NAME{{Hoekstra} H., {Yee} H.~K.~C., {Gladders} M.~D., {Barrientos} L.~F.,
  {Hall} P.~B. \atque {Infante} L.}, \IN{\apj}{572}{2002}{55}.

\bibitem{Hoekstra06}
\NAME{{Hoekstra} H., {Mellier} Y., {van Waerbeke} L., {Semboloni} E., {Fu} L.,
  {Hudson} M.~J., {Parker} L.~C., {Tereno} I. \atque {Benabed} K.},
  \IN{\apj}{647}{2006}{116}.

\bibitem{Hoekstra04a}
\NAME{{Hoekstra} H.}, \IN{\mnras}{347}{2004}{1337}.

\bibitem{Hamana13}
\NAME{{Hamana} T., {Miyazaki} S., {Okura} Y., {Okamura} T. \atque {Futamase}
  T.}, \IN{ArXiv e-prints}{}{2013}{}.

\bibitem{Schechter11}
\NAME{{Schechter} P.~L. \atque {Levinson} R.~S.}, \IN{\pasp}{123}{2011}{812}.

\bibitem{Jee13}
\NAME{{Jee} M.~J., {Tyson} J.~A., {Schneider} M.~D., {Wittman} D., {Schmidt} S.
  \atque {Hilbert} S.}, \IN{\apj}{765}{2013}{74}.

\bibitem{Heymans12}
\NAME{{Heymans} C., {Van Waerbeke} L., {Miller} L., {Erben} T., {Hildebrandt}
  H., {Hoekstra} H., {Kitching} T.~D., {Mellier} Y., {Simon} P., {Bonnett} C.,
  {Coupon} J., {Fu} L., {Harnois-D'eraps} J., {Hudson} M.~J., {Kilbinger} M.,
  {Kuijken} K., {Rowe} B., {Schrabback} T., {Semboloni} E., {van Uitert} E.,
  {Vafaei} S. \atque {Velander} M.}, \IN{ArXiv e-prints}{}{2012}{}.

\bibitem{STEP2}
\NAME{{Massey} R., {Heymans} C., {Berg{\'e}} J., {Bernstein} G., {Bridle} S.,
  {Clowe} D., {Dahle} H., {Ellis} R., {Erben} T., {Hetterscheidt} M., {High}
  F.~W., {Hirata} C., {Hoekstra} H., {Hudelot} P., {Jarvis} M., {Johnston} D.,
  {Kuijken} K., {Margoniner} V., {Mandelbaum} R., {Mellier} Y., {Nakajima} R.,
  {Paulin-Henriksson} S., {Peeples} M., {Roat} C., {Refregier} A., {Rhodes} J.,
  {Schrabback} T., {Schirmer} M., {Seljak} U., {Semboloni} E. \atque {van
  Waerbeke} L.}, \IN{\mnras}{376}{2007}{13}.

\bibitem{GREAT08}
\NAME{{Bridle} S., {Balan} S.~T., {Bethge} M., {Gentile} M., {Harmeling} S.,
  {Heymans} C., {Hirsch} M., {Hosseini} R., {Jarvis} M., {Kirk} D., {Kitching}
  T., {Kuijken} K., {Lewis} A., {Paulin-Henriksson} S., {Sch{\"o}lkopf} B.,
  {Velander} M., {Voigt} L., {Witherick} D., {Amara} A., {Bernstein} G.,
  {Courbin} F., {Gill} M., {Heavens} A., {Mandelbaum} R., {Massey} R.,
  {Moghaddam} B., {Rassat} A., {R{\'e}fr{\'e}gier} A., {Rhodes} J.,
  {Schrabback} T., {Shawe-Taylor} J., {Shmakova} M., {van Waerbeke} L. \atque
  {Wittman} D.}, \IN{\mnras}{405}{2010}{2044}.

\bibitem{GREAT10}
\NAME{{Kitching} T.~D., {Balan} S.~T., {Bridle} S., {Cantale} N., {Courbin} F.,
  {Eifler} T., {Gentile} M., {Gill} M.~S.~S., {Harmeling} S., {Heymans} C.,
  {Hirsch} M., {Honscheid} K., {Kacprzak} T., {Kirkby} D., {Margala} D.,
  {Massey} R.~J., {Melchior} P., {Nurbaeva} G., {Patton} K., {Rhodes} J.,
  {Rowe} B.~T.~P., {Taylor} A.~N., {Tewes} M., {Viola} M., {Witherick} D.,
  {Voigt} L., {Young} J. \atque {Zuntz} J.}, \IN{\mnras}{423}{2012}{3163}.

\bibitem{Melchior12}
\NAME{{Melchior} P. \atque {Viola} M.}, \IN{\mnras}{424}{2012}{2757}.

\bibitem{Tyson84}
\NAME{{Tyson} J.~A., {Valdes} F., {Jarvis} J.~F. \atque {Mills}, Jr. A.~P.},
  \IN{\apjl}{281}{1984}{L59}.

\bibitem{Tyson90}
\NAME{{Tyson} J.~A., {Wenk} R.~A. \atque {Valdes} F.},
  \IN{\apjl}{349}{1990}{L1}.

\bibitem{Hoekstra13}
\NAME{{Hoekstra} H., {Bartelmann} M., {Dahle} H., {Israel} H., {Limousin} M.
  \atque {Meneghetti} M.}, \IN{\ssr}{177}{2013}{75}.

\bibitem{Mahdavi13}
\NAME{{Mahdavi} A., {Hoekstra} H., {Babul} A., {Bildfell} C., {Jeltema} T.
  \atque {Henry} J.~P.}, \IN{\apj}{767}{2013}{116}.

\bibitem{Leauthaud10}
\NAME{{Leauthaud} A., {Finoguenov} A., {Kneib} J.-P., {Taylor} J.~E., {Massey}
  R., {Rhodes} J., {Ilbert} O., {Bundy} K., {Tinker} J., {George} M.~R.,
  {Capak} P., {Koekemoer} A.~M., {Johnston} D.~E., {Zhang} Y.-Y., {Cappelluti}
  N., {Ellis} R.~S., {Elvis} M., {Giodini} S., {Heymans} C., {Le F{\`e}vre} O.,
  {Lilly} S., {McCracken} H.~J., {Mellier} Y., {R{\'e}fr{\'e}gier} A.,
  {Salvato} M., {Scoville} N., {Smoot} G., {Tanaka} M., {Van Waerbeke} L.
  \atque {Wolk} M.}, \IN{\apj}{709}{2010}{97}.

\bibitem{Hoekstra12}
\NAME{{Hoekstra} H., {Mahdavi} A., {Babul} A. \atque {Bildfell} C.}, \IN{ArXiv
  e-prints}{}{2012}{}.

\bibitem{Okabe10}
\NAME{{Okabe} N., {Takada} M., {Umetsu} K., {Futamase} T. \atque {Smith}
  G.~P.}, \IN{\pasj}{62}{2010}{811}.

\bibitem{Okabe13}
\NAME{{Okabe} N., {Smith} G.~P., {Umetsu} K., {Takada} M. \atque {Futamase}
  T.}, \IN{\apjl}{769}{2013}{L35}.

\bibitem{Applegate12}
\NAME{{Applegate} D.~E., {von der Linden} A., {Kelly} P.~L., {Allen} M.~T.,
  {Allen} S.~W., {Burchat} P.~R., {Burke} D.~L., {Ebeling} H., {Mantz} A.
  \atque {Morris} R.~G.}, \IN{ArXiv e-prints}{}{2012}{}.

\bibitem{Kravtsov06}
\NAME{{Kravtsov} A.~V., {Vikhlinin} A. \atque {Nagai} D.},
  \IN{\apj}{650}{2006}{128}.

\bibitem{Johnston07}
\NAME{{Johnston} D.~E., {Sheldon} E.~S., {Wechsler} R.~H., {Rozo} E., {Koester}
  B.~P., {Frieman} J.~A., {McKay} T.~A., {Evrard} A.~E., {Becker} M.~R. \atque
  {Annis} J.}, \IN{ArXiv e-prints}{}{2007}{}.

\bibitem{Courteau13}
\NAME{{Courteau} S., {Cappellari} M., {de Jong} R.~S., {Dutton} A.~A.,
  {Emsellem} E., {Hoekstra} H., {Koopmans} L.~V.~E., {Mamon} G.~A., {Maraston}
  C., {Treu} T. \atque {Widrow} L.~M.}, \IN{ArXiv e-prints}{}{2013}{}.

\bibitem{Brainerd96}
\NAME{{Brainerd} T.~G., {Blandford} R.~D. \atque {Smail} I.},
  \IN{\apj}{466}{1996}{623}.

\bibitem{Mckay01}
\NAME{{McKay} T.~A., {Sheldon} E.~S., {Racusin} J., {Fischer} P., {Seljak} U.,
  {Stebbins} A., {Johnston} D., {Frieman} J.~A., {Bahcall} N., {Brinkmann} J.,
  {Csabai} I., {Fukugita} M., {Hennessy} G.~S., {Ivezic} Z., {Lamb} D.~Q.,
  {Loveday} J., {Lupton} R.~H., {Munn} J.~A., {Nichol} R.~C., {Pier} J.~R.
  \atque {York} D.~G.}, \IN{ArXiv Astrophysics e-prints}{}{2001}{}.

\bibitem{Mandelbaum05}
\NAME{{Mandelbaum} R., {Hirata} C.~M., {Seljak} U., {Guzik} J., {Padmanabhan}
  N., {Blake} C., {Blanton} M.~R., {Lupton} R. \atque {Brinkmann} J.},
  \IN{\mnras}{361}{2005}{1287}.

\bibitem{vanUitert11}
\NAME{{van Uitert} E., {Hoekstra} H., {Velander} M., {Gilbank} D.~G.,
  {Gladders} M.~D. \atque {Yee} H.~K.~C.}, \IN{\aap}{534}{2011}{A14}.

\bibitem{Schneider97}
\NAME{{Schneider} P. \atque {Rix} H.-W.}, \IN{\apj}{474}{1997}{25}.

\bibitem{Hoekstra04b}
\NAME{{Hoekstra} H., {Yee} H.~K.~C. \atque {Gladders} M.~D.},
  \IN{\apj}{606}{2004}{67}.

\bibitem{Hoekstra05}
\NAME{{Hoekstra} H., {Hsieh} B.~C., {Yee} H.~K.~C., {Lin} H. \atque {Gladders}
  M.~D.}, \IN{\apj}{635}{2005}{73}.

\bibitem{Hoekstra02b}
\NAME{{Hoekstra} H., {van Waerbeke} L., {Gladders} M.~D., {Mellier} Y. \atque
  {Yee} H.~K.~C.}, \IN{\apj}{577}{2002}{604}.

\bibitem{Sheldon04}
\NAME{{Sheldon} E.~S., {Johnston} D.~E., {Frieman} J.~A., {Scranton} R.,
  {McKay} T.~A., {Connolly} A.~J., {Budav{\'a}ri} T., {Zehavi} I., {Bahcall}
  N.~A., {Brinkmann} J. \atque {Fukugita} M.}, \IN{\aj}{127}{2004}{2544}.

\bibitem{Reyes10}
\NAME{{Reyes} R., {Mandelbaum} R., {Seljak} U., {Baldauf} T., {Gunn} J.~E.,
  {Lombriser} L. \atque {Smith} R.~E.}, \IN{\nat}{464}{2010}{256}.

\bibitem{Guzik01}
\NAME{{Guzik} J. \atque {Seljak} U.}, \IN{\mnras}{321}{2001}{439}.

\bibitem{Seljak00}
\NAME{{Seljak} U.}, \IN{\mnras}{318}{2000}{203}.

\bibitem{Cooray02}
\NAME{{Cooray} A. \atque {Sheth} R.}, \IN{\physrep}{372}{2002}{1}.

\bibitem{Limousin07}
\NAME{{Limousin} M., {Kneib} J.~P., {Bardeau} S., {Natarajan} P., {Czoske} O.,
  {Smail} I., {Ebeling} H. \atque {Smith} G.~P.}, \IN{\aap}{461}{2007}{881}.

\bibitem{Cacciato13}
\NAME{{Cacciato} M., {van Uitert} E. \atque {Hoekstra} H.}, \IN{ArXiv
  e-prints}{}{2013}{}.

\bibitem{Tasitsiomi04}
\NAME{{Tasitsiomi} A., {Kravtsov} A.~V., {Wechsler} R.~H. \atque {Primack}
  J.~R.}, \IN{\apj}{614}{2004}{533}.

\bibitem{Velander13}
\NAME{{Velander} M., {van Uitert} E., {Hoekstra} H., {Coupon} J., {Erben} T.,
  {Heymans} C., {Hildebrandt} H., {Kitching} T.~D., {Mellier} Y., {Miller} L.,
  {Van Waerbeke} L., {Bonnett} C., {Fu} L., {Giodini} S., {Hudson} M.~J.,
  {Kuijken} K., {Rowe} B., {Schrabback} T. \atque {Semboloni} E.}, \IN{ArXiv
  e-prints}{}{2013}{}.

\bibitem{Leauthaud12}
\NAME{{Leauthaud} A., {Tinker} J., {Bundy} K., {Behroozi} P.~S., {Massey} R.,
  {Rhodes} J., {George} M.~R., {Kneib} J.-P., {Benson} A., {Wechsler} R.~H.,
  {Busha} M.~T., {Capak} P., {Cort{\^e}s} M., {Ilbert} O., {Koekemoer} A.~M.,
  {Le F{\`e}vre} O., {Lilly} S., {McCracken} H.~J., {Salvato} M., {Schrabback}
  T., {Scoville} N., {Smith} T. \atque {Taylor} J.~E.},
  \IN{\apj}{744}{2012}{159}.

\bibitem{Dubinski91}
\NAME{{Dubinski} J. \atque {Carlberg} R.~G.}, \IN{\apj}{378}{1991}{496}.

\bibitem{Jing02}
\NAME{{Jing} Y.~P. \atque {Suto} Y.}, \IN{\apj}{574}{2002}{538}.

\bibitem{Hayashi07}
\NAME{{Hayashi} E., {Navarro} J.~F. \atque {Springel} V.},
  \IN{\mnras}{377}{2007}{50}.

\bibitem{Kazantzidis04}
\NAME{{Kazantzidis} S., {Kravtsov} A.~V., {Zentner} A.~R., {Allgood} B.,
  {Nagai} D. \atque {Moore} B.}, \IN{\apjl}{611}{2004}{L73}.

\bibitem{vanUitert12}
\NAME{{van Uitert} E., {Hoekstra} H., {Schrabback} T., {Gilbank} D.~G.,
  {Gladders} M.~D. \atque {Yee} H.~K.~C.}, \IN{\aap}{545}{2012}{A71}.

\bibitem{Hoekstra04}
\NAME{{Hoekstra} H., {Yee} H.~K.~C. \atque {Gladders} M.~D.},
  \IN{\apj}{606}{2004}{67}.

\bibitem{Mandelbaum06a}
\NAME{{Mandelbaum} R., {Hirata} C.~M., {Broderick} T., {Seljak} U. \atque
  {Brinkmann} J.}, \IN{\mnras}{370}{2006}{1008}.

\bibitem{Parker07}
\NAME{{Parker} L.~C., {Hoekstra} H., {Hudson} M.~J., {van Waerbeke} L. \atque
  {Mellier} Y.}, \IN{\apj}{669}{2007}{21}.

\bibitem{Bacon00}
\NAME{{Bacon} D.~J., {Refregier} A.~R. \atque {Ellis} R.~S.},
  \IN{\mnras}{318}{2000}{625}.

\bibitem{Kaiser00}
\NAME{{Kaiser} N., {Wilson} G. \atque {Luppino} G.~A.}, \IN{ArXiv Astrophysics
  e-prints}{}{2000}{}.

\bibitem{Waerbeke00}
\NAME{{Van Waerbeke} L., {Mellier} Y., {Erben} T., {Cuillandre} J.~C.,
  {Bernardeau} F., {Maoli} R., {Bertin} E., {McCracken} H.~J., {Le F{\`e}vre}
  O., {Fort} B., {Dantel-Fort} M., {Jain} B. \atque {Schneider} P.},
  \IN{\aap}{358}{2000}{30}.

\bibitem{Wittman00}
\NAME{{Wittman} D.~M., {Tyson} J.~A., {Kirkman} D., {Dell'Antonio} I. \atque
  {Bernstein} G.}, \IN{\nat}{405}{2000}{143}.

\bibitem{Euclid}
\NAME{{Laureijs} R., {Amiaux} J., {Arduini} S., {Augu{\`e}res} J.~.,
  {Brinchmann} J., {Cole} R., {Cropper} M., {Dabin} C., {Duvet} L., {Ealet} A.
  \atque et~al.}, \IN{ArXiv e-prints}{}{2011}{}.

\bibitem{Hoekstra08}
\NAME{{Hoekstra} H. \atque {Jain} B.}, \IN{Annual Review of Nuclear and
  Particle Science}{58}{2008}{99}.

\bibitem{Munshi08}
\NAME{{Munshi} D., {Valageas} P., {van Waerbeke} L. \atque {Heavens} A.},
  \IN{\physrep}{462}{2008}{67}.

\bibitem{Hilbert09}
\NAME{{Hilbert} S., {Hartlap} J., {White} S.~D.~M. \atque {Schneider} P.},
  \IN{\aap}{499}{2009}{31}.

\bibitem{Kilbinger13}
\NAME{{Kilbinger} M., {Fu} L., {Heymans} C., {Simpson} F., {Benjamin} J.,
  {Erben} T., {Harnois-D{\'e}raps} J., {Hoekstra} H., {Hildebrandt} H.,
  {Kitching} T.~D., {Mellier} Y., {Miller} L., {Van Waerbeke} L., {Benabed} K.,
  {Bonnett} C., {Coupon} J., {Hudson} M.~J., {Kuijken} K., {Rowe} B.,
  {Schrabback} T., {Semboloni} E., {Vafaei} S. \atque {Velander} M.},
  \IN{\mnras}{430}{2013}{2200}.

\bibitem{Crittenden02}
\NAME{{Crittenden} R.~G., {Natarajan} P., {Pen} U.-L. \atque {Theuns} T.},
  \IN{\apj}{568}{2002}{20}.

\bibitem{Schneider02}
\NAME{{Schneider} P., {van Waerbeke} L. \atque {Mellier} Y.},
  \IN{\aap}{389}{2002}{729}.

\bibitem{Schneider98}
\NAME{{Schneider} P., {van Waerbeke} L., {Jain} B. \atque {Kruse} G.},
  \IN{\mnras}{296}{1998}{873}.

\bibitem{Benjamin07}
\NAME{{Benjamin} J., {Heymans} C., {Semboloni} E., {van Waerbeke} L.,
  {Hoekstra} H., {Erben} T., {Gladders} M.~D., {Hetterscheidt} M., {Mellier} Y.
  \atque {Yee} H.~K.~C.}, \IN{\mnras}{381}{2007}{702}.

\bibitem{Newman13}
\NAME{{Newman} J., {Abate} A., {Abdalla} F., {Allam} S., {Allen} S., {Ansari}
  R., {Bailey} S., {Barkhouse} W., {Beers} T., {Blanton} M., {Brodwin} M.,
  {Brownstein} J., {Brunner} R., {Carrasco-Kind} M., {Cervantes-Cota} J.,
  {Chisari} E., {Colless} M., {Comparat} J., {Coupon} J., {Cheu} E., {Cunha}
  C., {de la Macorra} A., {Dell'Antonio} I., {Frye} B., {Gawiser} E., {Gehrels}
  N., {Grady} K., {Hagen} A., {Hall} P., {Hearin} A., {Hildebrandt} H.,
  {Hirata} C., {Ho} S., {Honscheid} K., {Huterer} D., {Ivezic} Z., {Kneib}
  J.-P., {Kruk} J., {Lahav} O., {Mandelbaum} R., {Marshall} J., {Matthews} D.,
  {M{\'e}nard} B., {Miquel} R., {Moniez} M., {Moos} W., {Moustakas} J.,
  {Papovich} C., {Peacock} J., {Park} C., {Rhodes} J., {Ricol} J., {Sadeh} I.,
  {Slozar} A., {Schmidt} S., {Stern} D., {Tyson} T., {von der Linden} A.,
  {Wechsler} R., {Wood-Vasey} W. \atque {Zentner} A.}, \IN{ArXiv
  e-prints}{}{2013}{}.

\bibitem{Schrabback10}
\NAME{{Schrabback} T., {Hartlap} J., {Joachimi} B., {Kilbinger} M., {Simon} P.,
  {Benabed} K., {Brada{\v c}} M., {Eifler} T., {Erben} T., {Fassnacht} C.~D.,
  {High} F.~W., {Hilbert} S., {Hildebrandt} H., {Hoekstra} H., {Kuijken} K.,
  {Marshall} P.~J., {Mellier} Y., {Morganson} E., {Schneider} P., {Semboloni}
  E., {van Waerbeke} L. \atque {Velander} M.}, \IN{\aap}{516}{2010}{A63}.

\bibitem{Waerbeke05}
\NAME{{Van Waerbeke} L., {Mellier} Y. \atque {Hoekstra} H.},
  \IN{\aap}{429}{2005}{75}.

\bibitem{Hu99}
\NAME{{Hu} W.}, \IN{\apjl}{522}{1999}{L21}.

\bibitem{Ma06}
\NAME{{Ma} Z., {Hu} W. \atque {Huterer} D.}, \IN{\apj}{636}{2006}{21}.

\bibitem{Bernardeau02}
\NAME{{Bernardeau} F., {Colombi} S., {Gazta{\~n}aga} E. \atque {Scoccimarro}
  R.}, \IN{\physrep}{367}{2002}{1}.

\bibitem{Crocce06}
\NAME{{Crocce} M. \atque {Scoccimarro} R.}, \IN{\prd}{73}{2006}{063519}.

\bibitem{Crocce12}
\NAME{{Crocce} M., {Scoccimarro} R. \atque {Bernardeau} F.},
  \IN{\mnras}{427}{2012}{2537}.

\bibitem{Taruya12}
\NAME{{Taruya} A., {Bernardeau} F., {Nishimichi} T. \atque {Codis} S.},
  \IN{\prd}{86}{2012}{103528}.

\bibitem{Huterer05}
\NAME{{Huterer} D. \atque {Takada} M.}, \IN{Astroparticle
  Physics}{23}{2005}{369}.

\bibitem{Semboloni11b}
\NAME{{Semboloni} E., {Hoekstra} H., {Schaye} J., {van Daalen} M.~P. \atque
  {McCarthy} I.~G.}, \IN{\mnras}{417}{2011}{2020}.

\bibitem{Castro05}
\NAME{{Castro} P.~G., {Heavens} A.~F. \atque {Kitching} T.~D.},
  \IN{\prd}{72}{2005}{023516}.

\bibitem{Kitching11b}
\NAME{{Kitching} T.~D., {Heavens} A.~F. \atque {Miller} L.},
  \IN{\mnras}{413}{2011}{2923}.

\bibitem{Heitmann10}
\NAME{{Heitmann} K., {White} M., {Wagner} C., {Habib} S. \atque {Higdon} D.},
  \IN{\apj}{715}{2010}{104}.

\bibitem{PD96}
\NAME{{Peacock} J.~A. \atque {Dodds} S.~J.}, \IN{\mnras}{280}{1996}{L19}.

\bibitem{Smith03}
\NAME{{Smith} R.~E., {Peacock} J.~A., {Jenkins} A., {White} S.~D.~M., {Frenk}
  C.~S., {Pearce} F.~R., {Thomas} P.~A., {Efstathiou} G. \atque {Couchman}
  H.~M.~P.}, \IN{\mnras}{341}{2003}{1311}.

\bibitem{Lawrence10}
\NAME{{Lawrence} E., {Heitmann} K., {White} M., {Higdon} D., {Wagner} C.,
  {Habib} S. \atque {Williams} B.}, \IN{\apj}{713}{2010}{1322}.

\bibitem{Semboloni07}
\NAME{{Semboloni} E., {van Waerbeke} L., {Heymans} C., {Hamana} T., {Colombi}
  S., {White} M. \atque {Mellier} Y.}, \IN{\mnras}{375}{2007}{L6}.

\bibitem{Kiessling11}
\NAME{{Kiessling} A., {Taylor} A.~N. \atque {Heavens} A.~F.},
  \IN{\mnras}{416}{2011}{1045}.

\bibitem{Hartlap07}
\NAME{{Hartlap} J., {Simon} P. \atque {Schneider} P.},
  \IN{\aap}{464}{2007}{399}.

\bibitem{Springel05}
\NAME{{Springel} V., {White} S.~D.~M., {Jenkins} A., {Frenk} C.~S., {Yoshida}
  N., {Gao} L., {Navarro} J., {Thacker} R., {Croton} D., {Helly} J., {Peacock}
  J.~A., {Cole} S., {Thomas} P., {Couchman} H., {Evrard} A., {Colberg} J.
  \atque {Pearce} F.}, \IN{\nat}{435}{2005}{629}.

\bibitem{Crocce10}
\NAME{{Crocce} M., {Fosalba} P., {Castander} F.~J. \atque {Gazta{\~n}aga} E.},
  \IN{\mnras}{403}{2010}{1353}.

\bibitem{vanDaalen11}
\NAME{{van Daalen} M.~P., {Schaye} J., {Booth} C.~M. \atque {Dalla Vecchia}
  C.}, \IN{\mnras}{415}{2011}{3649}.

\bibitem{McCarthy10}
\NAME{{McCarthy} I.~G., {Schaye} J., {Ponman} T.~J., {Bower} R.~G., {Booth}
  C.~M., {Dalla Vecchia} C., {Crain} R.~A., {Springel} V., {Theuns} T. \atque
  {Wiersma} R.~P.~C.}, \IN{\mnras}{406}{2010}{822}.

\bibitem{Schaye10}
\NAME{{Schaye} J., {Dalla Vecchia} C., {Booth} C.~M., {Wiersma} R.~P.~C.,
  {Theuns} T., {Haas} M.~R., {Bertone} S., {Duffy} A.~R., {McCarthy} I.~G.
  \atque {van de Voort} F.}, \IN{\mnras}{402}{2010}{1536}.

\bibitem{Kitching11a}
\NAME{{Kitching} T.~D. \atque {Taylor} A.~N.}, \IN{\mnras}{416}{2011}{1717}.

\bibitem{Zentner08}
\NAME{{Zentner} A.~R., {Rudd} D.~H. \atque {Hu} W.},
  \IN{\prd}{77}{2008}{043507}.

\bibitem{Waerbeke99}
\NAME{{van Waerbeke} L., {Bernardeau} F. \atque {Mellier} Y.},
  \IN{\aap}{342}{1999}{15}.

\bibitem{Vafaei10}
\NAME{{Vafaei} S., {Lu} T., {van Waerbeke} L., {Semboloni} E., {Heymans} C.
  \atque {Pen} U.-L.}, \IN{Astroparticle Physics}{32}{2010}{340}.

\bibitem{Semboloni11a}
\NAME{{Semboloni} E., {Schrabback} T., {van Waerbeke} L., {Vafaei} S.,
  {Hartlap} J. \atque {Hilbert} S.}, \IN{\mnras}{410}{2011}{143}.

\bibitem{Semboloni13b}
\NAME{{Semboloni} E., {Hoekstra} H. \atque {Schaye} J.},
  \IN{\mnras}{434}{2013}{148}.

\bibitem{Huterer06}
\NAME{{Huterer} D., {Takada} M., {Bernstein} G. \atque {Jain} B.},
  \IN{\mnras}{366}{2006}{101}.

\bibitem{Erben13}
\NAME{{Erben} T., {Hildebrandt} H., {Miller} L., {van Waerbeke} L., {Heymans}
  C., {Hoekstra} H., {Kitching} T.~D., {Mellier} Y., {Benjamin} J., {Blake} C.,
  {Bonnett} C., {Cordes} O., {Coupon} J., {Fu} L., {Gavazzi} R., {Gillis} B.,
  {Grocutt} E., {Gwyn} S.~D.~J., {Holhjem} K., {Hudson} M.~J., {Kilbinger} M.,
  {Kuijken} K., {Milkeraitis} M., {Rowe} B.~T.~P., {Schrabback} T., {Semboloni}
  E., {Simon} P., {Smit} M., {Toader} O., {Vafaei} S., {van Uitert} E. \atque
  {Velander} M.}, \IN{\mnras}{433}{2013}{2545}.

\bibitem{Hildebrandt12}
\NAME{{Hildebrandt} H., {Erben} T., {Kuijken} K., {van Waerbeke} L., {Heymans}
  C., {Coupon} J., {Benjamin} J., {Bonnett} C., {Fu} L., {Hoekstra} H.,
  {Kitching} T.~D., {Mellier} Y., {Miller} L., {Velander} M., {Hudson} M.~J.,
  {Rowe} B.~T.~P., {Schrabback} T., {Semboloni} E. \atque {Ben{\'{\i}}tez} N.},
  \IN{\mnras}{421}{2012}{2355}.

\bibitem{Benjamin13}
\NAME{{Benjamin} J., {Van Waerbeke} L., {Heymans} C., {Kilbinger} M., {Erben}
  T., {Hildebrandt} H., {Hoekstra} H., {Kitching} T.~D., {Mellier} Y., {Miller}
  L., {Rowe} B., {Schrabback} T., {Simpson} F., {Coupon} J., {Fu} L.,
  {Harnois-D{\'e}raps} J., {Hudson} M.~J., {Kuijken} K., {Semboloni} E.,
  {Vafaei} S. \atque {Velander} M.}, \IN{\mnras}{431}{2013}{1547}.

\bibitem{Simpson13}
\NAME{{Simpson} F., {Heymans} C., {Parkinson} D., {Blake} C., {Kilbinger} M.,
  {Benjamin} J., {Erben} T., {Hildebrandt} H., {Hoekstra} H., {Kitching} T.~D.,
  {Mellier} Y., {Miller} L., {Van Waerbeke} L., {Coupon} J., {Fu} L.,
  {Harnois-D{\'e}raps} J., {Hudson} M.~J., {Kuijken} K., {Rowe} B.,
  {Schrabback} T., {Semboloni} E., {Vafaei} S. \atque {Velander} M.},
  \IN{\mnras}{429}{2013}{2249}.

\bibitem{Blake12}
\NAME{{Blake} C., {Brough} S., {Colless} M., {Contreras} C., {Couch} W.,
  {Croom} S., {Croton} D., {Davis} T.~M., {Drinkwater} M.~J., {Forster} K.,
  {Gilbank} D., {Gladders} M., {Glazebrook} K., {Jelliffe} B., {Jurek} R.~J.,
  {Li} I.-h., {Madore} B., {Martin} D.~C., {Pimbblet} K., {Poole} G.~B.,
  {Pracy} M., {Sharp} R., {Wisnioski} E., {Woods} D., {Wyder} T.~K. \atque
  {Yee} H.~K.~C.}, \IN{\mnras}{425}{2012}{405}.

\bibitem{Choi12}
\NAME{{Choi} A., {Tyson} J.~A., {Morrison} C.~B., {Jee} M.~J., {Schmidt} S.~J.,
  {Margoniner} V.~E. \atque {Wittman} D.~M.}, \IN{\apj}{759}{2012}{101}.

\bibitem{LSST}
\NAME{{LSST Science Collaboration}, {Abell} P.~A., {Allison} J., {Anderson}
  S.~F., {Andrew} J.~R., {Angel} J.~R.~P., {Armus} L., {Arnett} D., {Asztalos}
  S.~J., {Axelrod} T.~S. \atque et~al.}, \IN{ArXiv e-prints}{}{2009}{}.

\bibitem{Cropper13}
\NAME{{Cropper} M., {Hoekstra} H., {Kitching} T., {Massey} R., {Amiaux} J.,
  {Miller} L., {Mellier} Y., {Rhodes} J., {Rowe} B., {Pires} S., {Saxton} C.
  \atque {Scaramella} R.}, \IN{\mnras}{431}{2013}{3103}.

\end{thebibliography}


\end{document}